\documentclass[10point]{article}
\usepackage{amsmath} 
\usepackage{amssymb}  
\usepackage{mathrsfs} 
\usepackage{stmaryrd} 
\usepackage{psfig} 

\title{\LARGE \bf A Predictive Theory of Games }

\author{David H. Wolpert
\thanks{D. Wolpert is with NASA Ames Research Center, Moffett Field, CA, 94035
        {\tt\small dhw@ptolemy.arc.nasa.gov}}%
}

\begin{document}

\maketitle

\begin{abstract}
Conventional noncooperative game theory hypothesizes that the joint
(mixed) strategy of a set of reasoning players in a game will
necessarily satisfy an ``equilibrium concept''. The number of joint
strategies satisfying that equilibrium concept has measure zero, and
all other joint strategies are considered impossible. Under this
hypothesis the only issue is what equilibrium concept is ``correct''.
 
This hypothesis violates the first-principles arguments underlying
probability theory. Indeed, probability theory renders moot the
controversy over what equilibrium concept is correct --- while in
general there are joint (mixed) strategies with zero probability, in
general the set \{strategies with non-zero probability\} has measure
greater than zero.  Rather than a first-principles derivation of an
equilibrium concept, game theory requires a first-principles
derivation of a distribution over joint strategies.

However say one wishes to predict a single joint strategy from that
distribution. Then decision theory tell us to first specify a loss
function, a function which concerns how we, the analyst/scientist
$external$ to the game, will use that prediction. We then predict that
the game will result in the joint strategy that is Bayes-optimal for
that loss function and distribution over joint strategies. Different
loss functions --- different uses of the prediction --- give different
such optimal predictions. There is no more role for an ``equilibrium
concept'' that is independent of the distribution and choice of loss
function.  This application of probability theory $to$ games, not just
within games, is called Predictive Game Theory (PGT).
 
This paper shows how information theory provides a first-principles
argument for how to set a distribution over joint strategies.  The
connection of this distribution to the bounded rational Quantal
Response Equilibrium (QRE) is elaborated. In particular, taking the
QRE to be an approximation to the mode of the distribution, correction
terms to the QRE are derived. In addition, some Nash equilibria are
not approached by any limiting sequence of increasingly rational QRE
joint strategies. However it is shown here that every Nash equilibrium
is approached with a limiting sequence of joint strategies all of
which have non-zero probability. (In general though not all strategies
in those sequences are modes of the associated distributions over joint
strategies.)

It is also shown that in many games, having a probability distribution
with support restricted to Nash equilibria --- as stipulated by
conventional game theory --- is impossible.  So the external analyst
should never predict a Nash outcome for such games. PGT is also used
to derive an information-theoretic (and model-independent)
quantification of the degree of rationality inherent in a player's
behavior. This quantification arises from the close formal
relationship between game theory and statistical physics. That close
relationship is also leveraged to extend game theory to situations
with stochastically varying numbers of players. This extension can be
viewed as providing corrections to the replicator dynamics of
conventional evolutionary game theory.

\end{abstract}

\section{Introduction}

Consider any scientific scenario, in which one wishes to predict some
characteristic of interest $y$ concerning some physical system.  To
make the prediction one starts with some information/data/prior
knowledge ${\mathscr{I}}$ concerning the system, together with known
scientific laws.  One then uses probabilistic inference to transform
${\mathscr{I}}$ into the desired prediction.  In particular, in Bayesian
inference we produce a posterior probability distribution $P(y
\mid {\mathscr{I}})$.   

Such a distribution is a far more informative structure than a single
``best prediction''.  However if we wish to synopsize the
distribution, we can distill it into a single prediction.  One way to
do that is to use the mode of the posterior as the prediction. This is
called the Maximum A Posterior (MAP) prediction. Alternatively, say we
are given a real-valued loss function, $L(y, y')$ that quantifies the
penalty we incur when we predict $y'$ and the true value is $y$.  The
Bayes optimal prediction then is the value of $y'$ that minimizes the
posterior expected loss, $\int dy L(y, y') P(y
\mid {\mathscr{I}})$. As an example, say that $y \in {\mathbb{R}}$ and
$L(y, y') = (y - 
y')^2$. Then the Bayes-optimal prediction is the posterior expected
value of $y$, $\int dy \; y P(y \mid {\mathscr{I}})$.  Formally, to
predict any other value than the Bayes-optimal prediction violates
Cox's and Savage's axioms concerning the need to use probability
theory when doing science (see~\cite{jabr03,gull88,lore90} and
Sec.~\ref{sec:cox} below).

As a technical comment, in practice evaluating the Bayes-optimal
prediction may be computationally difficult. In addition doing so
often requires that the practitioner specify ``prior probabilities'',
which when done poorly can lead to bad results. Finally, some
non-Bayesian axiomatizations of inference have been
offered~\cite{zell04}. 

Due to these reasons, when $\mathscr{I}$ consists of experimental
data, in practice often non-Bayesian techniques (e.g., Fisherian or
Waldian minimax) are used instead of pure Bayesian techniques. For
example, one might use an unbiased estimate of the data's mean and an
associated confidence interval rather than use a Bayes-optimal
predicted mean and associated posterior. More sophisticated
non-Bayesian techniques might use the bootstrap~\cite{wolp95c}
procedure, stacking~\cite{brei96}, etc.{\footnote{Note though that
often such techniques can be cast as approximations to Bayesian
techniques~\cite{jabr03,gull88,lore90}.}}

The general relationship between such non-Bayesian techniques and
purely Bayesian techniques is exceedingly subtle~\cite{wolp96c}.
However in the numeric sciences, even with non-Bayesian techniques, to
analyze experimental data the broad approach is to use $\mathscr{I}$
(the experimental data) to generate a probability distribution over
the quantity of interest and (if desired) generate an associated
single prediction.  (This is why numeric data is presented with
``error bars''.)  No one has ever suggested why this broad approach
would be appropriate when one is analyzing a physical system without
humans and $\mathscr{I}$ is experimental data, but not appropriate
when one is analyzing a physical system of a set of human players
engaged in a game and $\mathscr{I}$ is the game structure.

Indeed, the Bayesian approach can be motivated in purely
game-theoretic terms. Say we have a game $\gamma$ of some sort. Let
$\Delta_{\cal{X}}$ indicate the set of all possible joint mixed
strategies in $\gamma$. Now consider a ``meta-game'' $\Gamma$ that
consists of a scientist ($S$) playing against Nature ($N$). In this
meta-game $N$'s space of possible moves is $\Delta_{\cal{X}}$, i.e.,
the set of all possible joint mixed strategies in $\gamma$. The move
of the scientist $S$ is a prediction of what element of
$\Delta_{\cal{X}}$ player $N$ will adopt.  As usual in games against
Nature, $N$ has no utility function. However $S$ has a utility
function, given by the negative of a loss function that quantifies how
accurate her move is as a prediction of $N$'s move. So to maximize her
expected utility the scientist wants to choose her move(s) --- her
prediction of the joint mixed strategy that governs the game $\gamma$
--- to minimize her expected loss. To do that she needs $N$'s mixed
strategy $P(q \in \Delta_{\cal{X}})$.

Now $S$ only has partial information about $N$ (e.g., she only knows
the utility functions of the players in $\gamma$ and their move
spaces). Therefore her first task is to translate that information,
$\mathscr{I}$, into a distribution over $N$'s possible moves (i.e.,
over the possible mixed strategies of the game $\gamma$). Intuitively,
she must translate her partial information into an inference of $N$'s
mixed strategy. How to do this is the crux of PGT. Having done this,
just as in conventional non-cooperative game theory, the scientist $S$
choose her move to minimize expected loss (maximize expected utility)
in $\Gamma$, i.e., she predicts that the joint mixed strategy of the
game $\gamma$ is the one that, under expectation, is as close to the
actual one as possible. That is the scientist's assessment of the
game's ``equilibrium''.

Note that for the same $\gamma$ and the same inference by $S$ of the
mixed strategy $P(q)$, if we change $S$'s loss function, we change her
prediction. This is a game-theoretic example of how changing the loss
function of the scientist external to the game will in general change
the associated equilibrium concept mapping $\gamma$ to $S$'s
prediction for $\gamma$'s mixed strategy.

The broad approach of converting $\mathscr{I}$ to a distribution and
then --- if one has a loss function --- converting that distribution
to a final prediction is the one that will be adopted in this
paper. This paper is about how to use this approach to analyze
games. In other words, it about how to infer the mixed strategy $P(q)$
of a Nature whose moves are joint mixed strategies of a game $\gamma$.

As a particular example of the implications of this approach, suppose
that our prior information $\mathscr{I}$ concerning the game $\gamma$
does $not$ explicitly tell us that the players in $\gamma$ are all
fully rational. Then in general, probabilistic inference will produce
a non-delta function distribution over the ``rationalities'' of the
players (however that term is defined).  In this way, applying
probabilistic inference to games intrinsically results in bounded
rationality.

To be more concrete, note that mixed strategies in a non-cooperative
game are themselves probability distributions. Therefore probabilistic
inference concerning mixed strategies involves probability density
functions over probability distributions.  Now in Shannon's
information theory \cite{coth91,mack03,jabr03, stwo94} the fundamental
physical objects under consideration are probability distributions, in
the form of stochastic communications channels. Accordingly,
probabilistic inference in information theory also involves
probabilities of probabilities. This makes the mathematical tools for
probabilistic analysis in information theory contexts --- a topic
already well-researched --- well-suited to a probabilistic analysis of
noncooperative games.

More precisely, a central concept in information theory is a measure
of the amount of information embodied in a probability distribution
$p$, known as the Shannon entropy of that distribution,
$S(p)$. Amongst its many other uses, Shannon entroy can be used to
formalize Occam's razor based on first-principles arguments. This
formalization is known as the minimum information (Maxent) principle,
and its Bayesian formulation is embodied in what is known as the
entropic prior. This can serve as the foundation of a first-principles
formalism for probabilistic inference over probability distributions.

Using entropy to perform probabilistic inference this way has proven
extraordinarily successful in an extremely large number of
applications, ranging from signal processing to machine learning to
statistics. Recently it has also been realized that the mathematics
underlying this type of inference can be used to do distributed
control and/or optimization. In that context the mathematics is known
as Probability Collectives (PC). Preliminary experiments validate PC's
power for control of real-world (hardware) systems, especially when
the system is large (See collectives.stanford.edu
and\cite{wolp03b,wolp04b,mabi04,lewo04b,wobi04a,biwo04a,biwo04b,anbi04}.)

As another example of the successes of entropy-based inference,
consider the problem of predicting the probability distribution $p$
over the joint state of of a huge number of interacting
particles. This is the problem addressed by statistical physics. As
first realized by Jaynes, such prediction is an exercise in
probabilistic inference of exactly the sort Maxent can be applied to.
Accordingly statistical physics can be addressed --- and in fact
derived in full --- using Maxent \cite{jayn57,jabr03}.  In light of
all the tests physicists have done of the predictions of statistical
physics, this means that there are (at least) tens of thousands of
experimental confirmations of that principle in domains with a very
large number of interacting particles.

We can similarly use Shannon's entropy to do inference of the (the
distribution governing the) joint mixed strategy $q(x) = \prod_{i=1}^N
q_i(x_i)$ in any game involving $N$ players with pure strategies
\{$x_i$\}.  Probabilistic inference applied to game theory this way is
known as {\bf{Predictive Game Theory}} (PGT). In PGT, the whole point
is to apply probability theory in general and Bayesian analysis in
particular $to$ games and their outcomes. This contrasts with their
use in conventional game theory *within* the structure of individual
games (e.g., in correlated equilibria~\cite{auma87}).

\subsection{The relation between PGT and
conventional game theory: a first look}

Before presenting PGT in detail, this section  illustrates some of its
connection to other  work   in game  theory and  statistical  physics.

Statistical physics provides a unifying mathematical formalism for
the physics of many-particle systems. Any question related to that
type of physics can be analyzed, in principle at least, simply by
casting it in terms of that formalism, and then performing the
associated calculations.  There is no need to introduce any new
formalism for new questions, new Hamiltonians, etc.

PGT arises from information theory similarly to how statistical
physics arises from information theory, only in a different
context. Accordingly, PGT can play an unifying role for games
analogous to the one statistical physics plays for many-particle
systems. All questions related to games can be analyzed, in principle
at least, simply by casting them in terms of PGT, and then performing
the associated calculations. There is no need to create new formalisms
for new game theory issues, new presumptions about the way humans
behave in games, etc; one simply casts them in terms of PGT.

Now in PGT the idea of an ``equilibrium concept'', so central to
conventional game theory, does not directly arise. Let $q(x)$ indicate
a joint mixed strategy over joint move $x$. Generically in PGT, the
support of the probability density function over $q$'s, $P(q)$, has
non-zero measure. In this, one does not allow only a single
``equilibrium'' $q$, or even a countable set of $q$'s comprising the
``equilibrium'' joint strategies; the number of allowed $q$'s is uncountable.

The fact that the support of $P$ has non-zero measure typically
ensures a built-in ``bounded rationality'' to PGT. This is because
typically there will be $q$ that are allowed (i.e., that have non-zero
probability) in which one or more of the players is not fully
rational. This aspect of the measure of $P$ has other consequences as
well. For example, it means that rather than consider the values of
economic quantities of interest at a single (or at most countable set
of) equilibrium $q$, as is conventionally done, one should consider
the expected value of such quantities under $P(q)$. This means that
attributes of those quantities like how nonlinear they are (which is
crucial to approximating the integrals giving their expectation
values) have consequences when PGT is used to analyze economics
issues, consequences that they do not have when conventional game
theory is used.

Are there quantities in PGT that are analogous to equilibrium
concepts, even if $P$'s support has non-zero measure?  One possible
interpretation of what an ``equilibrium concept'' could mean in PGT is
the Bayes-optimal $q$. Note though that the Bayes-optimal $q$ in
general depends of the loss function of the external scientist making
a prediction about the physical system (i.e., about the game). So
consider two scenarios, both concerning the exact same game, with the
exact same knowledge concerning the game, and therefore the exact same
distributions over joint mixed strategies. However have the loss
function of the external scientist (reflecting how their prediction
will be used) differ between the two scenarios. Then the Bayes-optimal
prediction will also differ between the two scenarios. So the very
choice of ``equilibrium concept'' is determined (in part) by the
external scientist analyzing the game; the ``equilibrium'' joint mixed
strategy is not purely a function of the game itself, but rather also
involves the external scientist making predictions about the system.

This dependence on the external scientist of PGT's (analogue of the
conventional) notion of a game's equilibrium is not a philosophical
preference. It is not something that we have discretion to adopt or
not. Rather it is intrinsic to our analyzing games with human players
the same we analyze other physical systems in the Bayesian paradigm:
by deriving distributions over truths based on partial information,
and then (if needed) making single predictions based on that
distribution together with an external loss function. Under this
interpretation of equilibrium concept, we {\it{have no choice}} but to
accept the dependence of point predictions on the external scientist
making the prediction.

Another possible interpretation of the ``equilibrium'' of a game is as
the posterior 
\begin{eqnarray}
P(x \mid {\mathscr{I}}) &=& \int dq \; P(x \mid q, {\mathscr{I}}) P(q
\mid {\mathscr{I}}) \nonumber \\
&=& \int dq \; P(x \mid q) P(q \mid {\mathscr{I}}) \nonumber \\
&=& \int dq \; q(x) P(q \mid {\mathscr{I}}) .
\end{eqnarray}
Just like the Bayes-optimal $q$, $P(x \mid {\mathscr{I}})$ reflects
the ignorance of us, the external scientists concerning the game and
its players, as well as the intrinsic noise/randomness in how the
players choose their moves. Unlike the Bayes-optimal $q$ though, $P(x
\mid {\mathscr{I}})$ does not depend on the loss function of the
external scientist. 

On the other, in general $P(x \mid {\mathscr{I}})$ will not be a
product distribution, i.e., it will not have the moves of the players
be independent. This is true even though $P(q \mid {\mathscr{I}})$ is
restricted to such distributions (a linear combination of product
distributions typically is not a product distribution). In addition,
say that $P(q
\mid {\mathscr{I}})$ is restricted to Nash equilibria $q$. Typically,
if there are more than one such equilibria (i.e., the support of $P$
contains more than one point), then under $P(x \mid {\mathscr{I}})$
none of the players is playing an optimal response to the mixed
strategy over the other players. In other words, even though we might
know that all the players are in fact perfectly rational, {\it{our
prediction}} of their moves has ``cross-talk'' among the multiple
equilibria and does not have perfect rationality.

Typically for any $P(q \mid {\mathscr{I}})$ there is only one (perhaps
difficult to evaluate) Bayes-optimal prediction (e.g., for quadratic
loss functions that prediction is the posterior mean, $\int dq \; q
P(q \mid {\mathscr{I}})$). Similarly $\int dq \; q(x) P(x \mid
{\mathscr{I}})$ is always unique. So under either of this
interpretations of ``equilibrium concept'', every game has a unique
equilibrium. In this, whichever of these PGT-based interpretations of
equilibrium we adopt, all work in conventional game theory that
attempts to ``fix'' the possible multiplicity of conventional concepts
of equilibrium (e.g., the many proposed refinements of the Nash
equilibrium concept) is rendered moot.  The same fate obtains for the
different equilibrium concepts that have been proposed in cooperative
game theory.

As a practical matter, often calculating the exact Bayes-optimal $q$
can be quite difficult. As a substitute, even if it is not
Bayes-optimal, we can calculate the MAP $q$. When $P(q \mid
{\mathscr{I}})$ is peaked the MAP $q$ should be a good approximation
to the Bayes-optimal $q$. Indeed, it is common in Bayesian analysis to
approximate the Bayes-optimal prediction by expanding the posterior as
a Gaussian centered on the MAP prediction.

This MAP $q$ is the minimizer of a Lagrangian functional
${\mathscr{L}}(q)$. In general this MAP $q$ is a bounded rational
equilibrium rather than a Nash equilibrium.  As shown below, this MAP
bounded rational equilibrium can often be approximated by
simultaneously having each player $i$'s mixed strategy $q_i(x_i)$ be a
{\bf{Boltzmann distribution}} over the values of its expected utility
for each of its possible moves:
\begin{equation}
q_i(x_i) \propto e^{\beta_i E_q(u^i \mid x_i)} \; \forall i
\label{eq:qre}
\end{equation}
where the joint distribution $q(x) = \prod_i q_i(x_i)$ and $u^i(x)$ is
player $i$'s utility function. 

In general there may be more than one solution to the set of coupled
equations Eq.~\ref{eq:qre}. (See ~\cite{wolp04c} for examples of
closed-form solutions to this set of coupled equations.)  In
conventional game theory, the set of all such solutions is sometimes
called the (logit response) Quantal Response Equilibrium (QRE)
\cite{goho99,mcpa95,chfr97}. It has been used as a convenient way to
encapsulate bounded rationality.  Typically approximating the MAP
mixed strategy with the QRE should incur less and less error the more
players there are in the game. However as discussed below, for small
games the QRE may be a poor approximation to the MAP (which itself is
an approximation to the Bayes-optimal prediction). Below the
correction terms of the QRE (as an approximator of the MAP
distribution) are calculated.

Another relation between the QRE and PGT, one that doesn't involve
approximations, starts with the fact that at Nash equilibrium each
player $i$ sets its strategy $q_i$ to maximize its expected utility
$E_{q_i, q_{{-i}}}(u^i)$ for fixed $q_{-i}$.{\footnote{Throughout this
paper the minus sign before a symbol specifying a particular player
indicates the set of all of the other players, and similarly for a
minus sign before a set of player symbols.}}  Consider instead having
each player $i$ set $q_i$ to optimize an associated functional, the
``maxent Lagrangian'':
\begin{equation}
{\mathscr{L}}_i(q_i) \triangleq E_{q_i, q_{{-i}}}(u^i) - T_iS(q_i, q_{-i}).
\end{equation}
For all $T_i \rightarrow 0$ the equilibrium $q$ that simultaneously
minimizes ${\mathscr{L}}_i \; \forall i$ is a Nash
equilibrium~\cite{wolp04a,mcpa95,megi76, fukr93,fule93, luce59}. For $T_i > 0$
one gets bounded rationality. Indeed, under the identity $T_i
\triangleq \beta_i^{-1} \; \forall i$ the solution to this modified Nash
equilibrium concept turns out to be the QRE.


As discussed in ~\cite{wolp04a}, the maxent Lagrangian also arises in
statistical physics, where it is called (a mean field approximation
to) the ``free energy''.  This formal connection between PGT/QRE and
statistical physics can be exploited in several ways.  As an example,
consider the case where one's prior information consists of the
expected energy of a set of interacting particles with joint state
$r$, a scenario known as the ``canonical ensemble'' in statistical
physics (CE). In this situation the MAP estimate of the density
function $p(r)$ using an entropic prior is the minimizer of
${\mathscr{L}}(p) = E_p(H) - TS(p)$, where $H$ is the energy of the
system of particles.  In light of the formula for the maxent
Lagrangian, this suggests tha bounded rational players in a game can
be made formally identical to the particles in the CE. Under this
identification, the moves of the players play the roles of the states
of the particles, and particle energies are translated into player
utilities. Particles are distributed according to a Boltzmann
distribution over their energies, and mixed strategies are Boltzmann
distributions over expected payoffs.{\footnote{Note that having the
probability density over mixed strategies follow a Boltzmann
distribution does not mean that functionals of that density are
Boltzmann-distributed. In particular, the distribution over values of
the utility function need not be Boltzmann-distributed.}}

This connection between PGT and statistical physics raises the
potential of transferring some of the powerful mathematical techniques
that have been developed in the statistical physics community into
game theory. As an example, in the ``Grand Canonical Ensemble'' (GCE)
the number of particles of various types is variable rather than being
pre-fixed. One's prior information is then extended to include the
expected numbers of particles of those types. This corresponds to
having a variable number of players of various types in a bounded
rational game. This suggests how to extend game theory to accommodate
games with statistically varying numbers of players. Among other
applications, this provides us with a new framework for analyzing
games in evolutionary scenarios, different from evolutionary game
theory. (A different type of ``GCE game'' is analyzed below.) Even

There are many other aspects of statistical physics that might carry
over to PGT. For example, even in the CE, often there are regimes
where as some parameter of the system is changed an infinitesimal
amount, the character of the system changes drastically. These are
known as ``phase transitions''. The connection between the math of
PGT and that of statistical physics suggests that similar phenomena
may arise in games with human players.

PGT has many other advantages in addition to providing a way to
exploit techniques from statistical physics in the context of
noncooperative games. For example, as illustrated below it provides a
natural way to quantify the rationality of experimentally observed
behavior of human subjects.  One can then, for example, empirically
observe the dynamic relationship coupling the rationalities of real
players as they play a sequence of games with one another.  (Since
such correlations are inherently a property of distributions across
mixed strategies, they are not readily analyzed using conventional
non-distribution-based game theory.) 

Another strength of PGT arises if we change the coordinates of the
underlying space of joint pure strategies
\{$x$\} . After such a change, our mathematics 
describes a type of bounded rational cooperative game theory in which
the moves of the players become binding contracts they all offer one
another\cite{wobi04b, mawo05}. In this sense, PGT provides a novel
relation between cooperative and noncooperative game theory.

% ... 
%

\subsection{Roadmap}

The purpose of this paper, like that of the original work on game
theory, is to elucidate a framework for analyzing the reasonably
imputed consequences about the behavior of the players when all one
knows is the game structure. If possible, this framework should be
able to accommodate extra knowledge concerning the game and/or the
players if it is available. Loosely speaking, the goal is to provide
for game theory the analog of what the canonical and grand canonical
ensembles provide for statistical physics: a first-principles
mathematical scaffolding into which one inserts one's knowledge
concerning the system one is analyzing, to make predictions concerning
that system. (See the future work section below for further discussion
of this point.)

To do this, the next section starts by cursorily reviewing
noncooperative game theory, Bayesian analysis and the entropic prior
arising in information theory. In an appendix that prior is
illustrating by showing how it can be used to derive statistical
physics.  In the following section foundational issues of PGT and
associated mathematical tools are presented.

The next two sections form the core of the player. The first of them
applies the entropic prior to infer mixed strategies of coupled
players in a game $\gamma$. This application can be viewed as a
prescription for how to infer the mixing strategy $P(q \mid
{\mathscr{I}})$ adopted by a Nature involved in a meta-game with a
scientist, where the moves $q$ of Nature are mixed strategies in
$\gamma$. This section then relates this coupled-players analysis to
the QRE.  The section after this considers independent players,
leveraging the analysis for coupled players.

The following section illustrates some of the breadth of PGT. It is
shown there how bounded rationality arises formally as a cost of
computation for the independent players scenario.  We then present
rationality functions. These are a model-independent way to quantify
the (bounded) rationality of the mixed strategies followed by
real-world players. This section ends by showing how to apply PGT to
games with stochastically varying numbers of players.

An appendix discusses the relation between PGT and previous work, and
more generally the history of attempts to apply information theory
within game theory.

\section{Preliminaries}

This section first reviews noncooperative game theory. It then reviews
information theory and the associated Bayesian analysis. It ends by
illustrating that analysis with a review of how it can be used to
derive statistical physics. It is recommended that those already
familiar with these concepts still read the middle subsection on
Bayesian analysis.

\subsection{Review of noncooperative game theory}
\label{sec:pdthy}

In conventional noncooperative normal form
game~\cite{futi91,auha92,baol99,binm92,lura85} theory one has a set of
$N$ independent {\bf{players}}, indicated by the natural numbers
\{1, 2, $\ldots, N$\}. Each player $i$ has its own finite
set of allowed {\bf{pure strategies}}, each such pure strategy written
as $x_i \in X_i$. We indicate the the size of that space of possible
pure strategies by player $i$ as $|X_i|$. The set of all possible
joint strategies is $X \triangleq X_1 \times X_2 \times \ldots \times
X_N$ with cardinality $|X| \triangleq \prod_{i=1}^N |X_i|$, a generic
element of $X$ being written as $x$.

A {\bf {mixed strategy}} is a distribution $q_i(x_i)$ over player
$i$'s possible pure strategies, \{$x_i$\}.  In other words, it is a
vector on the $|X_i|$-dimensional unit simplex, $\Delta_{X_i}$. Each
player $i$ also has a {\bf{utility function}} (sometimes called a
``payoff function'') $u^i$ that maps the joint pure strategy of all
$N$ of the players into a real number.

As a point of notation, we will use curly braces to indicate an entire
set, e.g., \{$\beta_i$\} is the set of all values of $\beta_i$ for all
$i$. We will also write $\Delta_{{\cal{X}}}$ to refer to the Cartesian
product of the simplices $\Delta_{X_i}$, so that mixed joint
strategies (i.e., product distributions) are elements of
$\Delta_{{\cal{X}}}$.  We will sometimes refer to $u^i$ as player
$i$'s ``payoff function'', and to player $i$'s pure strategy $x_i$ as
its ``move''. $x$ is the joint move of all $N$ players. As mentioned
above, we will use the subscript $-i$ to indicate all moves /
distributions / utility functions, etc., other than $i$'s. We will use
the integral symbol with the measure implicit, so that it can refer to
sums, Lebesgue integrals, etc., as appropriate. In particular, given
mixed strategies of all the other players, we will write the expected
utility of player $i$ as $E(u^i) =
\int dx \; \prod_j q_j(x_j) u^i(x)$. As a final point of notation, we
will write $\vec{a}$ to mean a finite indexed set all of whose
components are either real numbers are infinite (greater than any real
number). We will then write ${\vec{a}} \succeq {\vec{b}}$ to indicate
the generalized inequality that $\forall i$, either $a_i$ and $b_i$
are real numbers and $a_i \ge b_i$, both $a_i$ and $b_i$ are infinite,
or $b_i$ is a real number and $a_i$ is infinite. Also, in the
interests or expository succinctness, we will be somewhat sloppy in
differentiating between probability distributions, probability density
functions, etc.; generically, ``$P(\ldots)$'' will be one or the other
as appropriate.

Much of noncooperative game theory is concerned with {\bf{equilibrium
concepts}} specifying what joint-strategy one should expect to result
from a particular game. In particular, in a {\bf{Nash equilibrium}} every
player adopts the mixed strategy that maximizes its expected utility,
given the mixed strategies of the other players. More formally,
$\forall i, q_i = {\mbox{argmax}}_{q'_i} \int dx \; q'_i \prod_{j \ne i}
q_j(x_j) \; u^i(x)$~\cite{grei99,futi91,baol99}.

One problem with the Nash equilibrium concept is its assumption of
{{\bf full rationality}}. This is the assumption that every player $i$
can both calculate what the strategies $q_{j \ne i}$ will be and then
calculate its associated optimal distribution.{\footnote{Here we use
the term ``bounded rationality'' in the broad sense, to indicate any
mixed strategy that does not maximize expected utility, regardless of
how it arises.}}  This requires in particular that each player
calculate the entire joint distribution $q(x) = \prod_j q_j(x_j)$. If
for no other reasons than computational limitations of real humans,
this assumption is essentially untenable. This problem is just as
severe if one allows statistical coupling among the players
~\cite{auma87,futi91}.

For simplicity, throughout each analysis presented in this paper we
will treat $N$, the pure strategy spaces, the associated utility
functions, and the statistical independence of the pure strategies
chosen by the players, as fixed parts of the problem definition rather
than random variables.  Further we will impose no {\it{a priori}}
restrictions about whether the players have encountered one another
before, what information they have about one another and the game
they're playing, whether they have engaged in the game before, what
their information sets are, whether there are any social norms at work
on them, etc. We do not even require, {\it{a priori}}, that the
players be prone to human psychological idiosyncracies. To incorporate
any information of this sort into the analysis would mean modifying
the priors and/or likelihoods considered below in a (mostly)
straightforward way, but is beyond the scope of this paper.

\subsection{Review of Bayesian analysis and decision theory}
\label{sec:cox}

Consider any scenario in which we must reason about attributes of a
physical system without knowing in full all salient aspects of that
system. This is the basic problem of inductive inference. How should
we do this reasoning? Many different desiderata, arising from work by
De Finetti, Cox, Zellner and many others, lead to the same conclusion:
if our goal is to assign real-valued numbers to the different
hypotheses concerning the system at hand, we should use the rules of
Probability theory\cite{besm00,berg85, wolp97,
zell04,pari94,vanh03,jabr03,gull88,lore90}. In particular, this
implies that we should use Bayes' theorem to calculate what we want to
know from what we are told/assume/observe/know:
\begin{eqnarray}
P({\mbox{truth }} z \mid {\mbox{data }} d) &\propto& P(d \mid z)
P(z)
\end{eqnarray}
where the proportionality constant is set by the requirement that
$P({\mbox{truth }} z \mid {\mbox{data }} d)$ be normalized, and ``data''
means everything we are told/assume/observe/know concerning the
system. $P({\mbox{truth }} z \mid {\mbox{data }} d)$ is called the
{\bf{posterior}} probability, $P({\mbox{data }} d \mid {\mbox{truth }} z)$
is called the {\bf{likelihood}}, and $P(z)$ is called the
{\bf{prior}}.

Say that rather than a full posterior distribution, for some reason we
must predict a single one of the candidate hypotheses $z$. According
to Savage's axioms, to do this we must be provided with a {\bf{loss
function}} $L(y, z)$ that maps any pair of a truth $z$ and a
prediction $y$ to a real-valued loss (see~\cite{besm00} and various
chapters in~\cite{auha92}). Then the associated {\bf{Bayes optimal}}
prediction is argmin$_{y} E_{P}(L(y, z))$ where the expectation is
over the posterior distribution $P({\mbox{truth }} z
\mid {\mbox{data }} d)$.{\footnote{There is controversy about the precise
details of Savage's axioms and their implications, the precise way
priors should be chosen, and even the precise physical meaning of
``probability''\cite{wolp96c}. Such details are not important for
current purposes. Other choices can be made, based on other
desiderata. However typically the broad outlines of any approach based
on such alternatives is the same: to do inference one constructs a
probability distribution over possible truths and then, if needed,
distills that distribution into a single prediction.}}

Note that the loss function is determined by the scientist external to
the system who is making the prediction; it is not specified in the
definition of the system under consideration.

According to the foregoing, to do statistical inference for a
particular physical scenario our first task is to translate the
``particular physical scenario'' we're considering into a mathematical
formulation of possible truths $z$, data $d$, etc. Having done that,
we can employ mathematical tools like Bayes' theorem, approximation
techniques for finding Bayes-optimal predictions, etc. to analyze our
mathematical formulation. After doing this we use our translation to
convert all this back into the physical scenario. This translational
machinery is how we couple the abstract mathematical structure of
probability theory to our particular physical inference problem.

To assist us in making this translation, we imagine an infinite set of
instances of our physical scenario. All of those instances share the
physical characteristics of our scenario that fix our statistical
inference problem.  {\it{Every}} other physical characteristic is
allowed to vary across those instances.  In this way the set of all of
those instances define our statistical inference problem~\cite{wolp96c}.
Formally, we define the {\bf{invariant}} of our inference problem as
the set of exactly those characteristics of the physical scenario, and
no others, that would necessarily be the same if we were presented
with a novel instance of the exact same inference
problem. Equivalently, we can define the invariant as the set of all
the physical instances consistent with those characteristics, and no
instance that is inconsistent with those characteristics.  By
explicitly delineating an inference problem's invariant, we can
mathematically formalize that problem. 

As discussed in the introduction, this Bayesian perspective is
inherent in much of conventional game theory. Most obviously, all the
work on Bayesian games, correlated equilibria, etc.  adopts elements
of the Bayesian perspective. In addition, we can define a meta-game
$\Gamma$ of a Scientist $S$ playing against Nature $N$ in which the
possible states of $N$ are the possible mixed strategies $q$ of the
game $\gamma$ we wish to analyze. The move of $S$ in $\Gamma$ is
interpreted as a prediction of $S$'s move, i.e., of the mixed strategy
of $\gamma$. The utility function of $S$ is interpreted as the
(negative) of the loss function of $S$. Accordingly, for $S$ to do
exactly what is prescribed by conventional game theory --- choose a
strategy that maximizes her expected utility --- she must make the
Bayes-optimal prediction of the outcome of the game. To do this she
must infer $N$'s mixed strategym, $P(q)$. However she has only limited
information concerning $N$, e.g., the specification of
$\gamma$. Accordingly, the crucial issue in game theory is how, based
on her limited information concerning $\gamma$, the external scientist
should infer $N$'s mixed strategy, $P(q)$. This is a problem of how to
infer a distribution over distributions.

This inference, done using invariants, is the topic of PGT.  In this
paper we will only explore such use of invariants in conjunction with
the entropic prior, since that prior is directly concerned with
inferring distributions over distributions. However other priors also
merit investigation.

\subsection{Review of the entropic prior}
\label{sec:ent_prior}

Shannon was the first person to realize that based on any of several
separate sets of very simple desiderata, there is a unique real-valued
quantification of the amount of syntactic information in a
distribution $P(y)$. He showed that this amount of information is (the
negative of) the Shannon entropy of that distribution, $S(P) = -\int
dy \; P(y) {\mbox{ln}}[\frac{P(y)}{\mu(y)}]$.\footnote{$\mu$ is an {\it{a
priori}} measure over $y$, often interpreted as a prior probability
distribution. Unless explicitly stated otherwise, here we
will always assume it is uniform, and not write it explicitly. See
\cite{jayn57,jabr03,coth91}.} Note that for a product distribution
$P(y) = \prod_i P_i(y_i)$, entropy is additive: $S(P) = \sum_i
S(P_i)$.

So for example, the distribution with minimal information is the one
that doesn't distinguish at all between the various $y$, i.e., the
uniform distribution. Conversely, the most informative distribution is
the one that specifies a single possible $y$. 

Say that the possible values of the underlying variable $y$ in some
particular probabilistic inference problem have no known {\it{a
priori}} stochastic relationship with one another. For example, $y$
may not be numeric, but rather consist of the three symbolic values,
\{red, dog, Republican\}. Then simple desiderata-based counting
arguments can be used to conclude the prior probability of any
distribution $p(y)$ is proportional to the {\bf{entropic prior}},
$\exp{(-\alpha S(p))}$, for some associated non-negative constant
$\alpha$.{\footnote{The issue of how to choose $\alpha$ --- or better
yet how to integrate over it --- is quite subtle, with a long
history. See in particular work on ML-II~\cite{berg85}and the
``evidence procedure''~\cite{stwo94}.)}}

Intuitively, absent any other information concerning a
particular distribution $p$, the larger its entropy the more {\it{a
priori}} likely it is.{\footnote{Note that this is different
from saying that the larger $s$ is, the more {\it{a priori}} likely it
is that the entropy of $p$ is larger: 
\begin{eqnarray*}
P_S(s) &=& \int dp \; \delta(S(p) - s) P(p) \\
&=& \frac{\int dp \; \delta(S(p) - s) {\exp{(-\alpha S(p))}}}
{\int dp \; {\exp{(-\alpha S(p))}}} .
\end{eqnarray*}
}}

If the possible $y$ have a more overt mathematical relationship with
one another, the situation is often not so clear-cut. For example,
symmetry group arguments are often invoked in such situations, and can
give more refined predictions. Despite this, for the most important
scenarios it considers, scenarios where it has had such great
successes, statistical physics simply uses the entropic prior, as
described below. In accord with this, in this paper attention will be
restricted to the entropic prior.

Say we have some information $\mathscr{I}$ concerning $p$. Then by
Bayes' theorem, the posterior probability of distribution $p$ is
\begin{equation}
P(p \mid {\mathscr{I}}) \propto \exp{(-\alpha S(p))} P({\mathscr{I}}
\mid p) .
\label{eq:bayes}
\end{equation}
The associated MAP prediction of $p$ based on $\mathscr{I}$ is
argmax$_p P(p  \mid {\mathscr{I}})$. 

Intuitively, Eq.~\ref{eq:bayes} pushes us to be conservative in our
inference.  Of all hypotheses $p$ equally consistent
(probabilistically) with our provided information, we are led to prefer
those that contain minimal extra information beyond that which is
contained in the provided information. This is a formalization of
Occam's razor.

Physically, $\mathscr{I}$ is all characteristics of the system that
would not change if we were presented with a novel instance of the
exact same inference problem. From a frequentist perspective, it is an 
invariant across a set of experiments:
$\mathscr{I}$ delineates what characteristics of the system are fixed
in those experiments, while all characteristics not in $\mathscr{I}$
are allowed to vary. In essence, $\mathscr{I}$ is the invariant that
defines the inference problem.

In particular, ${\mathscr{I}}$ includes any functions of $p$, $F(p)$,
such that we know (by the specification of the precise inference
problem at hand) that $F(p)$ would not change if we confronted a novel
instance of the same inference problem. In general we may not know the
actual value of $F(p)$ that is shared among the instances specified by
$\mathscr{I}$; we may only know that that value is the same in all of
those instances.{\footnote{Relating this back to the mathematics of
probability theory, in such a case that value of $F(p)$ is known as a
{\bf{hyperparameter}}. Formally, hyperparameters have their own
priors. To get a final posterior over what we wish to infer --- $p$
--- we must marginalize over possible values of all hyperparameters.
Implicitly, the reason that here we simply choose one value of a
hyperparameter and discard all others is that we expect the posterior
distribution of the hyperparameter to be highly peaked, so that we do
not need to carry out such marginalization. See the discussion of
ML-II in
\cite{berg85,besm00,stwo94}, and also~\cite{stwo94}.}}  

Note that ${\mathscr{I}}$ cannot specify $p$, the precise state of the
system --- there must be some salient characteristics of the system
that are not fixed by ${\mathscr{I}}$. If this were not the case the
likelihood $P({\mathscr{I}} \mid p)$ would be a delta function, and
therefore the prior would be irrelevant.  In such a case, statistical
inference would reduce to the truism ``whatever happens
happens''. Accordingly, we never have $\mathscr{I}$ contain a set of
functions \{$F_i(p)$\} whose values jointly fix $p$ exactly.

An important example of the foregoing occurs in statistical physics,
where $\mathscr{I}$ is the observed temperature $T$ of a physical
system. $T$ is taken to fix a function $F(p)$, namely the expected
energy $H(x)$ of the system under distribution $p(x)$: $F(p) = \int dx
\; H(x) p(x)$ is fixed by $\mathscr{I}$. It is the
application of the entropic prior to this situation that results in
the canonical ensemble mentioned above. The number of experiments
validating this application is extraordinary; including experiments in
high school labs, it is probably on the order of $10^8$ (at least). In
this paper that application serves as a touchstone for how to
translate $\mathscr{I}$ into a distribution over distributions, and
therefore as the primary analogy mentioned in the derivation of
PGT. However in the interests of expediting the flow of this paper,
that application is relegated to an appendix.

\section{Predictive Game Theory - general considerations}

\subsection{The two types of game theory}
\label{sec:two_types}

Say we are presented with a noncooperative normal form game for $N$
players other than us, and a set of $N$ subjects who will fill the
roles of the players in a fixed manner.  We wish to make predictions
concerning the outcome of that game when played by that set of players
other than us.  As discussed above, for us to obey Cox's axioms, when
making those predictions we must use probability theory. If we wish to
distill the probability distribution over outcomes into a single
prediction, then to obey Savage's axioms we must have a loss function
and use decision theory.

Note that these normative axioms differ fundamentally from those that
can be used to derive various equilibrium concepts. The axioms
underlying equilibrium concepts concern external physical reality,
namely the players of the game. They concern something outside of our
control. In other words, such axioms are hypothesized physical
laws. Like any other such laws, they can be contradicted or affirmed
by physical experiment, at least in theory. In fact, behavioral game
theory~\cite{came03} essentially does just that, experimentally
determining what such hypothesized physical laws are valid. (See also
all the work on behavior economics~\cite{kahn03a,kahn03b}.)

In contrast, Cox's and Savage's axioms are normative. They tell us how
best to make predictions about the physical world. They make no
falsifiable claims about the real world; it is under our control
whether they will be followed, not the external world's
control. Violating them in our analysis is akin to performing an
analysis in which we violate the axioms defining the integers.

Before showing some ways to arrive at a distribution over outcomes for
this situation, we must clarify what the space of ``outcomes'' of the
game is. There are two broad types of such spaces to consider, with
associated types of game theory.

In type I game theory, what a player chooses in any particular
instance of the game is its move in that instance.  (This is the
analog of the variable $y$ in App.\ref{sec:physics}.)  In general, in
the real world a particular human's choice of move will vary depending
on their mood, how distracted they are, etc.  Physically, this
variability arises from variability in the dynamics of
neurotransmitter levels in the synapses in their brain during their
decision-making, associated dynamical variability in the firing
potentials of their neurons during that process, etc.

Due to this variability the choice of each player is governed by a
probability distribution. (This is the analog of the variable $q$ in
App.\ref{sec:physics}). So the joint choice of the players is also
described by a distribution. We write that joint distribution as
\begin{eqnarray}
q(x) &=& P(x \mid q) \nonumber \\
&=& P(x_N \mid q) \prod_{i=1}^{N-1} P(x_i \mid q, x_{i+1},
x_{i+2}, \ldots) \nonumber \\
&=& P(x_N \mid q) \prod_{i=1}^{N-1} P(x_i \mid q) \nonumber \\
&=& \prod_{i=1}^{N} P(x_i \mid q) \nonumber \\
&=& \prod_{i=1}^N q_i(x_i),
\label{eq:prod_dist}
\end{eqnarray}
where the third equality follows from the statistical independence of
the players' choices. So the probability of a particular joint move
$x$ is given by a product distribution $q(x) \triangleq \prod_i
q_i(x_i)$.{\footnote{Loosely speaking, when used as an approximation
in statistical physics, such product distributions are called ``mean
field theory''. See~\cite{wolp04aa}.}} If the interaction between the
humans is not, physically, a conventional normal form noncooperative
game, then the space of allowed $q$ must be expanded to allow $q$ that
statistically couple the moves. Such extensions are beyond the scope
of this paper.

$q$ incorporates the subconscious biases of the players, the
day-to-day distribution of their moods, and more generally the full
physical stochastic nature of their separate decision-making
algorithms. It also reflects what the players know about each other,
whether they have directly interacted before, what they know about the
game structure, their information sets, etc. In short, $q$ {\it{is}}
the physical nature of the game setup, {\it{in toto}}.  Note that we
cannot examine the precise states of all the neurons and
neurotransmitters in the brains of the players, and even if we could
we cannot precisely evaluate the associated stochastic
dynamics. Accordingly, while it is physically real, in practice it is
impossible for us to ascertain a distribution $q_i$ exactly.

In contrast, $x$ is the quantity that the players consciously
determine, and it is observable. $q$ is instead the physical process
that specifies how the players select that observable.  Both of these
quantities differ from our (limited) knowledge about $q$. That
knowledge is embodied in probability distributions $P(q)$. $P$
reflects $us$ as much as the players.

We imagine an infinite set of instances of our setup, i.e., an
infinite set of $q$'s.  The invariant $\mathscr{I}$ specifies all
characteristics of the system --- and only those aspects --- such that
if they had been different in some particular instance, we would know
it.  In going from one instance to the next, we assume the problem is
``reset'', i.e., there is no information conveyed from one instance to
the next. (In particular the players' minds are ``wiped clean''
between instances.)  Our inference problem is to predict $q$ based on
such an invariant, i.e., formulate the posterior $P(q \mid
{\mathscr{I}})$. So an ``entropic prior'' concerns the probability of
any such joint mixed strategy $q$, our likelihood must concern $q$,
etc.

In type II game theory, what player $i$ chooses in any particular
instance of a game is a mixed strategy, $q_i(x_i)$.  Each player $i$'s
mixed strategy is separately randomly sampled, ``by Nature'', to get
the player's move $x_i$. So as in type I games, $q(x)
\triangleq \prod_i q_i(x_i)$.

In general, in real world type II games, a particular human's choice
of mixed strategy will vary depending on their mood, how distracted
they are, etc.  Accordingly the joint choice of the players is
described by a distribution $\pi(q)$.  $\pi$ reflects the stochastic
nature of the players, what they know about each other, what they know
about the game structure, etc. In short, $\pi$ {\it{is}} the physical
nature of the setup. In contrast, our (limited) knowledge about $\pi$
is embodied in a distribution $P(\pi)$.

As usual, we imagine an infinite set of instances of our setup, i.e.,
an infinite set of $\pi$'s, with no information conveyed between
instances.  The invariant $\mathscr{I}$ specifies all characteristics
of the system --- and only those aspects --- such that we would know
if they had been different in some particular instance. Our inference
problem is to predict $\pi$, i.e., formulate the posterior $P(\pi \mid
{\mathscr{I}})$.  So an ``entropic prior'' concerns the probability of
any such joint distribution $\pi$, our likelihood must concern $\pi$,
etc. Note that the invariant setting the likelihood, $P({\mathscr{I}}
\mid \pi)$, is a characteristic of an entire distribution over joint
mixed strategies (namely $\pi$), not (directly) of the joint mixed
strategies themselves.

In both game types, since $q$ is a product distribution, {\it{if one 
is given $q$}} then knowing one player's move provides no extra 
information about another player's move. (Formally, $P(x_i \mid q, 
x_j) = P(x_i \mid q)$.) However in the absence of knowing $q$ in full, 
knowing just the $q_i$ of some player $i$ may provide information 
about other the $q_j$ of other players $j$. For example, this could be 
the case if those mixed strategies $q_i$ and $q_j$ are determined in 
part based on previous interactions between the players.  Similarly, 
even if the players have never previously interacted, if there is 
overlap in what they each know about the game (e.g., they each know 
the utility functions of all players), that might couple members of 
the set \{$q_i$\}. Accordingly, in  type I game theory $P(q 
\mid {\mathscr{I}})$ need not be a product distribution (over the
$q_i$) in general, and in type II game theory $\pi$ need not be a
product distribution.

In general, which game type one uses to cast the problem is set by the
problem at hand. If the players all consciously choose mixed
strategies --- if that's how their thought processes work --- then we
have a type II game. If the players choose moves, we have a type I
game.  One can even have mixed game types, in which some players
choose moves, and some choose mixed strategies. Our lacking knowledge
of what scenario we face is analogous to lack of knowledge concerning
the payoff structure: our inference problem is not fully
specified.{\footnote{Of course, the lack of knowledge underlying both
game types can in principle be addressed by setting a prior
probability distribution over the underlying unknown and defining an
associated likelihood function. Here that would mean distributions
over whether each player chooses moves or chooses mixed strategies. No
such analysis which would essentially mix the two game types is
considered in this paper.}}

For the reasons elaborated above, we will adopt the entropic prior for
both game types. Note that for either game type, the entropic prior
evaluated for a product distribution is itself a product, i.e., if
$q(x) = \prod_i q_i(x_i)$, then $e^{\alpha S(q)} = \prod_i e^{\alpha
S(q_i)}$. As a result, by symmetry the associated marginal over $x$,
\begin{equation}
\int dx \; q(x) P(x) \propto \int dx \; \prod_i q_i(x_i) e^{\alpha
S(q_i)},
\end{equation}
must be uniform over $x$.

In some situations $P(q \mid {\mathscr{I}})$ will not be of interest,
but rather the associated posterior
\begin{eqnarray}
P(x \mid {\mathscr{I}}) = \int dq  \; P(q \mid {\mathscr{I}}) \prod_i
q_i(x_i)
\end{eqnarray}
will be. Now for the entropic prior, we know what the associated prior
$P(x)$ is (it's uniform). This suggests one formulate a
likelihood $P({\mathscr{I}}
\mid x)$. One could then use Bayes' theorem with the uniform $P(x)$ to arrive at the
posterior $P(x \mid {\mathscr{I}})$ directly, rather than arrive at it
via the intermediate variable $q$. This would constitute a third type
of game, in addition to the other two presented above, in which
instances would be $x$'s rather than $q$'s or $\pi$'s.

Unfortunately, it is hard to see how to formulate the likelihood
$P({\mathscr{I}} \mid x)$ without employing $q$ or $\pi$. Recall that
an invariant $\mathscr{I}$ is the set of all physical instances that
can occur in our inference problem, and no other instances. However in
formulating $P({\mathscr{I}} \mid x)$ instances are specified by
values of $x$, and for almost any inference problem, all $x$'s may
occur.  So in any such inference problem, the associated
$P({\mathscr{I}} \mid x)$ does not exclude any $x$ at all, i.e., it is
vacuous, as far as inference of $x$ is concerned. It is also hard to
see what might be gained by using such an alternative game
type. Indeed, since it conflates the distribution $q$ concerning
physical reality with the distribution $P$ concerning our (lack of)
knowledge about that reality, one would expect substantial losses of
insight if one used such an alternative game type for one's analysis.

As a final notational comment, we will use the following shorthand for
each $i$'s ``effective utility'', sometimes called $i$'s {\bf{environment}}:
\begin{equation}
U^i(x_i) \triangleq E(u^i \mid x_i).
\end{equation}
In type I game theory this reduces to
\begin{eqnarray}
U^i(x_i) &=& E_{q_{_{-i}}}(u^i \mid x_i) = \int dx_{-i} \;
q_{-i}(x_{-i} \mid x_i) u^i(x_i, x_{-i}) \nonumber \\
&=& \int dx_{-i} \;  u^i(x) \prod_{j \ne i} q_{j}(x_{j}) \; \; \;
{\mbox{(since}} \; q \; {\mbox{is a product distribution)}}.
\end{eqnarray}
We will write $E_{q_{_{-i}}}(u^i \mid
x_i)$ as $U^i_{q_{_{-i}}}(x_i)$ when the $q_{-i}$ defining $U^i$ needs
to be made explicit. We will also write 
\begin{equation}
E_q(u^i) = q_i \cdot U^i
\end{equation}
when working with type I games.  (The expansion of $U^i$ for type II
games proceeds analogously.)

\subsection{Needed Mathematical Tools}
\label{sec:eff_inv}

This section presents some preliminary mathematical tools from
statistical physics that are useful for performing Bayesian analysis
of games using entropic priors.  In essence, these tools amount to a
suite of relationships involving the Boltzmann distribution, entropy,
and optimization. Though it is a bit laborious to work through these
tools, they are crucial for understanding bounded rational players in
general, and for understanding the QRE in particular.  We will focus
on type I games; as usual, similar considerations apply for type II
games.

$ $

{\bf{(1)}} Start by noting that if we take its logarithm, any distribution
$q_i(x_i)$ can be expressed as an exponential of some function over
$x_i$. So in particular we can write any MAP $q_i$ that way:
\begin{equation}
{\mbox{argmax}}_{q_i} P(q_i \mid {\mathscr{I}}) \propto e^{\beta_i
f_i(x_i)}
\label{eq:ind_players}
\end{equation}
for some appropriate function $f_i$ and constant
$\beta_i \ge 0$.{\footnote{Of course, there is always freedom to absorb some
portion of any $\beta_i$ into the associated $ f_i$, but that is
irrelevant for current purposes.}}

$ $

{\bf{(2)}} We now relate the formulation of an MAP $q_i$ in
exponential form (as in (1)) to a particular choice for our game
theory problem's invariant. 

Say we associate with each player $i$ a ``guess'' she makes
(potentially explicitly, potentially not) for her environment
function, $U^i_{q_{_{-i}}}$. Write that guessed function as
$f_i(x_i)$. We presume that we can view the player's behavior as
though she were trying to perform well for that (guessed)
environment. Formally, we presume that in each instance of our
inference problem, the mixed strategy of player $i$ results in the
same (invariant) value $K_i$ for what $E(U^i)$ would be if player
$i$'s guess for her environment $U^i$ were correct, i.e., if $U^i$
equalled $f_i$. So the invariant of the game for player $i$ is
\begin{equation}
q_i \cdot f_i = K_i.
\label{eq:counterfact}
\end{equation}
Intuitively, with this invariant, as one goes from one instance of the
inference problem to the next, we presume that player $i$ is always
just as smart, as measured with the (potentially counterfactual)
environment function $f_i$.

For this invariant, the likelihood $P({\mathscr{I}} \mid q_i)$
restricts $q_i$ to lie on the hyperplane of distributions obeying
Eq.~\ref{eq:counterfact}:
\begin{equation}
P({\mathscr{I}} \mid q_i) = \delta(q_i \cdot f_i - K_i).
\label{eq:eff_like}
\end{equation}
So given our use of the entropic prior, the posterior for player $i$
is
\begin{equation}
P(q_i \mid {\mathscr{I}}) \propto e^{\alpha S(q_i)} \delta(q_i \cdot
f_i - K_i). 
\label{eq:eff_post}
\end{equation}
Accordingly the MAP $q_i$ is given by the $q$ maximizing the
so-called {\bf{maxent Lagrangian}},
\begin{equation}
{\mathscr{L}}(q_i) \triangleq  S(q_i) + \beta_i [q_i \cdot f_i - K_i] 
\label{eq:eff_lag}
\end{equation}
where in the usual way the $\beta_i$ are the Lagrange parameters, here
divided by $\alpha$.{\footnote{ As an aside, say that we replaced
Eq.~\ref{eq:counterfact} with the inequality constraint $q_i \cdot f_i
> K_i$. The entropy function is concave, and so is this inequality
constraint. Accordingly, by Slater's theorem, there is zero duality
gap~\cite{bova03} and we can apply the KKT conditions to get a
solution. In other words, for this modified invariant the maxent
Lagrangian still applies, and therefore so does the solution of
Eq.~\ref{eq:ind_players}.}} 

Solving for the $q_i$ minimizing this Lagrangian, $q_i^{\beta_i}$, we
get the distribution of Eq.~\ref{eq:ind_players}, with $\beta_i$ a
function of $K_i$. Equivalently, we can take $K_i$ to be a function of
$\beta_i$. This is what we will do below.

Now parallel the conventionaly nomenclature of statistical physics,
and define the {\bf{partition function}}
\begin{equation} 
Z_{f_i}(\beta_i) \triangleq \int dx_i e^{\beta_i f_i(x_i)}.
\end{equation}
(Note that the partition function is the normalization constant of
Eq.~\ref{eq:ind_players}.) Then using Eq.~\ref{eq:ind_players} to
express the Boltzmann distribution $q_i^{\beta_i}$, our constraint
Eq.~\ref{eq:eff_like} means that
\begin{eqnarray}
K_i(\beta_i)  = f_i \cdot q_i = 
\frac{d {\mbox{ln}}(Z_{f_i}(\beta_i))}{d \beta_i}
\label{eq:K_i}
\end{eqnarray} 
as is readily verified by evaluating the derivative.  As shorthand,
sometimes we will absorb $\beta_i$ into $f_i$, and simply write $Z(V)
\triangleq Z_V(1)$.

So say we are given some distribution $q_i$. Take its logarithm to get
a function $f_i$ and exponent $\beta_i$. Then use Eq.~\ref{eq:K_i} to
translate that $f_i$ and $\beta_i$ into a value of $K_i$. Using these
choices of $K_i$, $f_i$ and $\beta_i$, formulate the associated
invariant $\mathscr{I}$ given by Eq.~\ref{eq:counterfact}. As shown by
Eq.~\ref{eq:ind_players}, the MAP distribution for that $\mathscr{I}$
is our starting distribution $q_i$. In this way we can view that
$\mathscr{I}$ as the ``effective'' invariant for this (arbitrary)
starting $q_i$.  We can translate $any$ $q_i$ into an MAP distribution
by choosing an appropriate $\mathscr{I}$ this way.

$ $

{\bf{(3)}} We will refer to the function $K_i(.)$ arising in
Eq.~\ref{eq:K_i}, which maps $\beta_i$ to the expected value of $f_i$
under $q_i^{\beta_i}(.)$, as the {\bf{Boltzmann utility}} for player
$i$, where $f_i$ is implicit. We now present some general
characteristics of the Boltzmann utility, characteristics that are
particularly important for understanding the QRE.

First, with slight abuse of terminology, we will sometimes write the
Boltzmann utility with $f_i$ explicitly listed as the first argument
and the subscript $i$ dropped, i.e., as $K(f, \beta \in
{\mathbb{R}}^+)$.  In this case it is the domain of the first argument
of the Boltzmann utility $K(f,
\beta)$ that (implicitly) sets the space $X_i$ to be integrated over to
evaluate $K(f, \beta)$.{\footnote{Note that despite the terminology,
the Boltzmann utility is not a ``utility function'' in the sense of a
mapping from $x$ to $\mathbb{R}$. Rather it's what expected utility
would be for a particular type of mixed strategy, in a particular
environment, as a function of parameters of that mixed strategy.}}

As is readily verified, the variance (over $f_i$ values) of the
Boltzmann distribution of Eq.~\ref{eq:ind_players} is given by the
derivative of $K_i(\beta_i)$ with respect to $\beta_i$. Since
variances are non-negative, this means that $K_i(\beta_i)$ is a
non-decreasing function. In fact, for fixed $f_i$, so long as $f_i$ is
not a constant-valued function (i.e., not independent of its
argument), the associated Boltzmann utility $K_i(.)$ is a
monotonically increasing bijection with domain $\beta_i
\in [0, \infty)$ and associated range $[\frac{\int dx_i
\; f_i(x_i)}{|X_i|}, {\mbox{max}}_{x_i} f_i(x_i))$.\footnote{To see
this, note that the variance is non-zero for all $\beta_i < \infty$,
so long as $f_i(x_i)$ is not a constant. Accordingly, under such
circumstances $K_i(\beta_i)$ is invertible.}

We extend the domain of definition of $K_i$ by adding to it the special
value ``$\infty^*$'', and defining $K_i(\infty^*) = {\mbox{max}}_{x_i}
f_i(x_i)$. This makes $K_i$ a bijection whose domain is $\beta_i \in [0,
\infty) \; \cup \; \infty^*$ when
$f_i(x_i)$ is not a constant, and is the singleton \{$\infty^*$\}
otherwise. In both cases the range is $[\frac{\int dx_i
\; f_i(x_i)}{|X_i|}, {\mbox{max}}_{x_i} f_i(x_i)]$. 

With some abuse of notation from now on we
will extend the meaning of the linear ordering ``$\ge$'' to have $\infty^*
\ge k \; \forall k \in {\mathbb{R}}$. We will
will also drop the asterisk superscript from ``$\infty^*$'', relying
on the context to make the meaning of ``$\infty$'' clear. We will
engage in more abuse by writing ``$\vec{b}$'' even if some component
$b_i = \infty$ (so that $\vec{b}$ is not a Euclidean vector, properly
speaking).

Just as expected $f_i$ cannot decrease as the Boltzmann exponent
rises, the entropy of a Boltzmann distribution $e^{\beta_i f_i(x)} /
Z_{f_i}(\beta_i)$ cannot increase as its Boltzmann exponent $\beta_i$
rises.\footnote{To see this say we replace the invariant $q_i
\cdot f_i = K_i(\beta_i)$ with $q_i \cdot f_i
\ge K_i(\beta_i)$. Then for fixed $q_{-i}$, the MAP $q_i$ is the $q_i$
that maximizes $S(q_i)$ subject to that inequality constraint that
$q_i \cdot f_i \ge K_i(\beta_i)$. The entropy is a concave function
of its argument, as is this inequality constraint, so our problem is
concave. Therefore the critical point of the associated Lagrangian is
the MAP $q_i$.  Now if we increase $\beta_i$, and therefore increase
$K_i$, the feasible region for our new invariant decreases. This means
that when we do that the maximal feasible value of $S$ cannot
increase. So the entropy of the critical point of the Lagrangian for
our new invariant cannot increase as $\beta_i$ does. However that
critical point is just the Boltzmann distribution $q_i(x_i) \propto
exp(-\beta_i f_i(x_i))$, i.e., it is the MAP $q_i$ for original
equality invariant, $q_i \cdot f_i = K_i$.  So the property that
increasing $\beta_i$ cannot increase the entropy must also hold for
the original equality invariant.} So the picture that emerges is that
as $\beta_i$ increases, the Boltzmann distribution gets more peaked,
with lower entropy. At the same time, it also gets higher associated
expected value of $f_i$.

$ $

{\bf{(4)}} We now extend the discussion to allow $q_i$ not to be
$q^{\beta_i}_i$, the Boltzmann distribution over values of $f_i$ with
exponent $\beta_i$ (the distribution in
Eq.~\ref{eq:ind_players}). This extension will prove important in
quantifying the rationality of a player based solely on their mixed
strategy and environment (Sec.~\ref{sec:rat} below).

First expand $S(q_i)$ for the case where in fact $q_i$ does equal the
Boltzmann distribution $q^{\beta_i}_i$ . Then using
Eq.~\ref{eq:counterfact}, we see that for this Boltzmann distribution
$q_i$,
\begin{equation}
q_i \cdot f_i + \frac{S(q_i)}{\beta_i} = \frac{{\mbox{ln}(Z_{f_i}(\beta_i))}}{\beta_i}.
\label{eq:free_utility}
\end{equation}
Comparing with Eq.~\ref{eq:free_energy} in App.~\ref{sec:physics}, we
see that the quantity on the left-hand side is essentially identical
to the free energy of statistical physics.{\footnote{Free energy has a
different sign on the entropy. This just reflects the fact that
players work to raise utility whereas physical systems work to
minimize energy.}}  Accordingly, we call it the {\bf{free utility}} of
the player.

Note that the free utility is a function of $q_i$ and $\beta_i$, and
is defined even when $q_i$ is $not$ a Boltzmann distribution over
values of $f_i$. In contrast, the quantity on the right-hand side of
Eq.~\ref{eq:free_utility} is only a function of $\beta_i$.  For fixed
$\beta_i$, that quantity on the right-hand side of
Eq.~\ref{eq:free_utility} is an upper bound on the free utility of the
player.{\footnote{This follows from the fact that the $q_i$ that
maximizes the free utility for our $\beta_i$ is just the associated
Boltzmann distribution.}}  For that fixed $\beta_i$, the {\bf{free
utility gap}} of player $i$ is defined as the difference between its
actual free utility and the maximum possible at that $\beta_i$,
$\frac{{\mbox{ln}}(Z_{f_i}(\beta_i))}{\beta_i}$. That gap is zero ---
player $i$'s free utility is maximized --- at player $i$'s associated
equilibrium (Boltzmann) mixed strategy. Intuitively, player $i$
``tries to'' maximize free utility rather than expected utility,
insofar as it ``tries to'' achieve its MAP mixed strategy.

$ $

{\bf{(5)}} Finally, we present a restriction that simplifies our
discussion below of how the support of $P(q \mid {\mathscr{I}})$
covers $\Delta_{{\cal{X}}}$ (Sec.~\ref{sec:coverage} below).

We say that a particular $q$ is {\bf{benign}} for utilities \{$u^i$\}
if for all players $i$, we can write the associated expected utility
$q_i \cdot U^i_{q_{-i}}$ as $K(U^i_{q_{-i}}, \beta_i)$ for a $\beta_i
> 0$ with $U^i_{q_{-i}}$ defined in terms of $u^i$ and $q_{-i}$ in the
usual way.  In this paper, for simplicity we will only consider benign
$q$'s. This means in particular that we assume that for all
players, their expected utility is not worse than the one they would
get for a uniform mixed strategy (which corresponds to infinite
$\beta_i$). While the analysis can be extended to allow negative
$\beta_i$ (where player $i$ adopts a worse-than-uniform mixed
strategy), there is no need for such considerations here.

\subsection{Effective invariants and the QRE}

As discussed above, every product distribution $q$ can be specified by
saying that each of its marginalizations $q_i$ is an MAP prediction
for some associated ``guessed environment'' $f_i$ and $\beta_i$. But
not every $q$ can be expressed this way if we demand that the guessed
environments $f_i$ for each player are their actual environments. In
other words, only for a subset of all $q$'s will it be the case that
$f_i(x_i)
\triangleq U^i_{q_{-i}}(x_i) \; \forall i$. 

Demanding such self-consistency in $q$ results in a coupled set of
nonlinear equations for $q$. This is the set of equations that specifies
the QRE, Eq.~\ref{eq:qre}.{\footnote{In
\cite{mcpa95}, $U^i_{q_{-i}}$ is called ``a statistical reaction function'',
and the set of coupled equations giving that solution is called the
``logit equilibrium correspondence''.}}  It was first derived in this
manner, as the self-consistent solution to a set of MAP inferences,
in~\cite{wolp03b,wolp04a,wolp04c}.

Note that there is no particular decision-theoretic significance to
the QRE derived in this manner. In particular, it is not the
Bayes-optimal solution to any inference problem. Nor is it derived as
the MAP solution to any (single) problem. Rather it is given by a set
of MAP solutions, each for a separate inference problem. There is one
such problem for each separate player. We then posthoc ``tie
together'' those separate problems, by requiring that our solutions to
them are consistent with one another.

Unfortunately, such a two-stage process has no clear justification in
terms of Cox's and Savage's axioms. More generally, it is hard to
formally justify the approach of enforcing consistency among a set of
separate inference problems rather than considering a single aggregate
inference problem.  (Recall that the inference is being done by the
external scientist, and that that scientist is external to the
system.) To address that single inference problem, we must use a
single invariant that concerns the entire joint system.

In such an alternative to the QRE's posthoc ``tying together'' one
analyzes all players' distributions simultaneously, from the very
beginning. This means that we analyze the distribution over full joint
strategies that involve all the players. In this approach, to get any
particular player $i$'s distribution we would marginalize the
distribution over joint strategies, rather than (as in the QRE) start
with those marginal distributions and try to tie them together.

The natural invariant for this aggregate inference problem is the
``aggregate invariant'' that $q_i \cdot f_i = K_i \; \forall i$.
However as shown below, the QRE is not even the MAP of the posterior
over the space of joint strategies under this invariant (never mind
being Bayes-optimal.)  An analysis of this aggregate problem is the
subject of the next few sections. A discussion of the historical
context of the QRE can be found in an appendix.

\section{Coupled players}
\label{sec:coupled-players}

Recall that the posterior is given by the prior and the
likelihood. Since (for both game types) we've chosen the prior, our
next task is to set the ($\mathscr{I}$-based) likelihood.  We want
that likelihood to have the same form as the likelihood underpinning
the CE of statistical physics: a Heaviside theta function that
restricts attention to a subset of all possible systems, with the
distribution across that subset then set by our prior (see the
appendix). However as elaborated below, the likelihood theta function
appropriate for games is more complicated than the one that arises in
statistical physics. This is because there are multiple payoff
functions in games, each with its own effect on the system's theta
function, whereas there is only one Hamiltonian in a statistical
physics system.

In this section we illustrate how to set the likelihood in a
scenario where the players may have knowingly interacted with each
other before the current game. (In the next section we use these
results to address the case where the players have not previously
knowingly interacted.) In general those previous interactions are
allowed to vary from one instance to another; the invariant
restricting our instances will also be what restrictsfg the possible
previous interactions.

\subsection{Invariants of human players}
\label{sec:rat_invs}

In general of course, $\mathscr{I}$ does not specify everything about
our inference problem, and in particular it does not specify the value
of that which we want to infer. Here what we wish to infer is the
actual joint strategy of the players. (See Sec.~\ref{sec:ent_prior}.)
So the joint strategy is not specified by $\mathscr{I}$. Therefore a
player's payoff is also not speciefied for any particular one of its
moves, since that payoff will depend on the moves of the other players
in general; that payoff may vary between instances.

Instead, here we stipulate that any player will try to maximize her
expected utility, to the best of her computational abilities, the best
of her insights into the other players and the game structure,
etc.{\footnote{This is not the case in situations like Allais'
paradox; see below.}}  Intuitively, this means we assume a
``pressure'' embodied in the distribution over $q$'s biasing the
distribution to have $q_i$ that achieve high values of $U_q^i
\cdot q_i$.  This pressure is matched by counterpressures from the
other players affecting $U^i_q$.

What we consider invariant is that from one instance to the next
player $i$ does not change, and therefore how insightful player $i$ is
into the other players (based on her previous interactions with them),
how computationally powerful $i$ is, etc., does not change.  In other
words, how well player $i$ performs, {\it{in light of her (varying)
environment of possible payoffs}} (i.e., in light of $U^i$) is the
same in all instances. In short,  ``how smart'' every player
$i$ is does not change from one instance to the next. As in the case of
statistical physics though, here our invariant need not specify
precisely how smart each player is {\it{a priori}}, only that how smart
each of them is doesn't vary from one instance to the next.

As an example, consider the situation where the players knowingly are
repeatedly playing the game with each other, forming a sequence of
games.  Say we are considering the distribution over joint mixed
strategies $q$ at some fixed (invariant) sequence index $t$. In this
scenario it is an entire sequence of games leading up to game $t$ that
constitutes ``an instance of our inference problem''.  We must
determine what is invariant from one instance of that problem to
another.

Note that in game $t$ of any instance the players' actual moves are
independent, tautologically. (This is reflected in having $q$ be a
product distribution.) However in general the $q$ at $t$ will change
from one instance to the next. Indeed, consider any time $t' \le t$.
At that time, in every sequence of games, each player modifies its
mixed strategy based on the history of move-payoff pairs in that
sequence for times previous to $t'$, i.e., each player tries to learn
what strategy is best based on its history and adapts its strategy
accordingly.  Since the move-payoff pairs of that history are formed
by statistical sampling (of the joint mixed strategy), they will not
be the same in all sequences. Accordingly, in general the modification
$i$ makes to its mixed strategy at $t'$ will not be the same in all
sequences. Therefore the final joint mixed strategy $q$ will vary from
one sequence to the next.

As a result of this sampling, across the set of all instances (i.e.,
all sequences) there will be some statistical coupling between the
time $t$ mixed strategies of the separate players. This means in
particular that in general the time $t$ MAP $q$, argmax$_{q^t} P(q^t
\mid {\mbox{invariant}})$, is not a product of the individual time $t$
MAP $q_i$, 
\begin{eqnarray}
\prod_i {\mbox{argmax}}_{q^t_i} P(q_i^t \mid {\mbox{invariant}}) &=&
\prod_i {\mbox{argmax}}_{q^t_i} \int dq^t_{-i} P(q^t_i, q^t_{-i} \mid
{\mbox{invariant}}) \nonumber \\
&\ne& {\mbox{argmax}}_{q} P(q^t \mid {\mbox{invariant}})
\end{eqnarray}
(This is in contrast to the case with independent players considered
in Sec.~\ref{sec:ind_players}).

Since the final $q$ varies across the instances, in general we can't
expect that for each player $i$, its environment $U^i$ will be the
same at the end of each sequence. Indeed, even consider the case where
play evolves to a Nash equilibrium at time $t$. If the game has
multiple such Nash equilibria, then in general which one holds for a
particular sequence of games will depend on the history of moves and
payoffs in that sequence. Accordingly, $U^i$ will depend on that
sequence.

Formalizing all this means formalizing our invariant $\mathscr{I}$ of
``how smart'' a player is.  Here we consider how to do this for
type I game theory, where inference is of $q$. The discussion for type
II game theory proceeds {\it{mutatis mutandi}}.

Consider just those instances of our inference problem in which player
$i$ is confronted with some particular vector of move-conditioned
expected utility values, $U^i$.  We say that that $i$ is ``as smart''
in any one of those instances as another if in each of them
separately, on average, the move $i$ chooses has the same payoff. In
other words, how smart $i$'s is is the same in all of those instances
if $i$'s expected utility, $q_i \cdot U^i$, has the same (potentially
unknown) value in all of them. We write that value as
$\epsilon_i(U^i)$.  As an example, at a Nash equilibrium
$\epsilon_i(U^i) = {\mbox{max}}_{x_i} U^i(x_i) \;
\forall i$. 

Note how conservative this restriction on $q_i$ is. In particular, so
long as $\epsilon_i(U^i) < {\mbox{max}}_{x_i} U^i(x_i)
\; \forall i$, then we are guaranteed that multiple $q_i$ satisfy this
restriction. This is true even if the game has only a single Nash
equilibrium. 

Our invariant is simply that the functions \{$\epsilon_i$\} are the
same in all instances. This invariant does not concern the joint
choices (moves) of the players across the instances (which is given by
the $x$'s). Rather it concerns $q$, which is the physical nature of
the process driving the players to make those choices.  However the
invariant does not specify that process.  In particular it does not
stipulate how the players reason concerning each other. For example,
it does not stipulate how many levels of analysis of the sort ``I know
that you know that I know that you prefer ...''  any of the players go
through (if any levels at all). All that $\mathscr{I}$ stipulates is
that certain high-level encapsulations of that decision-making, given
by the \{$\epsilon_i$\}, are the same in all instances.

As a result of this invariance, even though the moves
\{$x_i$\} of the players are independent in any particular instance
(since $q$ is a product distribution), our (!) lack of knowledge
concerning the set of all the instances might result in a posterior
$P(q \mid {\mathscr{I}})$ in which the distributions \{$q_i$\} are
statistically coupled.  (Recall that $q$ reflects the players, and $P$
reflects our inference concerning them.) 

Note that for the entropic prior $P(q)$ there is no statistical
coupling between $x_i$ and $x_j$ in the prior distribution
$P(x)$. (Recall that for that prior, $P(x) = \int dq P(x \mid q) P(q)$
must be uniform, by symmetry.)  However the potential coupling between
the \{$q_i$\} means that in the posterior distribution, the moves are
$not$ statistically independent (assuming one doesn't condition on
$q$). Formally,
\begin{eqnarray}
P(x_i \mid {\mathscr{I}}) &=& \int dq_i \; P(x_i \mid q_i) P(q_i \mid {\mathscr{I}}) \nonumber \\
       &=& \int dq_i \; q_i(x_i) P(q_i \mid {\mathscr{I}})
\end{eqnarray}
so 
\begin{equation}
\prod_i P(x_i \mid {\mathscr{I}}) = \int dq \; \prod_i P(q_i \mid {\mathscr{I}}) q_i(x_i).
\end{equation}
On the other hand, recall that
\begin{equation}
P(x \mid {\mathscr{I}}) = \int dq  \; P(q \mid {\mathscr{I}}) \prod_i
q_i(x_i).
\label{eq:post_moves}
\end{equation}
So if $P(q \mid {\mathscr{I}})$ is not a product distribution, then in
general $P(x \mid {\mathscr{I}}) \ne \prod_i P(x_i \mid
{\mathscr{I}})$, i.e., in this situation $P(x \mid {\mathscr{I}})$ ---
which is the distribution over joint moves reflecting our
understanding of the system --- is not a product distribution
either. In such a situation, {\it{to us}}, $x_i$ and $x_j$ are
statistically coupled.

Such coupling also typically arises in the Bayes-optimal prediction
for the distribution over joint strategies. Indeed, say we adopt a
quadratic loss function, so that if we guess the joint distribution is
$q''$, when in fact it is $q'$, our loss is $(q'' - q')^2$. Then given
the posterior $P(q \mid {\mathscr{I}})$, the associated Bayes-optimal
prediction for $q$ --- the prediction that minimizes our posterior
expected quadratic loss --- is 
\begin{eqnarray}
p_{quad} &\triangleq& \int dq \; q P(q \mid
{\mathscr{I}}).
\end{eqnarray}
This is the same as the joint mixed strategy given by
Eq.~\ref{eq:post_moves}. (This is not the case for other loss
functions.)  Accordingly, our conclusion about coupling of the
\{$x_i$\} holds for this Bayes-optimal joint mixed
strategy.{\footnote{Note the slight abuse of terminology; the moves of
the players are statistically coupled in this ``joint mixed
strategy'', which is why we do not write that Bayes-optimal
distribution over $x$ as $q$ but a $p$.}}

Consider changing the cognitive process of some player $j \ne i$ in
way that does not change $\epsilon_j$. Also do not change anything
concerning all the other players. Do all this in such a way that how
that player $j$ chooses moves at time $t$ changes, but nothing else
changes about $j$'s behavior. So in particular, for any fixed vector
$U^i$, the $q_j$ that governs player $j$ at time $t$ and is consistent
with that $U^i$ will change.{\footnote{We mean ``consistent'' in the
sense that even though $q_{j \ne i}$ has changed, it is still true
that
\begin{eqnarray*}
U^i(x_i) &=& \int dx'_{-i} \; u^i(x_i, x'_{-i}) q_{-i}(x'_{-i})
\\
&=& \int dx'_{j} dx'_{-\{i,j\}} \; u^i(x_i, x'_j, x'_{-\{i,j\}}) q_j(x'_j)
q_{-\{i,j\}}(x'_{-\{i,j\}}).
\end{eqnarray*}}}
Now the distribution over possible $q_i$ at time $t$ is based on
behavior of player $i$ and of other players for times $t' \le
t$. Since those factors are unchanged by our change to $q_j$ at time
$t$, so is the distribution over possible $q_i$ then. Accordingly, the
change in $q_j$ will in general change the expected utility of player
$i$ at time $t$. In other words, changing player $j \ne i$ will in
general change $\epsilon_i$. This illustrates that our invariance is
implicitly determined by the set of players as a whole. This is in
addition to its reflecting how the players have interacted, the
structure of the game, etc.{\footnote{Note how problematic it would
be to try to encapsulate our invariance in a traditional,
non-Jaynesian ``bath''-based approach to statistical physics. In such
an approach, the invariants are the sums, across both the system under
consideration and an external ``bath'', of certain physical
quantities. In other words, the aggregate amount of those quantities
across the system and the external bath is taken to be conserved. For
example, the CE arises if one takes the aggregate energy to be
conserved. It is not at all clear how one could express our invariants
as the values of such conserved quantities.}}

\subsection{Specifying the function $\epsilon_i$}
\label{sec:epsilon}

Now in general for any player $i$, our invariant doesn't force all
instances to have the same vector $U^i$. So to complete the
quantification of how smart a player is we need to specify the
function $\epsilon_i$. To do this we use a Gedanken experiment; we
consider how player $i$ would behave in a counterfactual ``game
against Nature'' inference problem. In that new problem we focus on
just one player $i$, fixing the others. Formally, our invariant is
expanded from that of the original problem, to a new invariant
${\mathscr{I}}'$ that also include $U^i$. Since the invariant still
stipulates that $E(u^i) = \epsilon_i(U^i)$, having $U^i$ also
invariant means that the expected utility $E(u^i)$ does not change
between instances of this new problem.

Write the (potentially unknown) value of that invariant expected
utility as $v_i$. Since we use the entropic prior over $q_i$ this new
inference problem has the usual entropic posterior. Also as usual, the
MAP $q_i$ is given by a Boltzmann distribution:
\begin{equation}
q^*_i(x_i) \propto e^{b_i U^i(x_i)}
\label{eq:boltz_q_i}
\end{equation}
where the Lagrange parameter going into $b_i$ enforces the constraint,
namely that $q^*_i \cdot U^i = v_i$. (See Sec.~\ref{sec:eff_inv} for
the more general way that this constraint arises and some useful
equalities relating $b_i$, $v_i$, etc.)

We must now consider how $b_i$ changes as $U^i$ changes. The lowest
order case is where $b_i$ is a constant, independent of $U^i$. This
means that for real-valued $b_i$, $\epsilon_i$ is identical to the
Boltzmann utility $K_i$ discussed in Sec.~\ref{sec:eff_inv}, with
$U^i$ playing the role that $f_i$ does in the definition of Boltzmann
utility, and $b_i$ playing the role of $\beta_i$. Just as we extend
the domain of definition of $K_i(.)$ to include $\infty$, we do the
same for $\epsilon_i$ and for $q^*_i$: For $b_i = \infty$, $q^*_i$ is
the distribution that is uniform over the set argmax$_{x_i}
U^i(x_i)$, and zero elsewhere. 

Below we will use the shorthand $q^*(x) \triangleq \prod_i q^*_i(x_i)$
where for each $i$ the $U^i$ arising in the definition of $q^*_i$ is
understood to be based on the $q^*_{-i}$ (i.e., each $U^i$ means
$U^i_{q^*}$). So the definition of $q^*$ reflects coupling between the
player's mixed strategies (though not necessarily between their
moves): a change to $q^*_j$ for some particular $j$ in general will
modify the strategies $q^*_{k \ne j}$.  $q^*$ is the Quantal Response
Equilibrium (QRE) solution, discussed in Sec.~\ref{sec:eff_inv}.  In
general, for any particular game and $\vec{b}$, there is at least one,
and may be more than one associated $q^*$. This follows from Brouwer's
fixed point theorem~\cite{mcpa95, wolp04a}.

As a point of notation, the expression ${\mathscr{I}}_{\vec{b}}$ is
defined to be the invariant that $\forall i, q_i \cdot U^i_{q_{_{-i}}}
= K(U^i_{q_{_{-i}}}, b_i)$, where it is implicitly assumed that
${\vec{b}} \succeq {\vec{0}}$.  For any such $\vec{b}$ there is always
at least one $q$ that satisfies ${\mathscr{I}}_{\vec{b}}$ (e.g., the
QRE).

\subsection{The impossibility of a Nash equilibrium}
\label{sec:no_nash}

Say that for the coupled players invariant, $\mathscr{I}$, (the
support of) $P(q \mid {\mathscr{I}})$ is restricted to the Nash
equilibria of the underlying game, so that the players are perfectly
rational.  (See Sec.~\ref{sec:coverage} below.) Say that there are
multiple such equilibria, written $q^1, q^2,
\ldots$, with $P(q \mid {\mathscr{I}}) \triangleq \sum_j a^j \delta(q
- q^j)$. So the $a^i$ form a probability distribution over the
equilibria. Since the entropic prior extends over all $q \in
\Delta_{\cal{X}}$, in general none of the $a^i$ will equal zero exactly.

Since the equilibria are all product distributions, using
Eq.~\ref{eq:post_moves} we can write
\begin{eqnarray}
P(x \mid {\mathscr{I}}) &=& \sum_j a^j \prod_k q^j_k(x_k)
\end{eqnarray}
so that
\begin{eqnarray}
P(x_i \mid {\mathscr{I}}) &=& \int dx_{-i} \; P(x \mid {\mathscr{I}}) \nonumber \\
&=& \sum_j a^j q^j_i(x_i) .
\label{eq:sum_nashs}
\end{eqnarray}
Consider the case where the Nash equilibria are not exchangeable, so
$P(q \mid {\mathscr{I}})$ is not a product distribution. This means
that $P(x \mid {\mathscr{I}})$ is not a product distribution in
general, so that the players appear to be coupled, to us. (See the
discussion just below Eq.~\ref{eq:post_moves}.) 

At an intuitive level, such coupling is analogous to the
consistency-among-players coupling that underlies the concept of a
Nash equilibrium. However because it mixes the Nash equilibria with
each other, in general the sum in Eq.~\ref{eq:sum_nashs} is not a best
response mixed strategy for the product distribution $\prod_{j \ne i}
P(x_j \mid {\mathscr{I}})$. Formally, $p^i(x_i) \triangleq P(x_i
\mid {\mathscr{I}})$ does not maximize the dot product
\begin{equation}
\int dx_i \; p^i(x_i) \bigl[ \int dx_{-i} \; u(x_i, x_{-i}) \prod_j P(x_j \mid
{\mathscr{I}})\bigr].
\end{equation}
Similarly $P(x_i \mid {\mathscr{I}})$ is not a best-response mixed
strategy for $P(x_{-i} \mid {\mathscr{I}})$. So when the underlying
game has multiple non-exchangeable equilibria, then even if the
players are perfectly rational, in general we will not predict
distributions governing the moves of the players that are
best-response mixed strategies to each other.

Note that this conclusion does not depend critically on our choice of
$\epsilon_i$, or even on our choice of encapsulating $\mathscr{I}$ in
terms of such functions $\epsilon_i$. (After all, we're explicitly
allowing the case where $P(q \mid {\mathscr{I}})$ is restricted to
Nash equilibria.) Rather it comes from the fact that our prior allows
non-zero probability for all of the Nash equilibria.

\subsection{The QRE and $\epsilon_i$}
\label{sec:epsilon_i-disc}

Unlike the usual motivation of the QRE, the motivation for our choice
of $\epsilon_i$ does {\it{not}} say that $q_i$ must be a Boltzmann
distribution. It does not say that the probability distribution over
possible $q_i$ is a delta function about a Boltzmann distribution
$q_i$. Rather it says that $q^*_i$, the most likely $q_i$ for the
single-player inference problem, is a Boltzmann distribution. It then
uses that fact to motivate a functional form for $\epsilon_i$ in the
multi-player scenario.  Here we only assume that the relation between
$E_q(u^i)$ and $U^i$ given by that functional form is consistent with
$q_i = q^*_i$.  In general the invariant $E_q(u^i) = \epsilon_i(U^i;
b_i)$ holds for many $q_i$ in addition to the Boltzmann distribution.

Indeed, fix $q$, and consider any $i$ and the associated
$U^i_{q_{-i}}$.  Recall that we are restricting attention to benign
$q$'s (cf. the discussion at the end of Sec.~\ref{sec:eff_inv}). So no
matter what it is, our $q_i$ is consistent with our invariant for that
$U^i_{q_{-i}}$, for some $b_i$. Since this is true for all $i$, any
$q$ is consistent with our full invariant for some
$\vec{b}$. Furthermore, for any finite $\alpha$, the support of the
entropic prior is all $\Delta_{{\cal{X}}}$. This means that every $q$
has non-zero posterior probability $P(q \mid {\mathscr{I}}_{\vec{b}})$
for some $\vec{b}$.

In contrast not every $q_i$ is a Boltzmann distribution, i.e., not
every $q_i$ is part of a QRE. In other words, to assume a system is in
a QRE is to make a restrictive assumption about the physical system
$q$, an assumption that may or may not be correct. This is not the
case with our invariant.

Finally, it turns out that the QRE can be viewed as an approximation
to the MAP prediction for our $\mathscr{I}$. A detailed discussion of
this is presented in Sec.~\ref{sec:qre_map} below.

\subsection{The MAP $q$}

Given our invariant, our likelihood is
\begin{eqnarray}
P({\mathscr{I}} \mid q) &=& \prod_i \delta(E_q(u^i) -
\epsilon_i(U^i_q)) \nonumber \\
&=& \prod_i \delta(q_i \cdot U^i - \epsilon_i(U^i_q)).
\label{eq:like}
\end{eqnarray}
Recall that with the canonical ensemble the likelihood stipulates a
linear constraint on the underlying probability distribution. In
contrast, due to the nonlinearity of $\epsilon_i$, here the
likelihood stipulates a non-linear constraint on $q$.

As usual, if we wish we can distill the associated posterior into a
single prediction for $q$, e.g., into the MAP estimate.  Naively, one
might presume that $q^*$ is that MAP estimate. After all, $q^*$
respects our constraints that $E_q(u^i) = \epsilon_i(U_q^i) \;
\forall i$, and it maximizes the entropy of each player's strategy
considered in isolation of the others. However in general $q^*$ will
not maximize the entropy of the joint mixed strategy subject to our
constraints. In other words, while MAP for each individual player's
strategy, in general it is not MAP for the joint strategy of all the
players. The reason is that setting each separate $q_i$ to maximize
the associated entropy (subject to having $q$ obey our invariant), in
a sequence, one after the other, will not in general result in a $q$
that maximizes the sum of those entropies. So it will not in general
result in a $q$ that maximizes the entropy of the joint system.

Proceeding more carefully, the MAP estimate of the mixed strategy $q$
is given by the critical point of the Lagrangian
\begin{equation}
{\mathscr{L}}(q, \{\lambda_i\}) = S(q) + \sum_i \lambda_i(q_i \cdot U^i -
\epsilon_i(U^i))
\end{equation}
where the $\lambda_i$ are the Lagrange parameters enforcing the
constraints provided by the likelihood function of
Eq.~\ref{eq:like}. The critical point of this Lagrangian must satisfy
\begin{eqnarray}
0 &=& \frac{\partial {\mathscr{L}}}{\partial q_i(x_i)}  \nonumber\\
&=& -1 - {\mbox{ln}}[q_i(x_i)] + \lambda_i E(u^i \mid x_i) + \sum_{j \ne
i} \lambda_j [E(u^j \mid x_i) - \frac{\partial
\epsilon_j(U^j)}{\partial q_i(x_i)}] \nonumber \\
&=& -1 - {\mbox{ln}}[q_i(x_i)] + \lambda_i E(u^i \mid x_i) + \nonumber \\
&& \;\;\;\;\;\; \sum_{j \ne i} \lambda_j [E(u^j \mid x_i) -
\int dx_j \frac{\partial \epsilon_j(U^j)}{\partial U^j(x_j)} \frac{\partial
U^j(x_j)}{\partial q_i(x_i)}] \nonumber \\
&=& -1 - {\mbox{ln}}[q_i(x_i)] + \lambda_i E(u^i \mid x_i) + \nonumber
\\ 
&& \;\;\;\;\;\; \sum_{j \ne i} \lambda_j [E(u^j \mid x_i) - 
\int dx_j \frac{\partial \epsilon_j(U^j)}{\partial U^j(x_j)} E(u^j \mid x_i, x_j)].
\label{eq:collective_map}
\end{eqnarray}
Accordingly, at the MAP solution, for all players $i$, 
\begin{eqnarray}
q_i(x_i) &\propto& e^{\lambda_i E(u^i \mid x_i) + \sum_{j \ne i} \lambda_j [E(u^j \mid x_i) - 
\int dx_j \frac{\partial \epsilon_j(U^j)}{\partial U^j(x_j)} E_q(u^j
\mid x_i, x_j)]}
\label{eq:map_sol}
\end{eqnarray}

This is a set of coupled nonlinear equations.  The solution will
depend on the functional form of each $\epsilon_j$. The form being
investigated here is Boltzmann utility functions, so
we must plug that into Eq.~\ref{eq:collective_map} to evaluate
$\frac{\partial \epsilon_j(U^j)}{\partial U^j(x_j)}$. After doing that,
interchange the order of the two differentiations, to differentiate
with respect to $U^j(x_j)$ before differentiating with respect to
$b_j$. Carrying through the algebra one gets
\begin{eqnarray}
\frac{\partial \epsilon_j(U^j)}{\partial U^j(x_j)} &=& q^*_j(x_j)[1 \;+\;
b_j \{U^j(x_j) - E_{q^*_j}(U^j)\}] \nonumber \\
&=& q^*_j(x_j)[1 \;+\; b_j \{E_{q_{_{-j}}}(u^j \mid x_j) - E_{q^*_j,{q_{_{-j}}}}(u^j)\}]
\end{eqnarray}
We must now plug this into the integrals occurring in
Eq.'s~\ref{eq:collective_map} and~\ref{eq:map_sol}. Each such integral becomes
\begin{eqnarray}
\int dx_j && \!\!\!\!\!\!\!\!\!\!\!\! \frac{\partial \epsilon_j(U^j)}{\partial U^j(x_j)} E_q(u^j
\mid x_i, x_j) \nonumber \\
&& =\; \int dx_j  \; q^*_j(x_j)E_q(u^j
\mid x_i, x_j) \; [1 \;+\; b_j
\{E_{q_{_{-j}}}(u^j \mid x_j) - E_{q^*_j,{q_{_{-j}}}}(u^j)\}]
\nonumber \\
&&
\label{eq:partial_int_exact}
\end{eqnarray}

Together with the constraints \{$E_q(u^j) = \epsilon_j(U^j)$\},
Eq.~\ref{eq:map_sol} now gives us a set of coupled nonlinear equations
for the parameters \{$\lambda_j$\} and the $q_j$. The solution to this
set of equations gives our MAP $q$.

\subsection{The relation between the MAP $q$ and the QRE}
\label{sec:qre_map}

Ultimately the only free parameters in our solution for the MAP $q$
are ${\vec{b}}$.  The QRE solution $q^*$ is also a set of coupled
nonlinear equations parameterized by ${\vec{b}}$. In general there is
a very complicated relation between the MAP $q(x)$ and $q^*(x)$, one
that varies with ${\vec{b}}$ (as well as with the \{$u^j$\}, of
course). In particular, in general the two solutions differ.

Intuitively, the reason for the difference between the two solutions
is that each player $i$ does not operate in a fixed environment, but
rather in one containing intelligent players trying to adapt their
moves to take into account $i$'s moves. This is embodied in the
likelihood of Eq.~\ref{eq:like}. In contrast to that likelihood, the
likelihoods of the QRE each implicitly assume that the associated
player $i$ operates in a fixed environment. 

Formally, if we make a change to $q_i$, then the likelihood of
Eq.~\ref{eq:like} will induce a change to $q_{-i}$, to have the
invariant for the players other than $i$ still be satisfied. This
change to $q_{-i}$ will then induce a ``second order'' follow-on
change to $q_i$, to satisfy the invariant for player $i$. This
second-order effect will not arise in the likelihood associated with
the QRE $q^*_i$, which treats the other players as fixed.

Note that with the likelihood of Eq.~\ref{eq:like} the second-order
effect will induce a further change to $q_{-i}$, to ensure the
invariant is still satisfied, which will then cause a third order
change to $q_i$, and so on.  This back-and-forth is a direct
mathematical manifestation of the ``I know that you know that I know
that you prefer ...''  feature at the core of game theory. This is the
phenomenon that distinguishes game theory as a subject from decision
theory. The difference between the QRE and the MAP $q$ is an
encapsulation of this distinguishing feature.

There are other ways to view the intuitive nature of the relationship
between the QRE and the MAP $q$. For example, in deriving the MAP $q$
one follows standard probability theory and multiplies likelihoods
concerning the separate players to get a likelihood concerning the
full joint system. The mode of (the product of the prior and) that
joint likelihood gives the single most likely solution to our inference
problem. In contrast, the QRE $q$ starts by separately finding the
most likely solutions to each of many different inference problems
(one problem for each player). It then multiplies those solutions
concerning different problems together. It is not apparent what
justifying formal argument (i.e., one based on Savage's axioms) there
is for taking that product of solutions of different problems as one's
guess for the solution to a single joint problem.

The mathematical relationship between the QRE and the MAP $q$ is a
complicated one. Here we consider the simplifying approximation that
{\it{under the integral of Eq.}}~\ref{eq:partial_int_exact}, we can
equate $q(x)$ and $q^*(x)$.  In other words, assume we can use the
mean-field approximation within integrands. Exploiting this, we can
evaluate the integral in Eq.~\ref{eq:partial_int_exact}:
\begin{eqnarray}
\int dx_j && \!\!\!\!\!\!\!\!\!\!\!\! \frac{\partial \epsilon_j(U^j)}{\partial U^j(x_j)} E_{q}(u^j
\mid x_i, x_j) \nonumber \\
&&\approxeq\; \int dx_j  \; [{q}_j(x_j)E_{q}(u^j
\mid x_i, x_j)  \;+ \nonumber \\
&& \;\;\;\;\;\;\;\;\;\;\;\;\;\;\;\;\;\;\;\;\;\;
{q^*}_j(x_j)E_{q^*}(u^j \mid x_i, x_j) b_j
\{E_{q^*}(u^j \mid x_j) - E_{q^*}(u^j)\}]  \nonumber \\
&&=\;  E_{q}(u^j \mid x_i) \;\;- \nonumber \\
&& \;\;\; b_j [E_{q^*}(u^j)E_{q}(u^j \mid
x_i) -  \int dx_j
{q}_j(x_j) E_{q^*}(u^j \mid x_j) \; E_{q^*}(u^j \mid x_j, x_i)] \nonumber\\
&&
\end{eqnarray}
where we have used the fact that $q$ is a product distribution.
Plugging this into Eq.~\ref{eq:collective_map} gives
\begin{eqnarray}
0 &=&  -1 \;-\; {\mbox{ln}}[q_i(x_i)] \;+\; \lambda_i E_q(u^i \mid x_i) \; + \nonumber
\\ 
&& \;\;\;\;\;\;\; \sum_{j \ne i} \lambda_j b_j [E_{q^*}(u^j)E_{q^*}(u^j \mid
x_i) - \nonumber \\
&&  \;\;\;\;\;\;\;\;\;\;\;\;\;\;\;\;\;\;\;\;\;\;\;\; \int dx_j
{q^*}_j(x_j) E_{q^*}(u^j \mid x_j) \; E_{q^*}(u^j \mid x_j, x_i)] 
\label{eq:mean_field}
\end{eqnarray}
as our equation for $q_i$ in terms of $q_{_{-i}}$ and $q^*$.

So consider the situation where for all $j$,
\begin{eqnarray}
E_{q^*}(u^j)E_{q^*}(u^j \mid
x_i) &=&  \int dx_j
{q^*}_j(x_j) E_{q^*}(u^j \mid x_j) \; E_{q^*}(u^j \mid x_j, x_i)] .
\label{eq:correlationless}
\end{eqnarray}
In this situation, in light of Eq.~\ref{eq:map_sol}, we recover for
the MAP $q$ the very QRE solution that we assumed when we made the
mean-field approximation, where $b_i = \lambda_i \; \forall
i$. Accordingly, if the QRE solution obeys
Eq.~\ref{eq:correlationless}, it is an MAP solution.  If the QRE only
approximately obeys Eq.~\ref{eq:correlationless}, then the exact MAP
solution can be found by expanding around the QRE via
Eq.~\ref{eq:mean_field}.  

The difference between the two sides of Eq.~\ref{eq:correlationless}
is a covariance, evaluated according to $q^*_j$, between the random
variables $E_{q^*}(u^j \mid x_j, x_i)$ and $E_{q^*}(u^j \mid
x_j)$.{\footnote{Note that that second random variable is just the
average (according to $q^*_i$) of the first one. So we can rewrite the
covariance another way, as a covariance evaluated according to
$q^*_j(x'_j) q^*_i(x'_i)$, between the random variables $E_{q^*}(u^j
\mid x'_j, x_i)$ and $E_{q^*}(u^j \mid x'_j, x'_i)$.}}  Comparing
Eq.'s~\ref{eq:partial_int_exact} and ~\ref{eq:map_sol}, this provides
the following result concerning our mean-field approximation:

$ $

\noindent {\bf{Theorem 1:}} The QRE $q^*$ is the MAP of $P(q \mid
{\mathscr{I}})$ with the vector equality $\lambda = b$ iff $\forall i$,
\begin{eqnarray*}
&& \sum_{j \ne i} (b_j)^2 {\mbox{Cov}}_{q^*_j} [E_{q^*}(u^j \mid x_j,
x_i), E_{q^*}(u^j \mid x_j)]
\end{eqnarray*}
is independent of $x_i$, where Cov is the covariance operator:
\begin{eqnarray*}
{\mbox{Cov}}_p [a(y), b(y)] &\triangleq& \int dy \; p(y)a(y)b(y) \;-\;
\int dy \; p(y) a(y)\int dy \; p(y) b(y).
\end{eqnarray*}

Particularly for very large systems (e.g., a human economy), it may be
that $E_{q^*}(u^j \mid x_j, x_i) = E_{q^*}(u^j
\mid x_j)$  for almost any $i, j$ and associated moves $x_i,
x_j$. In this situation the move of almost any player $i$ has no
effect on how the expected payoff to player $j$ depends on $j$'s
move. If this is in fact the case for player $i$ and all other players
$j$, then the covariance for each $j, x_j, x_i$ reduces to the
variance of $E_{q^*}(u^j \mid x_j)$ as one varies $x_j$ according to
$q^*_j$.

This variance is given by the partition function:
\begin{eqnarray}
{\mbox{Var}}_{q^*_{j}}(E_{q^*}(u^j \mid x_j))&=& {\mbox{Var}}_{q^*_{j}}(U^j_{q^*}) \nonumber \\
&=& \frac{\partial^2
{\mbox{ln}}(Z_{U^j_{q^*}}(b'_j))}{\partial (b'_j)^2}|_{b'_j=b_j} .
\end{eqnarray}
In particular, for $b_j \rightarrow \infty$ --- perfectly rational
behavior on the part of agent $j$ --- the variance goes to 0. So if
every $i$ is ``decoupled'' from all other agents, then in the limit
that all such agents become perfectly rational, the expression in
Thm. 1 generically goes to 0.  (The $b_j$-dependence in the covariance
occurs in an exponent, and therefore generically overpowers the
$(b_j)^2$ multiplicative factor.) So the QRE approaches the MAP
solution in that situation.

On the other hand, if the players have bounded rationality, their
variances are nonzero. In this case the expression in Thm. 1 is
nonzero for each $i, x_i$. Typically for fixed $i$ the precise nonzero
value of that variance will vary with $x_i$. In this case, by Thm. 1,
we know that the QRE differs from the MAP solution.

There are many special game structures (e.g., zero-sum games) in which
one can make some arguments about the likely form of the sum in
Thm. 1. An elaboration of those arguments is the subject of ongoing
research.

\subsection{The posterior $q$ covers all Nash equilibria}
\label{sec:coverage}

Not all $q$ can be cast as a QRE for some appropriate $\vec{b}$. So in 
particular, a $q$ that occurs in the real world will in general
differ, even if  only slightly, from all possible QRE's. This can be
viewed as a shortcoming of the QRE (a shortcoming that applies to all
equilibrium concepts with a sufficiently small number of parameters). 

Now as ${\vec{b}} \rightarrow \infty$, the QRE reduces to some mixed
strategy Nash equilibrium. Different sequences of the $\vec{b}$ going
to the infinity vector can lead to different Nash equilibria. However
in general starting from the point where all $b_j = 0$ and
continuously increasing the components of $\vec{b}$ can only lead to
one particular equilibrium, and other Nash equilibria are not the
limit of such a sequence~\cite{mcpa95}. This too can be viewed as a
short-coming of the QRE.

However from the perspective of PGT, there is far more to the
posterior distribution specified by a particular vector $\vec{b}$ than
some single $q$ chosen using that posterior, be that $q$ the
associated Bayes-optimal $q$, the MAP $q$, or an approximation to the
MAP $q$ like the QRE. In this, the potential impossibility of one
particular sequence of such $q$'s approaching some particular one of
the game's Nash equilibria is not necessarily a reason for concern.

To formalize this we start with the following result:

$ $

\noindent {\bf{Proposition 1:}} Define ${\cal{Q}}({\vec{b}})
\triangleq \{ q \in \Delta_{{\cal{X}}} \; : \; \forall i, P(q_i \mid
{\mathscr{I}}_{{\vec{b}}'}) > 0$ for some ${{\vec{b}}'} \succeq
{\vec{b}}\}$. Let $B$ be some sequence of $\vec{b}$ values that
converges to $\vec{\infty}$, i.e., such that for all ${\vec{b}} \in B$
having no infinite components, $\exists \; {\vec{b}}' \in B$ where
${\vec{b}}' \succ b$. 
Then all members of ${\cap}_{{\vec{b}} \in B}
{\cal{Q}}({\vec{b}})$ are Nash equilibria of the game.

$ $

\noindent {\bf{Proof:}} 
Hypothesize $\exists \; {\tilde{q}} \in {\cap}_{{\vec{b} \in
B}}{\cal{Q}}({\vec{b}})$ which is not a Nash equilibrium. Then
$\exists \; i$ such that $U^i_{{\tilde{q}}_{_{-i}}}$ is not constant
valued. In addition, we know that ${\tilde{q}}_i \cdot
U^i_{{\tilde{q}}_{-i}}
\equiv v_i < {\mbox{max}}_{x_i} U^i_{{\tilde{q}}_{-i}}(x_i)$. 
However recall from Sec.~\ref{sec:eff_inv} that if $U^i_{{\tilde{q}}_{-i}}$ is
not constant-valued, the Boltzmann utility
$K(U^i_{{\tilde{q}}_{-i}}, .)$ is a monotonically increasing bijection with domain $[0,
\infty)$ and range $[\frac{\int dx_i \;
U^i_{{\tilde{q}}_{-i}}(x_i)}{\int dx_i \; 1}, {\mbox{max}}_{x_i} U^i_{
{\tilde{q}}_{-i} }(x_i))$. Since $v_i$ falls within that range, this
means that we can invert $K(, U^i_{{\tilde{q}}_{-i}})$ to get a unique
finite value ${\tilde{b}}_i$ that is consistent with
$\tilde{q}$. Accordingly, $P({\mathscr{I}}_{\vec{b}} \mid
{\tilde{q}})$ is non-zero only if $b_i = {\tilde{b}}_i$, and therefore 
so is $P({\tilde{q}} \mid {\mathscr{I}}_{\vec{b}})$.

However by definition $\tilde{q}$ must be a member of
${\cal{Q}}({\vec{b}})$ for all $\vec{b}$ in the limiting
sequence. That means in particular that it must be a member of
${\cal{Q}}({\vec{b}}')$ for some ${\vec{b}}'$ where $b'_i >
{\tilde{b}}_i$. But by definition, all members $q$ of
${\cal{Q}}({\vec{b}}')$ have $P(q \mid {\mathscr{I}}_{{\vec{b}}}) > 0$
for some $b$ such that $b_i \ge b'_i > {\tilde{b}}_i$. Since we know
$P({\tilde{q}} \mid {\mathscr{I}}_{\vec{b}})$ is non-zero only if $b_i
= {\tilde{b}}_i$, this means that ${\tilde{q}} \not \in
{\cal{Q}}({\vec{b}}')$, contrary to hypothesis. {\bf{QED}}.

$ $

Conversely, every $q$ has a non-infinitesimal posterior probability
(density), for some (potentially infinite) $\vec{b}$ that specifies
that posterior. More formally,

$ $

\noindent {\bf{Proposition 2:}} For any benign $q \in \Delta_{{\cal{X}}}$
there is a unique $\vec{b}$ and 
associated invariant ${\mathscr{I}}_{\vec{b}}$ such that $P(q \mid
{\mathscr{I}}_{\vec{b}}) \ne 0$. For that $\vec{b}$, for all $q' \in
\Delta_{{\cal{X}}}$,
\begin{equation}
\frac{P(q \mid {\mathscr{I}}_{\vec{b}})}{P(q' \mid
{\mathscr{I}}_{\vec{b}})} \; \ge \; 
|X|^{-\alpha}
\end{equation}
where $\alpha$ is the exponent of the entropic prior.

$ $

\noindent {\bf{Proof:}} First recall that any $q$ has non-zero
posterior probability $P(q \mid {\mathscr{I}}_{\vec{b}})$ for some
$\vec{b}$, assuming finite entropic prior constant $\alpha$. (See
Sec.~\ref{sec:no_nash}.) So to prove the first part of the proposition
we must establish the uniqueness of that $\vec{b}$.

Consider any $i$ and the given $q$. Say $q_i \cdot U^i_{q_{-i}} \equiv
v_i \ne {\mbox{max}}_{x_i} U^i_{q_{-i}}(x_i)$. This means that
$U^i_{q_{-i}}$ is not the constant function that is independent of its
argument. Now recall from Sec.~\ref{sec:eff_inv} that for any such
$v_i$ and fixed non-constant $U^i$, there is always a unique $b_i
\in [0, \infty)$ such that $K_i(b_i)$ equals $v_i$.
On the other hand, as explained in the discussion in that subsection,
if $v_i = {\mbox{max}}_{x_i} U^i_{q_{-i}}(x_i)$, then regardless of
whether $U^i_{q_{-i}}$ varies with its argument, $b_i
= \infty$. So there is a unique $b_i$ consistent with $q$, which we
write as $b_i^*$. Since this holds simultaneously for all $i$, the
entire vector ${\vec{b}}^*$ with components \{$b^*_i$\} is unique.

This means that the likelihood $P({\mathscr{I}}_{{\vec{b}}^*} \mid q)
= 1$. On the other hand, $P({\mathscr{I}}_{{\vec{b}}^*} \mid q') \le
1$ for any $q'$. Accordingly, the ratio in the proposition is bounded
above by the ratio of the exponential prior at $q$ to that at $q'$.
However the ratio of $e^{\alpha S(q'')}$ between any two points
$q''$ is bounded below by $\frac{exp(\alpha \cdot 0)}{exp(\alpha
{\mbox{ln}}(|X|))}$. {\bf{QED}}

$ $

\noindent In particular, this result holds for Nash equilibrium $q$;
such equilibria arise for ${\vec{b}} = {\vec{\infty}}$ The relative
probabilities of those Nash $q$ are given by the ratios of the
associated prior probabilities, i.e., by (the exponential of) the
associated entropies, $S(q)$. This reflects our presumption that it is
{\it{a priori}} more likely that the adaptation/learning processes
that couple the players results in a Nash equilibrium with broad $q$
that that it results (for example) in a ``golf hold'' pure strategy
$q$. (Generically, such golf hole solutions are more difficult to find 
for any broadly applicable learning process.)

Prop. 2 also holds for any particular $q$ infinitesimally close to one
of the Nash equilibria.  In this sense, the posterior probability is
arbitrarily tightly restricted to any one of the Nash equilibria for
some appropriate $\vec{b}$.

The picture that emerges then is that $\forall {\vec{b}}, \; \exists$
proper submanifold of $\Delta_{\cal{X}}$ that is the support of the
posterior. There is no overlap between those submanifolds (one for
each ${\vec{b}}$), and their union is all of $\Delta_{\cal{X}}$,
including the Nash equilibria $q$'s (for which ${\vec{b}} =
{\vec{\infty}}$).  Those Nash equilibria are the limit points of those
submanifolds (in a sequence of increasing $\vec{b}$). 

Within any single one of the submanifolds no $q$ has too small a
posterior (cf. Prop. 2).  This is because all $q$ within a single
submanifold have the same value (namely 1) of their
likelihoods. Accordingly, the ratios of the posteriors of the $q$'s
within the submanifold is given by the ratios of (the exponentials of)
the entropies of those $q$'s.

Finally, consider the case that the submanifolds get a unique maximum
as ${\vec{b}} \rightarrow \infty$. This means that the mode of the
posterior --- the MAP $q$ --- necessarily goes to a single one of the
Nash equilibria in that limit. In this sense, ``only one Nash
equilibrium is picked out by that limit''. In particular, this
limiting behavior holds for the QRE approximation to the MAP $q$. As
mentioned, this has been seen as a problematic aspect of the QRE
equilibrium concept. However from the prospect of PGT there is nothing
untoward about this behavior. After all, all of the Nash equilibria
have non-zero posterior in that limit (cf. Prop. 2); it just so
happens that the QRE ends up at a single one of those equilibria.

\subsection{Alternative choices of $\mathscr{I}$}
\label{sec:alt-epsilon_i}

Of course, one can always design ``learning'' algorithms for players to
follow in such a way that our assumed invariants don't hold. After
all, in the extreme case you can design ``learning'' algorithms that are
intentionally stupid, giving higher probability to moves with lower
expected utility. Less trivially, there are many algorithms that are
of interest in the game theory community even though they would never
be considered by anyone in the machine learning community applying
learning algorithms to real-world problems (e.g., ficticious play). It
may well be that such algorithms don't obey the assumed invariants
exactly for some \{$U^i$\}. 

However this issue also obtains, at least as strongly, for alternative
encapsulations of rationality like Nash equilibrium, trembling hand,
quantal response, etc. It is trivial to design ``learning algorithms''
that guarantee that those equilibria cannot arise. More generally, in
{\it{all}} statistical inference --- in other words, in all of science
--- any formalization of invariants may well have some error. This is
even true in statistical physics, and is an intrinsic feature to any
predictive science.

All of this notwithstanding, there are a number of alternative choices
of $\epsilon_i$ to the one considered here that should be investigated
in depth. To give a simple example, the ${\mathscr{I}}$ considered
here is only a ``lowest order'' choice for an invariant. In
particular, as mentioned above, our choice of $\mathscr{I}$ assumes
$b_i$ is independent of $U^i$.  A more sophisticated analysis than can
be fully developed here would consider possible couplings between
$b_i$'s and $U^i$'s; in this paper $b_i$'s and $U^i$'s are
independent.

However one does not need to couple $b_i$'s and $U^i$'s to get
reasonable alternative choices of $\epsilon_i$.  For example, due to
the $U^i$-independence of $b_i$ with our $\mathscr{I}$, whatever $b_i$
is, for some $q$ the associated likelihood $P({\mathscr{I}}_{\vec{b}}
\mid q) = 0$ (just like whatever the temperature of a physical system,
some phase space distributions are incompatible with that
temperature). To avoid this, in many scenarios we might want to allow
how smart a player $i$ is to vary from one instance to the next, even
without considering detailed mathematical structures relating
variations in $b_i$ with those in $U^i$.

One way to do this would mean allowing $b_i$ to vary in an
$U^i$-independent manner, with only its average value
fixed.{\footnote{Indeed, in practice each $b_i$ is at best loosely
known. So formally speaking it is a random variable with its own
distribution, and so even within a type I game it must be marginalized
out to get our posterior $P(q \mid {\mathscr{I}})$. This type of
random variable is known as a
``hyperparameter''~\cite{berg85,besm00}. (A more common example of a
hyperparameter is the typically unknown width of a Gaussian noise
process that corrupts some data.) In particular, in almost all of this
paper we are implicitly assuming that the posterior over each $b_i$ is
quite peaked, so that in our analysis we can simply set $b_i$ to a
constant, albeit an unknown one.}} This simple generalization of
$\mathscr{I}$ can be accommodated by switching the analysis to involve
type II games.  Although the details of that analysis (like all other
details of type II game theory) is beyond the scope of this paper, it
is worth making some broad comments on it.

That type II analysis starts by extending the definition of an
environment vector to type II games in the obvious way: indexed by
$q'_i$, the type II environment is defined by
\begin{eqnarray}
U^i_{\pi_{-i}}(q'_i) &\triangleq& \int dx' dq'_{-i} \;
\pi_{-i}(q'_{-i}) q'_{-i}(x'_{-i}) q'_i(x'_i) u^i(x'_i, x'_{-i})
\end{eqnarray}
so that the expected value of $u^i$ is given by
\begin{eqnarray}
\pi_i \cdot U^i_{\pi_{-i}} &=& \int dq'_i \; \pi_i(q'_i) U^i_{\pi_{-i}}(q'_i)
\nonumber \\
&=& E_{\pi_i, \pi_{-i}}(u^i)
\end{eqnarray}
The analysis also extends the definition of $K(., .)$ to type II games
in the obvious way: $K(U^i_{\pi_{-i}}, B_i)$ is what $\pi_i \cdot
U^i_{\pi_{-i}}$ would be if $\pi_i$ were the associated Boltzmann
distribution, $\pi_i(q'_i) \propto \exp(B_i 
U^i_{\pi_{-i}}(q'_i))$.  

The new invariant would then be that for all $i$,
\begin{eqnarray}
\pi_i \cdot U^i_{\pi_{-i}} &=& K(U^i_{\pi_{-i}}, B_i)
\label{eq:typeII-inv}
\end{eqnarray}
This invariant allows any $q_i$ to occur, thereby allaying the
potential shortcoming with the invariant this paper focuses on. It is
now certain $\pi_i$ that are excluded rather than certain
$q_i$.

To motivate our next alternative for the invariant $\mathscr{I}$,
consider the distribution induced by $q_{-i}(x_{-i})$ over player
$i$'s move-specified utilities $u^i(x_i, .)$ (one such distribution
for each $x_i$),
\begin{eqnarray}
P_{q_{_{-i}}}(u^i(x_i, .) = u)
&=& \int dx'_{-i} \; q_{-i}(x'_{-i}) \delta(u - u^i(x_i, x_{-i})).
\end{eqnarray}
${\mathscr{I}}$ implicitly assumes that those aspects of $i$'s
behavior that it is safe for us to presume are only those that involve
the first moments of these distributions,
\begin{eqnarray}
U^i_{q_{_{-i}}}(x_i) &=&
\int du \; P_{q_{_{-i}}}(u^i(x_i, .) = u) u \nonumber \\
&=& 
\int dx'_{-i} \; q_{-i}(x'_{-i}) u^i(x_i, x'_{-i}).
\end{eqnarray}
In this it simply emulates conventional game theory.

However in many real-world coupled-players scenarios the higher
moments, reflecting the breadth and overlaps of the distributions over
$u^i(x_i, .)$, will have a major impact on our inference of
$q_i$. Intuitively, if those distributions --- each a function purely
of $q_{-i}$ --- maintain the same mean but get broader with more
overlap between them, that will increase the variability of what
inferences $i$ makes concerning those means and their linear
ordering. (For example, that is the case if $i$ makes its inference of
those means based on empirical samples of the distributions.) This
will make our associated distribution over $q_i$ broader --- there are
more $q_i$ that we can conceive of $i$ arriving at.  Similarly, such
broadening of the distributions over the $u^i(x_i, .)$ would often be
evident to $i$. That might make $i$ realize it can have less
confidence in its inference of the ordering of the means of those
distributions. In such a situation, many real-world players $i$ would
become more conservative in formulating their mixed strategy,
$q_i$. So not only might the distribution over $q_i$ get broader, but
it may also shift, if $q_{-i}$ changes to cause this kind of
broadening of the distributions.

To be more quantitative, say the variances of the $U^i(x_i, .)$,
\begin{eqnarray}
V^i_{q_{_{-i}}}(x_i) &\triangleq& \{\int dx_{-i} \; q_{-i}(x_{-i}) [u^i(x_i,
x_{-i})]^2\} \;\; - \;\; [U^i_{q_{_{-i}}}(x_i)]^2 ,
\end{eqnarray}
are increased, and that the overlap between the distributions over each
$u_i(x_i, .)$ (measured for example via Kullback-Leibler distance
between those distributions) also are increased.  Then there is often
increased uncertainty on our part about the relationships between
$i$'s sample-driven preferences among the $x_i$. This often means we
are less sure in our inference of what $i$'s current mixed strategy
is, which means our posterior over $q_i$ should get broader.

In addition, under such broadening in the $u^i(x_i, .)$ there is
increased uncertainty about what $i$'s best move would be for the
actual move $x_{-i}$ that will be formed by sampling $q_{-i}(x_{-i})$.
Typically this means that the information that $i$ has gleaned via its
previous interactions with the other players is not as helpful to $i$
for determining its best move for the current game. Intuitively, when
these distributions are broader $i$ faces worse signal-to-noise in
discerning the relation between the $U^i(x_i)$ based on limited
data. This will often manifest itself by changes to what mixed
strategy $i$ is most likely to adopt.

A standard illustration of both of these effects arises if one
compares two extreme scenarios. The first is the ``US economy game''
that any particular US citizen $i$ repeatedly engages in with the 300
million other human players in the US. The second is a simple game
against Nature that $i$ repeatedly engages in where there is no
variance in Nature's choice of move. Our inference of $q_i$ is far
easier in the second scenario. Similarly, typically $i$ will have an
easier time discerning its best move in the second
scenario.{\footnote{See
\cite{wolp03b,wotu02c,wolp04a,wobi04a,wotu01a,tuwo03}, and references
therein to ``Collective Intelligence'' for a discussion of how this
second type of effect can be addressed for mechanism design and
distributed control.}}

One approach to incorporate such effects would be to have the set of
all \{$V^i(x_i)$\} (running over $i$ as well as the associated $x_i$)
and overlaps between the distributions over the \{$u^i(x_i, .)$\} help
specify $\vec{b}$. Such an approach could obviously address the second
of the effects we're concerned with, involving how much information
$i$ has managed to glean concerning the other players. It is not a
fully satisfactory approach to addressing the first effect however.
This is because once $\vec{b}$ is set --- however that is done ---
some $q$ are excluded, i.e., some $q$ have posterior probability equal
to 0. Typically to change $\vec{b}$ to allow those previously excluded
$q$ --- and thereby broaden the distribution over $q_i$ --- the
Bayes-optimal (or MAP) $q$ also changes. Moreover, such a modification
invariably excludes some $q$ that were previously allowed (see
Sec.~\ref{sec:coverage}).  Instead what we want is our increase of the
breadth of the posterior over $q$ to allow previously excluded $q$,
while still allowing all $q$ we did earlier.

The exclusionary character of the posterior over $q$ that is causing
this difficulty can be removed by casting the analysis in terms of
type II games rather than (as in the exposition above) type I
games. After all, in general the $\pi$ that obeys the conventional
type II game invariant has support extending over all $q$
(cf. Eq.~\ref{eq:typeII-inv}).  A detailed exploration of how to use
type II games to incorporate the effects of the \{$V^i(x_i)$\} and
overlaps between the \{$u^i(x_i, .)$\} into our posterior is beyond
the scope of this paper however.

The discussion so far has focused on variants of $\mathscr{I}$ that
are at most loosely based on empirical data. Those variants
incorporate none of the insight of behavior economics, prospect
theory, or behavioral game
theory~\cite{kahn03a,kahn03b,tvka92,came03}. Crucially important
future investigations involves incorporating that work, and more
generally the entire field of user-modeling and knowledge-engineering,
into our choice of $\mathscr{I}$.

Finally, it is worth noting that there are alternative $\mathscr{I}$'s
that don't involve the numeric values of the $u^i$'s, but rather only
require that each $u^i$ provides an ordering over the $q$. The idea
here is to consider what is invariant if $i$ stays ``just as
smart'', while $U^i$ undergoes a non-affine monotonically
increasing transformation, and $q^i$ changes accordingly. For example,
one might argue that $q_i$ would be ``just as smart'' after such a
transformation if the fraction of alternative $q'_i$ such that $q'_i
\cdot U^i > q_i \cdot U^i$ is the same  before and after the
transformation.  Formally, this would mean that $\int dq'_i \;
\Theta(U^i \cdot [q_i - q'_i])$ is a constant, rather than (as in the
choice considered in this paper) $U^i \cdot q_i$.  Intuitively, under
this choice, ``how smart'' $i$ is reflects how good she is at ruling
out some of the candidate $q'_i$ as inferior to the final $q_i$ she
uses.{\footnote{One obvious variation of this measure of how smart $i$
is is to replace the uniform measure in the integral $\int dq'_i \;
\Theta(U^i \cdot [q_i - q'_i])$ with a non-uniform one, for example
emphasizing those $q'_i$ having larger dot product with $U^i$. A
related variation would replace the Heaviside function in the
integrand with some smooth increasing function, e.g., a logistic
function.}} This formalization of how smart $i$ is is essentially
identical to what is called {\bf{intelligence}} in work on Collective
Intelligence~\cite{wolp03a,wotu01a,wotu02a}.

\section{Independent players}
\label{sec:ind_players}

\subsection{Basic formulation}

When the players have never previously knowingly interacted, there is
no statistical coupling between the associated mixed strategies,
\{$q_i$\}. In this case the setup for coupled players
(Sec.~\ref{sec:coupled-players}) does not apply.  Instead we must
separately specify likelihoods for each of the players. The joint
likelihood is then the product of those separate likelihoods.

Here for simplicity we consider a game of complete information, so
that every player knows the move spaces and utility functions for all
players. Intuitively, those players are not just unaware physical
particles, without any ``goals'' that they are ``trying'' to achieve.
Rather each is a reasoning entity, trying to maximize its own utility,
and it knows the same holds for the other players. This results in the
``I know that you know that I know that you prefer ...''  common
knowledge feature that lies at the core of many views of game
theory.{\footnote{See ~\cite{aubr95} for a fine-grained distinction
between such ``common knowledge'' and ``mutual knowledge''; such
distinctions are not important for current purposes. Also
see~\cite{kala82} for related, qualitative discussion.}

Intuitively, this overlap in knowledge among the players acts as a
``virtual coupling'' between the players. However it is not a formal
statistical coupling. After all, as mentioned above, $P({\mathscr{I}}
\mid q) = \prod_i P({\mathscr{I}} \mid q_i)$ for our independent players
invariant $\mathscr{I}$.  Therefore (for an entropic prior) the
posterior distributions over mixed strategies are statistically
independent: 
\begin{eqnarray}
P(q \mid {\mathscr{I}}) &=& \prod_i P(q_i \mid {\mathscr{I}}).
\label{eq:ind_moves}
\end{eqnarray}
Given this independence, how do we capture the ``virtual coupling'', so
crucial to noncooperative game theory, in the independent-players
invariant $\mathscr{I}$?

To answer this, concentrate on some particular player $i$. As a
surrogate for virtual coupling, say we had a game of actual coupling,
as in Sec.~\ref{sec:coupled-players}.  That would set up a
distribution over the joint moves of the players other than $i$,
\begin{eqnarray}
P(x'_{-i} \mid {\mathscr{I}}_c) &=& \int dx'_i \; P(x'_i, x'_{-i} \mid
{\mathscr{I}}_c) \nonumber \\
&=& \int dx'_i dq \; q(x'_i, x'_{-i}) P(q \mid {\mathscr{I}}_c) \nonumber 
\\
&=& \int dq \; [\int dx'_i q(x'_i, x'_{-i})] P(q \mid {\mathscr{I}}_c) \nonumber 
\\
&\propto& \int  dq \; q_{-i}(x'_{-i}) \prod_{j} e^{{a^i} S(q_j)}
\delta(q_j \cdot U^j_q - \epsilon^i_j(U^j_q)) 
\label{eq:post_others}
\end{eqnarray}
where the subscript $c$ on the invariant indicates it's the invariant
for a counterfactual coupled players scenario, $a^i$ is an associated
entropic prior constant for player $i$, and each $\epsilon^i_j$ is an
associated Boltzmann utility function, with (implicit) Boltzmann
constant $b^i_j$.

Now if player $i$ makes move $x_i$, and the remaining players make
move $x'_{-i}$, then the utility for player $i$ is $u^i(x_i,
x'_{-i})$. Accordingly, if the distribution over $x'_{-i}$ were actually 
given by Eq.~\ref{eq:post_others}, then the expected utility for
player $i$ for making move $x_i$ would be
\begin{eqnarray}
U_c^i(x_i) &\triangleq& \int dx'_{-i} \; u^i(x_i, x'_{-i}) P(x'_{-i} \mid
{\mathscr{I}}_c) \nonumber \\
&=& \frac{ \int dx'_{-i} dq \; u^i(x_i, x'_{-i}) q_{-i}(x'_{-i})
\prod_{j} e^{{a^i} S(q_j)}
\delta(q_j \cdot U^j_q - \epsilon^i_j(U^j_q))} 
{\int dq \; \prod_{j} e^{{a^i} S(q_j)}
\delta(q_j \cdot U^j_q - \epsilon^i_j(U^j_q)} \nonumber \\
&& \label{eq:surr_f_i}
\end{eqnarray}
Note that $U^i_c$ implicitly depends on an associated value ${a^i}$,
as well as on the values \{$b^i_j$\} parameterizing the set of
functions \{$\epsilon^i_j$\}. 

Say that in choosing its move player $i$ assumes that its actual
utility $U^i$ is well-approximated by $U^i_c$ for some appropriate
${a^i}$ and \{$b^i_j$\}. This means that the reasoning of player $i$
reflects the ``I know that you know ...''  common knowledge feature of
game theory; it makes its move under the presumption that the
counterfactual coupled players scenario gives a good approximation to
its actual environment. With this approach, there is no infinite
regress difficulty like that underlying other approaches to the issue
of common knowledge. (This reliance on counterfactual coupling to
formalize that common knowledge feature can be viewed as an
alternative to approaches like Aumann's epistemic knowledge
\cite{auma99}.)

Note (as discussed just below Eq.~\ref{eq:post_moves}) that $p_{quad}
\triangleq \int dq \; q P(q \mid {\mathscr{I}}_c)$ is the
Bayes-optimal distribution over joint moves under quadratic loss 
and the invariant ${\mathscr{I}}_c$. So the distribution $P(x_{-i}
\mid {\mathscr{I}}_c)$ underlying $U^i_c$ is the same as the distribution induced by
sampling that single Bayes-optimal distribution. Also recall that
$p_{quad}(x)$ is not a product distribution; under it the moves of the
players are not statistically independent. So we are modeling every
player $i$ as though she achieves a certain performance level for a
counterfactual game in which all the players (herself included) make
their moves according to the (coupled) distribution $p_{quad}$ --- but
in reality she is free to make moves according to a different
distribution.


Say that player $i$ makes the perfectly rational move for the
counterfactual game. In this situation, player $i$ chooses her moves
on the presumption that the other players all behave according to that
counterfactual game. The coupling in that counterfactual game can be
viewed as how player $i$'s implements the common knowledge reasoning
underlying much of conventional game theory. Our presumption,
formalized below, is that while the behavior of player $i$ will not
necessarily be perfectly rational for the counterfactual game, that
behavior can be approximated as though she is trying to behave that
way.

Say that all players $i$ go through the kind of counterfactual
reasoning outlined above for associated values of ${a^i}$ and
\{$b^i_j$\} that do not vary much between them.  Then they will all
have used very similar distributions $P(x \mid {\mathscr{I}}_c)$ to
choose their moves. This commonality in their reasoning will not
statistically couple their moves; Eq.~\ref{eq:ind_moves} will still
hold. However it will generate the virtual coupling inherent in the
``I know that you know ...''  feature. Intuitively, it is because they
all model the ``I know that you know ...''  phenomenon in terms of
similar statistical coupling scenarios that they are virtually
coupled.

Now in practice, no player $i$ will exactly evaluate such a
counterfactual coupling scenario to get a guess for $U^i$ (and indeed
may not even be able to, for example due to computational
limitations).  But we can presume that each such player will go
through reasoning not too different from such an evaluation, for some
particular ${a^i}$ and \{$b^i_j$\}. Accordingly, as a surrogate for
each player $i$'s actual reasoning, and the associated virtual
coupling among all the players, we can stipulate that each player's
reasoning results in a mixed strategy $q_i$ that is highly consistent
with a counterfactual statistical coupling scenario given by
Eq.~\ref{eq:surr_f_i}.

To formalize this we must define what it means to have $q_i$ be
``highly consistent'' with $U^i_c$. One natural way to do that is by
stipulating that $q_i \cdot U^i_c = K_i$ for some parameter $K_i$,
exactly as in the discussion of effective invariants in
Sec. ~\ref{sec:eff_inv}. In other words, we stipulate that
$E_{p_{quad}}(u^i)$ be an $i$-dependent constant. Plugging it in, this
definition of ``highly consistent'' gives us our invariant for player
$i$, i.e., it gives us the likelihood over $q$ for each player $i$.

In Sec.~\ref{sec:map_ind} we will replace each $K_i$ with an
equivalent parameter $\beta_i$ that is easier to work with. This
parameter will just be the parameter saying how smart $q_i$ is for
utility $U^i_c$, as in Eq.~\ref{eq:counterfact} and the associated
discussion of effective invariants in Sec.~\ref{sec:eff_inv}.  To have
the notation reflect this alternative parameterization we will
sometimes write $K(U^i_c, \beta_i)$ (again, just as in
Sec.~\ref{sec:eff_inv}).  One of the major advantages of
parameterizing the $i$'th likelihood with $\beta_i$ rather than $K_i$
is that $\beta_i$ always ranges from 0 to $+ \infty$, for any game,
for any player $i$, and independent of what $q_{-i}$ is. This is not
the case for $K_i$; its range of values will depend on $q_{-i}$ in
general. Intuitively, $\beta_i$ is simply $K_i$ normalized to account
for this.

As mentioned above, since the players are independent, the joint
likelihood is the product of the separate individual
likelihoods. Using our notation for $K_i$, we can write this
likelihood as
\begin{eqnarray}
P({\mathscr{I}} \mid q) &=& \prod_i P({\mathscr{I}} \mid q_i) \nonumber
\\
&=& \prod_i \delta(q_i \cdot U^i_c - K(U^i_c, \beta_i)),
\label{eq:ind_like}
\end{eqnarray}
with each $U^i_c$ given by Eq.~\ref{eq:surr_f_i}. 

Comparing this with Eq.~\ref{eq:like}, and recalling that
$\epsilon_i(U^i_q) = K(U^i_q, b_i)$, we arrive at an alternative
motivation for the choice of Eq.~\ref{eq:ind_like} for the
independent-players likelihood. Our presumption for the independent
players scenario is that each player is coupled to an environment in
the exact same way as in the coupled players scenario, via the
function $\epsilon^i$ for some appropriate Boltzmann exponent (labeled
$\beta_i$ for the independent players scenario, and $b_i$ for the
coupled players scenario). However in the coupled players scenario the
environment of each player $i$ is set by the actual $q_{-i}$. In
contrast, in the independent players scenario, each player $i$'s
environment is set by a counterfactual $q_{-i}$. Intuitively, we are
presuming that each player $i$ acts just as $we$ do, when we make
predictions for a coupled players scenario.

Plugging in, the posterior for the independent players scenario is
given by
\begin{eqnarray}
P(q \mid {\mathscr{I}}) &\propto& e^{\alpha S(q)} P({\mathscr{I}} \mid
q) \nonumber \\
&=& \prod_i e^{\alpha S(q_i)} \delta(q_i \cdot U^i_c - K(U^i_c,
\beta_i)) .
\label{eq:ind_post}
\end{eqnarray}

Plugging Eq.~\ref{eq:surr_f_i} into this result, we get the posterior
probability over $q$ for independent players: 
\begin{eqnarray}
P(q \mid {\mathscr{I}}) &\propto& \prod_i e^{\alpha S(q_i)}
\delta \bigl[ K(U^i_c, \beta_i) \; - \nonumber \\
&&\;\;\;\;q_i \cdot \frac{
\int dx_{-i} d{q'} \; u^i(., x_{-i}) {q'}_{-i}(x_{-i})
\prod_{j} e^{{a^i} S({q'}_j)}
\delta({q'}_j \cdot U^j_{q'} - \epsilon^i_j(U^j_{q'}))}
{\int d{q'} \; 
\prod_{j} e^{{a^i} S({q'}_j)}
\delta({q'}_j \cdot U^j_{q'} - \epsilon^i_j(U^j_{q'})} \bigr] . \nonumber \\
&&
\end{eqnarray}
Next we plug in the usual coupled players $\epsilon^i_j$:
\begin{eqnarray}
P(q \mid {\mathscr{I}}) &\propto& \prod_i e^{\alpha S(q_i)}
\delta \bigl[ K(U^i_c, \beta_i) \; - \nonumber \\
&&\;\;\;\;q_i \cdot \frac{
\int dx_{-i} d{q'} \; u^i(., x_{-i}) {q'}_{-i}(x_{-i})
\prod_{j} e^{{a^i} S({q'}_j)}
\delta({q'}_j \cdot U^j_{q'} - K(U^j_{q'}, b^i_j))}
{\int d{q'} \; 
\prod_{j} e^{{a^i} S({q'}_j)}
\delta({q'}_j \cdot U^j_{q'} - K(U^j_{q'}, b^i_j))} \bigr] . \nonumber \\
&&
\end{eqnarray}
where the $K$ function is as defined in Sec.~\ref{sec:eff_inv} with
$U^i_c$ given by Eq.~\ref{eq:surr_f_i} and parameterized by $a^i$ and
the set of values $\{b^i_j\}$.  
As usual, the posterior over $x$ is given by $\int dq \; q(x) P(q \mid 
{\mathscr{I}})$ and is identical to the Bayes-optimal $q$ under a
quadratic loss function.

Intuitively, for fixed $i$, the \{$b^i_j$\} are how smart player $i$
imputes the other players in the counterfactual game to be, which she
uses to encapsulate the common knowledge aspect of the game. So it
encapsulates how she thinks the other players will choose their
moves. In particular, she presumes that in formulating their mixed
strategies, the other players will consider how smart she is to be
$b^i_i$. Player $i$ then uses $a^i$ to set the relative probabilities
of the $q$'s that are all consistent with those \{$b^i_j$\}. More
carefully, $a^i$ and the \{$b^i_j$\} serve as our presumptions of the
values of these quantities inherent to player $i$. Properly speaking,
we do not really presume that she explicitly has such quantities and
uses them to calculate a counterfactual game. Rather we presume that
her behavior can be well-approximated by such a common-knowledge type
of reasoning by her.

In contrast, $\beta_i$ reflects our assessment of how well player $i$
carries out such reasoning. It measures how smart we believe she is
in evaluating the counterfactual game, and even the degree to which
that game really guides her choice of move. $\alpha$ then controls the
relative probabilities of the $q$'s that are all consistent with our
assessment of $a^i$ and the \{$b^i_j$\} for all players $i$.

Note that since $P(q \mid {\mathscr{I}})$ is a
product distribution, if $P(q_i \mid {\mathscr{I}})$ changes, there is
no effect on $P(q_{j \ne i} \mid {\mathscr{I}})$. This is true even if
the players are all fully rational, both in the actual game and the
counterfactual game, so that with probability 1 the system is at a
Nash equilibrium of the original game.  Accordingly, issues like
whether such a Nash equilibrium is ``stable'' do not occur in the
independent players scenario. Any changes to (the distribution
governing) player $i$'s mixed strategy has no effect on the
(distribution governing) the mixed strategy of any other player
$j$. This is because such a player $j$ is playing best-response to the
counter-factual game, not to the actual game.

\subsection{Independent players and the impossibility of a
Nash equilibrium}

Since $P(q \mid {\mathscr{I}})$ for independent players is a product,
\begin{eqnarray}
P(x \mid {\mathscr{I}}) &=& \int dq \; q(x) P(q \mid {\mathscr{I}})
\nonumber \\
&=& \prod_i \int dq_i q_i(x_i) P(q_i \mid {\mathscr{I}})
\label{eq:x_post_ind}
\end{eqnarray}
So our estimate of the joint distribution over moves, $P(x \mid
{\mathscr{I}})$, is a product distribution. This contrasts with the
coupled players scenario (see Sec.~\ref{sec:no_nash}).  

However just like in the coupled players scenario, in general the
distribution $P(x \mid {\mathscr{I}})$ need not be a Nash equilibrium,
even if the players are all fully rational. This can be the case even
if the players all agree exactly on the counter-factual game, and if
that agreed game is one in which everyone is perfectly rational. This
non-Nash result can even hold if our inference $P(x \mid
{\mathscr{I}})$ is a delta function (so that $P(q \mid {\mathscr{I}})$
is as well), one that exactly describes the actual joint move of the
players in the real (not the counter-factual game). So unlike the
coupled players scenario, there is no issue with how we, the external
scientists go about our inference; our inference is in fact exactly
correct.

Stated differently, assume the following:
\begin{enumerate}

\item The players all share the exact same ``common knowledge'',
namely that they all perform perfectly rationally;

\item They all perform perfectly rationally for that common knowledge
(so that common knowledge is in fact correct);

\item As a result they definitely make a particular joint move $x$
(i.e., their joint mixed strategy is actually a joint pure strategy).
\end{enumerate}

Then it may still be that that $x$ is not a Nash equilibrium. This is
illustrated in the first example.

$ $

\noindent {\bf{Example 1:}} As an example, say that all players agree
on the counterfactual game, and it's a game in which the players all
play perfectly rationally, i.e., the $b^i_j$ are all infinite. Also
have each player be perfectly rational, i.e, have all $\beta_i$ be
infinite. Say that the game has two non-exchangeable pure strategy
Nash equilibria, $x^*(1)$ and $x^*(2)$.

Evaluating, $U^i_c(x_i)$ for this scenario is the expected payoff to
player $i$ if she makes move $x_i$, and if the distribution of other
players' moves, $P(x_{-i} \mid {\mathscr{I}}_c)$, is given by the
uniform average of $\delta_{x_{-i}, x^*_{-i}(1)}$ and $\delta_{x_{-i},
x^*_{-i}(2)}$.{\footnote{See Eq.~\ref{eq:post_others}. Since the two
Nash equilibria are both pure strategies, they have the same entropy
(zero). This means they have the same value of their prior
probability, and therefore the same posterior probability in the
counterfactual game.}}  Now since player $i$ is perfectly rational
(for the counterfactual game), she will play a mixed strategy that is
payoff-maximizing for this environment, $U^i_c(x_i)$. More precisely,
her distribution $P(q_i \mid {\mathscr{I}})$ has its support
restricted to such mixed strategies $q_i$.{\footnote{As usual, the
relative probabilities of those $q_i$ will be given by (the
appropriate exponential of) their entropies
(cf. Eq.~\ref{eq:ind_post}), and the distribution over her moves that
we estimate, $P(x_i \mid {\mathscr{I}})$, is given by
Eq.~\ref{eq:x_post_ind} for this $P(q_i
\mid {\mathscr{I}})$.}} However that environment
is $not$ the one that arises if the other players are all playing
optimally, i.e., it equals neither the environment of player $i$ for
the Nash equilibrium $u(x_i, x^*_{-i}(1))$ nor the environment for the
Nash equilibrium $u(x_i, x^*_{-i}(2))$. Accordingly, in general the
optimal $q_i$ for the counterfactual game --- the mixed strategy
played by player $i$ --- is neither of the two associated Nash
equilibrium pure strategies, $\delta_{x_i, x^*_i(1)}$ nor
$\delta_{x_i, x^*_i(2)}$.

To illustrate this, say that player $i$ has three possible moves. Have
the payoff to player $i$ for those three moves, $x^*_i(1), x^*_i(2)$
and $x^*_i(3)$ be given by the vector $(10, 0, 9)$ when the other
players collectively make Nash move $x^*_{-i}(1)$. Have those payoffs
be $(0, 10, 9)$ when the other players collectively make Nash move
$x^*_{-i}(2)$. (So $x^*_i(1)$ is indeed best-response for
$x^*_{-i}(1)$ and $x^*_i(2)$ is best-response for $x^*_{-i}(2)$.)
However the distribution over the other players' moves that $i$
considers is 
\begin{eqnarray}
\frac{\delta_{x_{-i}, x^*_{-i}(1)} + \delta_{x_{-i},x^*_{-i}(2)}}{2}.
\end{eqnarray}
The best response mixed strategy player $i$ can play for this
distribution is $\delta_{x_i, x^*_i(3)}$, for which the expected
payoff is 9. (The expected payoff for the other two pure strategies
are both 5.) This is neither of the two original game Nash equilibrium
moves for player $i$, which establishes the claim.

So even the outcome of the pure rationality independent players
scenario need not be a Nash equilibrium of the original game. This is
quite reasonable. After all, unless there's collusion or some form of
(knowing) interaction between the players in the past (even if
mediated by intermediaries, e.g., via a social norm), then there's no
way they can coordinate.  Intuitively, each player $i$ must ``hedge
her bets''. She presumes that the other players will be playing a Nash
equilibrium, but since there is more than one such equilibrium, each
with a non-zero probability, she must take both into account in
choosing her move. This means that her move will not be optimal for
either one of the Nash equilibria considered by itself.{\footnote{A
similar phenomenon occurs in simply single-dimensional decision
theory. Under quadratic loss, if $P(z)$ is the actual distribution of
a random variable, the Bayes-optimal prediction --- the prediction
that minimizes expected loss under that $P$ --- is $y = E_P(z)$. That
expectation may even be a point where there is zero probability mass,
i.e., it may be that $P(y) = 0$.}}  This contrasts with the type of
situation that would prevent us from predicting a Nash equilibrium for
the coupled players scenario. There the difficulty can arise when
$we$, the external scientists making the prediction, are forced to
hedge our bets.

$ $

\noindent {\bf{Example 2:}} Now consider another scenario where again
the players and their counterfactual versions are all fully
rational. In this scenario say there is a single Nash equilibrium in
mixed strategies, an equilibrium under which it $not$ the case that
each player's mixed strategy is uniform over its
support.{\footnote{Some have worried that this scenario calls into
question the validity of the Nash equilibrium concept. The issue is
why a player $i$ should play a particular non-uniform mixed strategy
over its best response pure strategies, when the only ``advantage'' of
that mixed strategy is that it happens to make the mixed strategies of
other players be best-response. See for
example~\cite{hars73,hase88,auma90}.}}

As usual, player $i$ considers the counterfactual game to predict what
the other players are doing. Doing this gives her a set of moves that
she could make, all of which are best-response. Now by symmetry, our
estimate of $i$'s distribution, $P(x_i
\mid {\mathscr{I}})$, is uniform over those best-response moves of hers, and
zero elsewhere.{\footnote{It is interesting to consider this result in
light of experimental and theoretical work concerning risk-dominant
Nash equilibria.}} This uniformity will hold for all
players. Therefore the estimate we make of the joint mixed strategy is
$not$ the Nash equilibrium of the game (under which some players have
mixed strategies that are non-uniform over their support). This does
not mean that we claim that the Nash equilibrium is impossible. We
assign non-zero $P(q \mid {\mathscr{I}})$ to that Nash equilibrium $q$
in general. It is just that our estimate of the joint mixed strategy
will not be that Nash equilibrium.

This contrasts with the coupled players scenario. In that scenario, if
you are explicitly provided $\mathscr{I}$ saying that all players are
perfectly rational, then it is precisely that $\mathscr{I}$ that tells 
you that player $i$ must play the Nash equilibrium non-uniform
distribution. If you are $not$ provided that explicit prior
information, then in fact you should not assume that there is perfect
rationality.

\subsection{The MAP $q$ for independent players}
\label{sec:map_ind}

Since for independent players the posterior is a product distribution,
the MAP $q$ is also. So with some abuse of notation, we can write
\begin{eqnarray}
{\mbox{MAP}}(q) &\triangleq& {\mbox{argmax}}_q P(q \mid {\mathscr{I}})
\nonumber \\
&=&  {\mbox{argmax}}_q \prod_i e^{\alpha S(q_i)} \delta(q_i \cdot U^i_c
- K(U_c^i, \beta_i)) \nonumber \\
&=& \prod_i {\mbox{argmax}}_{q_{_{i}}} P(q_i \mid {\mathscr{I}})
\nonumber \\
&=& \prod_i {\mbox{MAP}}(q_i)
\label{eq:maps_ind}
\end{eqnarray}
where the index variable $x = (x_1, x_2, \ldots)$ is implicit, as is
the conditioning on the independent-players $\mathscr{I}$.  For
notational simplicity define
\begin{equation}
{\mbox{MAP}}(q_i) \; \triangleq \; {\tilde{q}}_i
\end{equation}
for each $i$,  So we can rewrite Eq.~\ref{eq:maps_ind} as ${\tilde{q}}
= \prod_i {\tilde{q}}_i$.

In the usual way, by maximizing entropy subject to the associated
equality constraint, each ${\tilde{q}}_i$ can be written as
$e^{\beta_i U^i_c(x_i)}$ up to an overall proportionality constant.
Recall that in writing ${\tilde{q}}_i$ this way that $\beta_i$ is the
Lagrange parameter enforcing our constraint that $U^i_c \cdot
{\tilde{q}}_i = K_i$, i.e., enforcing our restriction that the
${\tilde{q}}_i$ be ``well-consistent'' with $U^i_c$. Writing it out,
\begin{eqnarray}
U^i_c \cdot {\tilde{q}}_i &=& K(U_c^i, \beta_i) \nonumber \\
&=& \frac{\int dx_i \;  e^{\beta_i U^i_c(x_i)} U^i_c(x_i)}{\int dx_i \;
e^{\beta_i U^i_c(x_i)}}  .
\end{eqnarray}

Given this form for each ${\tilde{q}}_i$, we can write the value of
${\tilde{q}}$ for some arbitrary $x$ as
\begin{equation}
{\tilde{q}}(x) \;\propto\; \prod_i e^{\beta_i U^i_c(x_i)}.
\end{equation}
where the proportionality constant is independent of $x$.
Plugging in, this becomes 
\begin{eqnarray}
{\tilde{q}}(x)  &\propto&  \prod_i exp \bigl[ \beta'_i \int dx_{-i}dq
\; u^i(x_i, x_{-i}) q_{-i}(x_{-i}) 
\prod_{j \ne i} e^{{a^i} S(q_j)}
\delta(q_j \cdot U^j_{q_{_{-j}}} - \epsilon^i_j(U^j_{q_{_{-j}}})) \bigr] , \nonumber \\
&&
\end{eqnarray}
where for simplicity we have absorbed all proportionality constants
into $\beta_i$, writing that new value of $\beta_i$ as $\beta'_i$.

As an example, say that our game has a single Nash equilibrium over
pure strategies, $x^*$. Let the $b^i_j$ (implicit in the
$\epsilon^i_j$) all go to infinity, keeping the ${a^i}$ all finite, in
such a way that the posterior distribution over $q$ for the
counterfactual coupled game approaches a single $q$ which is a delta
function about $x^*$. So $U^i_c(.)$ approaches $u^i(., x^*_{-i})$.
Then $\tilde{q}$ approaches a product of (independent) Boltzmann
distributions:
\begin{equation}
{\tilde{q}}(x) \;\propto\; \prod_i e^{\beta_i u^i(x_i, x^*_{-i})} .
\label{eq:map_ind_ex}
\end{equation}
This is a product of mixed strategies, each of the form of the
Boltzmann distribution. As such it is similar to the QRE. Unlike the
QRE though, there is no coupling between the different mixed
strategies comprising $\tilde{q}$. This reflects the fact that, by
hypothesis, the players are independent of each other in how they form
their mixed strategies, as well as in the subsequent moves they
make. Whenever there is such independence --- which is the case in
much of conventional noncooperative game theory, implicitly or
otherwise --- the QRE is not an appropriate choice for what kind of
product of Boltzmann distributions to use to capture bounded
rationality.

Now say the counterfactual game has two pure strategy Nash equilibria,
$x^*(1)$ and $x^*(2)$, and that in evaluating the counterfactual
game agent $i$ gives them probabilities $c_i$ and $1 - c_i$,
respectively. Then rather than Eq.~\ref{eq:map_ind_ex}, we get
\begin{equation}
{\tilde{q}}(x) \;\propto\; [\prod_i e^{\beta_i c_i u^i(x_i, x^*_{-i}(1))}]
\times [\prod_i e^{\beta_i (1 - c_i)u^i(x_i, x^*_{-i}(2))}] ,
\end{equation}
i.e., a product of the kind of equilibria arising for the two Nash
equilibria taken separately. If $\beta_i \rightarrow \infty$, then
agent $i$ chooses the best response to either $x^*_{-i}(1)$ or
$x^*_{-i}(2)$, depending on which gives $i$ higher expected payoff
(where the expectation is evaluated according to the distribution
$(c_i, 1-c_i)$).

\section{Miscellaneous topics}
This section presents some illustrative extensions of the basic PGT
framework presented above.

\subsection{Cost of computation}

For a large range of games, the independent players scenario results
in a tradeoff between how smart a player is and the cost of the
computation they must engage in to determine their behavior. This
relation between the cost of computation and bounded rationality
emerges from the mathematics; it is not some {\it{ad hoc}} hypothesis
we make to explain the observed (bounded rational) behavior of real
human beings. In addition, using this mathematics, we can quantify the
tradeoff and when it occurs, and more generally determine what
characteristics of the game are most intimately related to the
tradeoff. (All of that analysis is the subject of future work.)

Say $\beta_i$ increases while all other parameters are fixed, so
$U^i_c$ doesn't change. Then the set of $q_i$ satisfying our invariant
$\mathscr{I}$ shifts (cf. Sec.~\ref{sec:eff_inv}). Typically such
shifts in that set arising from increases in $\beta_i$ also shrink
that set (i.e., its measure decreases).  Intuitively, the smarter
player $i$ is (for the counterfactual game), the more assured it is in
assessment of the counterfactual game, and therefore the more assured
it is in making its move.  As an example, say that ${a^i}$ and the
values $\{b^i_j\}$ restrict $P(q \mid {\mathscr{I}}_c)$ to one $q$
that is a Nash equilibrium of the game, an equilibrium which is a
joint pure strategy of the players. So $P(x_{-i} \mid
{\mathscr{I}}_c)$ is a delta function about the moves of the players
other than $i$ at that Nash equilibrium. Then for $\beta_i
\rightarrow \infty$, $q_i$ also becomes restricted to that equilibrium,
i.e., the support of the likelihood, $P({\mathscr{I}} \mid q_i)$ gets
restricted to a single $q_i$ (one that is a delta function about that
Nash equilibrium's $x_i$). Accordingly the measure of $q_i$ allowed by
the likelihood goes to 0 as $\beta_i$ approaches infinity.

When the set of $q_i$ allowed by the likelihood shrinks this way, the
set of $q_i$ allowed by the associated posterior, $P(q_i \mid
{\mathscr{I}})$ (i.e., the set of $q_i$ in the support of that
posterior) must also shrink.  Typically this mean that the entropy of
that posterior shrinks.  Usually this in turn means that the integral
of that posterior, $P(x_i \mid {\mathscr{I}}) = \int dq_i \; q_i(x_i)
P(q_i \mid {\mathscr{I}})$, also get a smaller entropy as $\beta_i$
increases.  We can illustrate this by returning to our single pure
strategy Nash equilibrium example. In that example, for $\beta_i
\rightarrow \infty$, the support of $P(x_{i} \mid {\mathscr{I}})$ gets restricted
to the Nash equilibrium $x_{i}$, and therefore its entropy goes to
zero, the smallest possible value. As another example, recall from
Sec.~\ref{sec:eff_inv} that since we have fixed $U^i_c$, the entropy
of the MAP $q_i$ cannot increase as $\beta_i$ increases.

In such situations, all these distributions with decreasing entropy
have more and more information as $\beta_i$ increases (recall that the
amount of information in a distribution is the negative of its
entropy). Now model agent $i$'s computational process (in deciding how
to move) as starting with the assumption that ${a^i},
\{{b^i_j}\}$ accurately describes the other agents, so that the
associated counterfactual game results in an accurate approximation of
$U^i$. Under this model, we can interpret the amount of information
in $P(x_{i} \mid {\mathscr{I}})$ as the amount of ``computational
effort'' $i$ expends to try to approximate $P(x_{-i} \mid
{\mathscr{I}}_c)$ accurately and guess accordingly.

As just argued, typically that amount of information in $P(x_{i} \mid
{\mathscr{I}}_c)$ --- the negative of its entropy --- increases as
$\beta_i$ does.  So under this model, the larger $\beta_i$ is, the
more computational effort $i$ expends. On the other hand, assume that
the ${a^i}, \{{b^i_j}\}$ going into $i$'s counterfactual game
calculation give an accurate approximation to the actual $U^i$.  In
this case, the expected payoff to $i$ rises as $\beta_i$ does. So when
the ${a^i}, \{{b^i_j}\}$ give an accurate approximation to $U^i$
(i.e., $i$'s modeling is accurate), rising $\beta_i$ both means more
expected payoff to $i$ and more computational effort by $i$.
Evidently $\beta_i$ controls a tradeoff between how smart $i$ is and
how much computational effort it expends.{\footnote{The analogous
argument for the coupled players scenario is more problematic.  This
is because as $i$ changes her distribution, for example by increasing
her (coupled players value) $b_i$, the distribution of the other
players must also change, due to the coupling between players. This
means that the effect on the entropy of $i$'s distribution and to her
expected payoff can be more complicated.}}

\subsection{Rationality functions}
\label{sec:rat}

In many situations it would be useful to have a way of quantifying the
rationality of a player $i$, based purely on its behavior, without any
model of its decision-making process (even as ill-specified a model as
saying that the player ``evaluates a counterfactual game to some given
degree of accuracy''). We would like to be able to do this for any
mixed strategy $q_i$ and for any environment $U^i$ (whether that mixed
strategy is the choice of player $i$, as in type II games, or instead
governs how $i$ makes choice, as in type I games). We would like
similar generality for judging potential moves $x_i$. 

In particular, we do not want to $require$ that the mixed strategy of
real-world players has some {\it{a priori}}-specified parameterized
form, e.g., a Boltzmann distribution over its environment. We do not
want to assume that our data is a (perhaps noise-corrupted) stochastic
realization of such a mixed strategy, and accordingly solve for the
best-fit values of the associated parameters to some experimental data
(as is done in much of the experimental work involving the QRE, e.g.,
\cite{goho99}).  After all, any requirement that the mixed strategy of
a real-world player is $exactly$ given by such a parametric function
will almost always be in error, at least to a degree.  This section
presents such a broader quantification of rationality.

Consider the situation where players $i$ has mixed strategy $q_i$ and
her environment is some fixed $U^i$. It is reasonable to say that two
choices of $q^i$ are equally rational if they have the same dot
product with $U^i$. However we will often want to do more than simply
say whether two $q_i$ are equally rational for some particular $U^i$;
we will often want to say whether a $q_i$ operating in environment
$U^i$ is more or less rational than a $q'_i$ operating in environment
$(U')^i$. To do this we need a scalar-valued function $R(V, p)$ that
measures how rational an arbitrary distribution $p(y)$ is for an
arbitrary utility function $V(y)$, i.e., that measures how peaked
$p(y)$ is about the maximizers of $V(y)$, argmax$_{y} V(y)$, and about
the other $y$ that have large $V(y)$ values.

Say that $p$ is a Boltzmann distribution over $V(y)$, $p(y) \propto
e^{\beta V(y)}$. Then we can use information theory in general, and
effective invariants and the functions $\epsilon_i$ discussed above in
particular, to motivate quantifying the rationality of $p$ for $V$ as
the value $\beta$. The larger $\beta$ is, the more peaked $p$ is about
the better mixed strategies, and therefore the more ``rational'' $p$
is. 

In addition, so long as $p''$ and $p'$ are Boltzmann distributions for
$V''$ and $V'$ respectively, this measure of the associated $\beta$
value can be used to compare the rationality of $p''$ for $V''$ with
the rationality of $p'$ for $V'$.  We can do this even if the range of
the function $V'$ differs from that of $V''$. This attribute of our
measures differs from other naive choices for measuring
rationality. In particular, it differs from the choice of measuring
rationality as $p \cdot V$, which not only reflects how peaked $p$ is
about $y$ that give large $V(y)$, but also reflects the range of
values of $V(.)$. (Indeed, simply translating the values of $V(.)$ by
a constant will modify the value of this alternative choice of
rationality function.)

In general though $p$ will not be a Boltzmann distribution. So we need
to extend our reasoning, to define an $R$ that we can reasonably view
as a quantifier of rationality for any $p$.  Formally, we make two
requirements of $R$:
\begin{enumerate}

\item If $p(y) \propto e^{\beta V(y)}$, for non-negative $\beta$,
then the peakedness of the distribution --- the value of $R(V, p)$ ---
is $\beta$.

\item Out of all $p$ satisfying $R(V, p) = \beta$, the one that has maximal
entropy is proportional to $e^{-\beta V(y)}$. In other words, we
require that the Boltzmann distribution maximizes entropy subject to a
provided value of the rationality/temperature.
\end{enumerate}
We call any such $R$ a {\bf{rationality function}}. 

Note that a rationality function can be applied to physical systems,
where $V(y)$ is interpreted as the Hamiltonian over microstates
$y$. Such a function is defined even for systems that are not at
physical equilibrium (and therefore aren't described by Boltzmann
distributions). In this, rationality functions are an extension of the
conventional definition of temperature in statistical physics.

As an illustration, a natural choice is to define $R(V, p)$ to be the
$\beta$ of the Boltzmann distribution that ``best fits" $p$. To
formalize this we must quantify how well any given Boltzmann
distribution ``fits'' any given $p$.  Information theory provides many
measures for how well a distribution $p_1$ is fit by a distribution
$p_2$. On such measure is the {\bf{Kullback-Leibler
distance}}~\cite{coth91,duha00,wolp04aa}:
\begin{equation}
KL(p_1 \; || \; p_2) \triangleq S(p_1 \; || \; p_2) - S(p_1)
\end{equation}
\noindent where $S(p_1 \; || \; p_2) \triangleq -\int dy \; p_1(y)
{\mbox{ln}}[\frac{p_2(y)}{\mu(y)}]$ is known as the {\bf{cross
entropy}} from $p_1$ to $p_2$ (and as usual we implicitly choose
uniform $\mu$). 

The KL distance is always non-negative, and equals zero iff its two
arguments are identical. In addition, $KL(\alpha p^1 + (1-\alpha)p^2
\; || \; p^2)$ is an increasing function of $\alpha \in [0.0, 1.0]$,
i.e., as one moves along the line from $p^1$ to $p^2$, the KL distance
from $p^1$ to $p^2$ shrinks.{\footnote{This follows from the fact that
the second derivative with respect to $\alpha$ is non-negative for all
$\alpha$, combined with the fact that KL distance is never negative
and equals 0 when $\alpha = 0$.}} The same is true for $KL(p^2 \; ||
\; \alpha p^1 + (1-\alpha)p^2)$. In addition, those two KL distances
are identical to 2nd order about $\alpha = 0$. However they differ as
one moves away from $\alpha = 0$ in general; KL distance is not a
symmetric function of its arguments. In addition, it does not obey the
triangle inequality, although it obeys a
variant~\cite{coth91}. Despite these shortcomings, it is by far the
most common way to measure the distance between two distributions.

Recall the definition of the partition function, $Z(V) \triangleq \int
dy \; e^{V(y)}$ (the normalization constant for the distribution
proportional to $e^{V(y)}$).  Using the KL distance and this
definition, we arrive at the rationality function
\begin{eqnarray}
R_{KL}(V, p) &\triangleq& {\mbox{argmin}}_\beta KL(p \; || \; \frac{e^{\beta
V}}{Z(\beta V)}) \nonumber \\
&=& {\mbox{argmin}}_\beta [-\beta \int dy\; p(y) V(y) + {\mbox
{ln}}(Z(\beta V)) - S(p)] \nonumber \\
&=& {\mbox{argmax}}_\beta [\beta \int dy\; p(y) V(y) - {\mbox
{ln}}(Z(\beta V))] .
\end{eqnarray}
\noindent In~\cite{wolp04a} it is proven that $R_{KL}$ respects the two
requirements of rationality functions.  Note that the argument of the
argmin is globally convex (as a function of the minimizing variable
$\beta$). In addition its second derivative is given by the variance
(over $y$) of the Boltzmann distribution $e^{\beta V(y)} / Z(\beta
V)$. This typically makes numerical evaluation of $R_{KL}$ quite fast.

Comparing the definition of $R_{KL}$ to Eq.~\ref{eq:free_utility}, we
see that the KL rationality of a distribution $p$ is just the value of
$\beta$ for which $p$ has minimal free utility gap.  When $p$ is a
Boltzmann distribution over the states of a statistical physics
systems, this $\beta$ is (the reciprocal of) what is called
temperature in the in App.~\ref{sec:physics}. Systems described by
such distributions are at physical equilibrium. In other words, the
physical temperature of a physical system at physical equilibrium is
(the reciprocal of) its KL rationality. KL rationality is also defined
for off-equilibrium systems however, unlike physical temperature.

%

To help understand the intuitive meaning of the KL rationality
function, consider fixing its value for agent $i$ to some value
$\rho_i$. Say $q_{-i}$ is also fixed (and therefore so is player $i$'s
environment, $U^i_{q_{-i}}$).  Then there is a value $a_i$ such that
the set of all $q_i$ having rationality value $\rho_i$ is identical to
the set of all $q_i$ for which $E_{q_i}(U^i_{q_{-i}}) = a_i$.  In
fact, $a_i$ is the expected value of $U^i$ that would arise if
$q_i(x_i)$ were a Boltzmann distribution (over $U^i_{q_{-i}}(x_i)$
values) with Boltzmann exponent $\beta_i =
\rho_i$.{\footnote{To see all this, note that by definition of KL
rationality function, 
\begin{eqnarray*}
-\frac{ \partial {\mbox{ln}} (Z(\beta U^i_{q_{-i}}))} {\partial
\beta}|_{\beta = R_{KL}(U^i_{q_{-i}}, q_i)} = \int dx_i \; q_i(x_i)
U^i_{q_{-i}}(x_i).
\end{eqnarray*}
However by the discussion in Sec.~\ref{sec:eff_inv}, we know that the
quantity on the left-hand side is just the Boltzmann utility evaluated
at the specified value of $\beta$, $K_i(\beta)$. So
$R_{KL}(U^i_{q_{-i}}, q_i) = R_{KL}(U^i_{q_{-i}}, q'_i)$ $\Rightarrow$
$K( R_{KL}(U^i_{q_{-i}}, q_i)) = K( R_{KL}(U^i_{q_{-i}}, q'_i))$
$\Leftrightarrow$ $E_{q_i}(U^i_{q_{-i}}) = E_{q'_i}(U^i_{q_{-i}})$. So any
two $q_i$'s with the same rationality must have the same expected
$U^i_{q_{-i}}$. To prove the other direction, recall that for fixed
$U^i_{q_{-i}}$, the Boltzmann utility is a bijection from values of
$\beta$ into $\mathbb{R}$. {\bf{QED.}}.}}

So knowing that player $i$ has KL rationality $\rho_i$ is equivalent
to knowing that the actual expected value of $U^i$ under $q_i$ equals
the ``ideal expected value'', in which $q_i$ is replaced by the
Boltzmann distribution over $U^i_{q_{-i}}(x_i)$ values with exponent
$\beta_i = \rho_i$. (However note that such a constraint on the value
of $\rho_i$ does not actually specify $q_{-i}$, so it does
not specify that ideal expected value of $U^i$.) The (loose) physical
analog of this result is that all distributions over states of a
physical system having the same (potentially non-equilibrium)
temperature also have the same expected value of the Hamiltonian.

Comparing with the discussion in Sec.~\ref{sec:coupled-players}, we
see that specifying the KL rationalities of all the players is exactly
the same as specifying that they all obey the coupled players
invariant, with the parameters of the functions $\epsilon_i$ given by
those specified rationality values. An $\mathscr{I}$ specifying the
one scenario is identical to an $\mathscr{I}$ specifying the other
one. Accordingly, all the discussion in
Sec.'s~\ref{sec:rat_invs},~\ref{sec:alt-epsilon_i} holds for making
predictions based on specified rationalities of the players. In
particular, as discussed in Sec.~\ref{sec:rat_invs}, the rationalities
of the players in a game reflects the structure of that game, as much
as it reflects the intrinsic characteristics of the players.

All of the foregoing was for quantifying the rationality of a
particular $q_i$. However we can view the rationality of a particular
$x_i$ as a special case, where the ``mixed strategy'' $q_i$ is a delta
function about one of its moves. ($behaviorally$, it makes no
difference if that $x_i$ is a sample of some preceding $q_i$ that $i$
chose, or instead is $i$'s choice directly.) Plugging that in to the
KL rationality function, we get the following definition of the
rationality of a move $x_i$:

Say that a player $i$ makes move $x_i$ when there is an environment
$U^i$.  Then the KL rationality of that move is the $\beta$ such that
if $i$ had instead chosen a Boltzmann mixed strategy with exponent
$\beta$, the resultant expected value of $u^i$ would have been the
same as $i$'s actual expected utility. Formally, the KL rationality
function is the mapping from $(x_i, U^i)$ to the $\beta$ such that
\begin{equation}
U^i(x_i) \;=\; \frac{\int dx'_i U^i(x'_i) e^{\beta U^i(x'_i)}} {\int
dx'_i e^{\beta U^i(x'_i)}}.
\end{equation}

\subsection{Variable numbers of players}

There are many statistical ensembles considered in statistical physics
in addition to the CE. In particular, in the Grand Canonical Ensemble
(GCE), the numbers of the particles of various types in the system is
itself a stochastic quantity, in addition to the states of those
particles. This is how one analyzes the statistics of physical systems
involving chemical and/or particle physics interactions that change
the particles of the system.

Recall that the CE can most cleanly be derived as an MAP distribution
with an entropic prior and an appropriate expectation value constraint
(App.~\ref{sec:physics}). The GCE can be derived the same way. Whereas
with the CE the expectation value constraint only concerns the
expected energy, in the GCE it also concerns the expected numbers of
particles of the various possible types \cite{jayn57}.

As pointed out in \cite{wolp03b,wolp04a,wolp04c}, the same same
approach used in the GCE can also be applied in a game theory
context. In such a context, rather than ``particles of various
types'', one has ``players of various types''. Broadly speaking, after
this substitution, the ensuing analysis for the game theory context
proceeds analogously to that of the statistical physics context.

To illustrate this we present a game theory scenario that roughly
parallels the GCE.{\footnote{One difference is that the GCE allows
arbitrary statistical coupling between all variables. In contrast,
here we impose numerous statistical independences among the variables,
e.g., statistical independence between the moves of the
players. Another difference is that there are multiple utility
functions in games, whereas there is only analogous quantity (the
Hamiltonian) in physical systems. This makes the formulas here more
complicated than those in the GCE.} We postulate some pre-fixed set of
{\bf{player types}}. All players of a given type have the same move
space and the same payoff function. At the beginning of each instance
of our scenario, a set of players is randomly chosen, and each is
assigned a rationality value randomly. Those players are then coupled
as discussed above in Sec.~\ref{sec:coupled-players}, e.g., via a
sequence of noncooperative games, and the instance ends with all of
the players making a move.

We know that the expected number of players of any one of the player
types is the same from one instance to the next, although we do not
necessarily know that expectation value. We similarly assume the
expected rationality for each player type (i.e., the expect value of
$b_i$, in the terminology of Sec.~\ref{sec:epsilon}) is the same from
one instance to the next, without necessarily knowing those
rationality values. These rationality values are statistically
independent from each other.

We formalize this with an encoding of our variables into $x$ modeled
on the scheme used to derive the GCE ~\cite{jayn57}. For all player
types $i$, $x_{i}^N$ indicates the number of players of that type. For
all integers $j > 0$, and all player types $i$, $x_{i,j}^M$ indicates
the move of the $j$'th player of type $i$, assuming there is such a
player (i.e., assuming that $j \le x^N_{i}$). The meaning of
$x^M_{i,j}$ for larger $j$ is undefined/irrelevant. Similarly
$x_{i,j}^R$ indicates the rationality of the $j$'th player of type
$i$, assuming there is such a player, and is undefined otherwise.

We write $x^N, x^M$, and $x^R$, respectively to indicate the vector of
all player-type cardinalities, the (countably infinite dimensional)
vector of the moves by all possible players (including those that do
not actually exist), and the (countably infinite dimensional) vector
of the rationalities of all possible players (including those that do
not actually exist). We also write the utility function of the type
$i$ players as $g_i(x^M, x^N)$, where $\forall i, \; g_i(x^M, x^N)$ is
independent of $x^M_{k,j} \; \forall j > x^N_k$.  Finally, we write
${\bar{N}}_i$ and ${\bar{R}}_i$ to indicate the (fixed but potentially
unknown) expected number of players of type $i$ and expected
rationality of those players, respectively.

As in Sec.~\ref{sec:epsilon}, the moves of our players are independent
once the characteristics of the game are fixed (i.e., we are dealing
with a conventional noncooperative game in which the moves are given
by sampling an associated joint mixed strategy). However here the
moves can be statistically dependent on those characteristics. For
example, if the rationality $x^R_{i,j} = 0$ for some $j < x^N_i$, then
we know that $q^M_{i, j}$ must be uniform, independent of the mixed
strategies of the other playres.

Reflecting this, we write
\begin{eqnarray}
q(x) &\triangleq&
\prod_{i,j} \bigl{[} q_{i}^N(x_{i}^N)
q^R_{i,j}(x^R_{i,j}) \prod_{i',j'}
q^M_{i',j'}(x^M_{i',j'} \mid x^N, x^R) \bigr{]} 
\label{eq:var_num_q}
\end{eqnarray}
where the products over $j$ and $j'$ both run from 1 to $\infty$.
When the argument makes clear what the superscript \{$M, N, R$\}
should be, we will sometimes leave that superscript implicit.
Note that in reflection of the statistical coupling of the components
of $x$, $q$ is $not$ a product distribution. So in particular the
entropy of $q$ is not a sum of the entropies of its marginalizations,
as it was above.

Writing it out,
\begin{eqnarray}
q(x^M_{i,j} \mid x^N, x^R) &=& \frac{\int dx' \;
q(x') \delta(x'^M_{i,j} - x^M_{i,j}) \delta(x'^R - x^R) \delta(x'^N -
x^N)} {\int dx' \; q(x') \delta(x'^R - x^R) \delta(x'^N - x^N)} .
\end{eqnarray}
With some abuse of notation, we will write ``$q^M_{i,j}(. \mid x^N,
x^R)$'' to mean the (infinite-dimensional) vector with component
$x^M_{i,j}$ given by $q(x^M_{i,j} \mid x^N, x^R)$.

Our invariant says that each $q^N_i$ must result in an average of
$x^N_i$ that equals ${\bar{N}}_i$, and similarly for each $q^R_i$ and
${\bar{R}}_i$. It also says that once $x^N$ and $x^R$ are fixed, $q^M$
must be the joint mixed strategy appropriate for an associated coupled
players type II game. To write out this latter condition, first define
``$-(i, j)$'' to mean all players other than $(i, j)$ (including
players of type $i$ other than the $j$'th one of that type). Next as
shorthand we will often take the distribution over all agents other
than $(i, j)$ implicit and write
\begin{eqnarray}
U^{i,j}(x^M_{i,j}, x^R, x^N) &\triangleq& U^{i,j}_{q^M_{_{-(i,j)}} (. \mid
x^R, x^N)} (x^M_{i,j}, x^N) \nonumber \\
&\triangleq&  \int dx_{-(i,j)} \; q(x^M_{-(i,j)} \mid x^R, x^N) \;
g^i(x^M_{i,j}, x^M_{-(i,j)}, x^N) \nonumber \\
&&
\end{eqnarray}
where we will write $U^{i,j}(., x^R, x^N)$ to mean the
(infinite-dimensional) vector with component $x^M_{i,j}$ given by
$U^{i,j}(x^M_{i,j}, x^R, x^N)$.  So the coupled players portion of our
invariant says that{\footnote{Unfortunately, even with this abusive
notation, book-keeping in the equations can get messy.}}
\begin{eqnarray}
q^M_{i,j}(. \mid x^R, x^N)  \cdot U^{i,j}(., x^N, x^R) &=&
K(U^{i,j}(., x^N, x^R), \; x^R_{i,j}) \;\;\;\; \forall i,j \nonumber
\\
&&
\end{eqnarray}

Combining these three separate aspects of the invariant and explicitly
expanding in full each instance that a component of $q$ occurs, we get
\begin{eqnarray}
P({\mathscr{I}} \mid q) &\triangleq& 
\prod_i \; {\bigl{[}} \; \delta({\bar{N}}_i - \int dx^N_i q^N(x^N_i) x^N_i) \;
\delta({\bar{R}}_i - \int dx^R_i q^R(x^R_i) x^R_i) \;\; \times 
\nonumber \\
&& \;\;\;\;\;\;\;\; \prod_{j} \int dx^N dx^R \; q^N(x^N) q^R(x^R) \; \times
\nonumber \\
&& \;\;\;\;\;\;\;\;\;\;\;\;\;\;\;\;\;\;\delta( \; q^M_{i,j}(. \mid x^R, x^N)
\cdot U^{i,j}_{q^M_{_{-(i,j)}}(. \mid x^R, x^N)}(., x^N) - \nonumber \\
&& \;\;\;\;\;\;\;\;\;\;\;\;\;\;\;\;\;\;\;\;\;\;\;\;\;\;\;\;\;\;\;\;\;\;\;\;
K(U^{i,j}_{q^M_{_{-(i,j)}}(. \mid x^R, x^N)}(., x^N), \;
x^R_{i,j}) \; ) \; {\bigr{]}}. \nonumber \\
&&
\label{eq:like_var}
\end{eqnarray}

We then combine this likelihood with an entropic prior over $q$. This
gives us the posterior $P(q \mid {\mathscr{I}})$. As usual, if we wish
to we can consider the MAP $q$ according to this posterior, various
Bayes-optimal $q$'s according to this posterior, etc., thereby getting 
a single distribution over $x$'s.

Again just like in the usual analysis, as an alternative to these
distributions, over $x$'s we can simply write $P(x \mid {\mathscr{I}})$
directly, getting the same answer as the Bayes-optimal $q$ under
quadratic loss:
\begin{eqnarray}
P(x \mid {\mathscr{I}}) &=& \int dq \; P(q \mid {\mathscr{I}}) q(x).
\end{eqnarray}
To evaluate this integral we must use Eq.~\ref{eq:var_num_q} to plug
in for $q(x)$, Eq.~\ref{eq:like_var} for the likelihood, and then use the
usual entropic prior. Also as usual we must be careful to calculate
the normalization constant for the posterior $ P(q \mid
{\mathscr{I}})$ and divide that into the product of the likelihood and
prior.

However arrived at, once we get a distribution over $x$, we can then
marginalize over various components of $x$ to get distributions over
the associated quantities of interest. For example, we can do this to
determine the typical move of a player of a particular type, the
typical number of players of some type conditional on a particular
move made by the first player of that type, etc..

\section{Discussion and Future Work}

It is worth comparing PGT to approaches based on models of actual
humans beings, like those using models of agent learning
~\cite{fule98} or models incorporating the mathematical structure of
statistical physics~\cite{durl99,brdu01}. Broadly speaking, PGT's
motivation is more like that of conventional game theory than that of
model-based approaches. Like conventional game theory, PGT
investigates what can be gleaned by careful consideration of the
abstract problem of interacting goal-directed agents, before the
introduction of experiment-based insight concerning the behavior of
those agents.

An even closer analogy to PGT's motivation than that provided
by conventional game theory is Bayesian statistics, and especially
Bayesian statistics using invariance-based arguments to set the
prior~\cite{jabr03}. Like such Bayesian statistics, PGT is a
first-principles-driven derivation of a framework for analyzing
systems, a framework into which one can ``slot in'' any kind of
experimental data as it becomes available.''

While the extraordinary success of statistical physics has been used
to choose the entropic prior for this paper, it is important to
emphasize that many other priors can also be motivated using
first-principles arguments, many of them also based on
information-theoretic arguments. Similarly, many other choices of
likelihood (the invariant) can be motivated (as discussed above). PGT
is $not$ restricted to the prior and likelihood considered in this
paper, any more than conventional game theory is restricted to some
particular refinement of the Nash equilibrium concept. The defining
characteristics of PGT is the application of such priors and
likelihoods to game outcomes rather than (or in addition to) within
games. The prior and likelihoods considered here are simply the
examples worked out in this initial paper.

Obviously, if you happen to know what algorithm the players are using,
then that should be reflected in the likelihood. PGT for various
simple choices of such algorithms/likelihoods is the subject of future
work. More generally, humans have lots of cognitive quirks presumably
arising due to evolution. Accordingly the precise priors and
likelihood investigated here may work best for computational agents
involved in a game with no foreknowledge of the game. Important future
work involves analysis with other priors and likelihoods incorporating
behavioral economics results, prospect theory, etc.. These
alternatives can be used for the external scientist's assessment of
the individual players and/or (in the independent players scenario)
for the ``models'' the players have of each other. 

Indeed, PGT can be seamlessly extended to encompass other kinds of
$\mathscr{I}$, even kinds that do not involve utility functions. In
particular, one or more observed samples of a mixed strategy $q_i$ can
naturally be incorporated into the likelihood term, $P(q_i \mid
{\mathscr{I}})$. As another example, we can remove from $\mathscr{I}$
the stipulation that our players' choices of pure strategy are
independent of one another, i.e., the stipulation that we use a
product distribution. Doing so naturally results in correlated moves
among the players, without any need for carefully designed ansatz's
like those behind correlated equilibria \cite{auma87}.

Similarly, there is a good deal of empirical evidence that human
players do not prefer to maximize expected utility functions $\int
dx_i q_i(x_i) U^i(x_i)$. Rather a long line of experiments starting
with Allais' paradox \cite{alla53} indicate that what is invariant in
the decision-making of a human $i$ is some non-linear functional of
its mixed strategy $q_i$. As more gets understood about such
psychological phenomena \cite{liha05} it should be straightforward to
incorporate that understanding into (Bayesian) PGT. One simply changes
what is considered invariant from one instance of the inference
problem to the next, from being a linear functional of $q_i$ to being
some other type of functional.

Related future work will integrate behavior modeling (``user
modeling'', belief nets, etc.)  with PGT, to get an empirical science
of human interactions. Such behavior modeling can run the gamut from
knowledge concerning humans in general (e.g., behavioral economics) to
knowledge concerning certain particular humans (psychological
profiling, and in particular ``games against nature'', i.e., the
decision-making belief net of a particular human, in a non-game theory
context~\cite{kuho05}).

In addition to the foregoring, there is a huge amount of future work
in PGT that carries over from conventional game theory. At the risk of
being glib, almost every aspect of conventional game theory can be
re-analyzed using PGT. This includes in particular cooperative game
theory, in which context PGT should cut the Gordian knot of what
equilibrium concept to adopt. Other broad topics that should be
investigated using PGT --- and therefore bounded rationality --- are
mechanism design, folk theorems, and signaling theory. It may also
prove profitable to have such investigations be extended to allow
varying number of players. Similarly, ``bounded rational''
evolutionary game theory, in particular for finite numbers of agents,
can be investigated using the ``GCE'' (variable number of players)
variant of PGT illustrated above. All of this is in addition to more
circumscribed game theory issues, like different types of
noncooperative games (Bayesian games, correlated equilibrium games,
differential games, etc.).

Other future work involves completing the analysis of the relationship
between QRE and the coupled players MAP (and Bayes-optimal)
$q$'s. This can also be extended to the independent
players. Similarly, coverage issues like those presented in Prop.'s 1
and 2 for the coupled players scenario bears investigating for the
independent players scenario.


Other future work will investigate what happens in the variable number
of players scenario if the random variable of the number of player of
type $i$ is not independent of the random variable of the total
utility accrued by all players of that type. One aspect of such an
investigation would see what happens if that random variable is
statistically coupled to ${\bar{U}}^i / x^N_i$, the average, of
players of type $i$, of the expected utility of those players. In
particular, it is interesting to see what happens if that variable is
coupled to $\frac{{\bar{U}}^i}{x^N_i \sum_j {\bar{U}}^j}$, the ratio
of total expected utility that is earned by players of type $i$,
divided out among those players.

All of this is in addition to the future work mentioned in the
preceding sections.

%
%

\section{Appendix 1 --- Historical context of PGT}

Despite its widespread and profound usefulness in other fields,
attempts to use Shannon entropy in game theory, psychology, and
economics has proven controversial (see for example~\cite{luce03} and
references therein). By and large though those attempts have
considered Shannon entropy as a physical quantity occuring
{\it{within}} the system under study, and then tried to relate that
physical quantity to other aspects of the system. In contrast, where
Shannon entropy has proven so successful in statistical physics,
statistics, signal processing, etc., is in guiding the external
scientist in his {\it{inference about}} the system under study. It is
in this latter sense that Shannon entropy is used in PGT.

The results in~\cite{wolp03b,wolp04a,wolp04c} can be viewed as the
first derivation of bounded rational equilibria using full
probabilistic reasoning.  (The arguments in \cite{mcpa95} concerned
equilibrium concepts rather than distributions over the space of all
possible mixed strategies.)  It should be noted though that the maxent
Lagrangian has a history far predating both the work
in~\cite{wolp03b,wolp04a,wolp04c} and that in \cite{mcpa95}. As the
free energy of the CE it has been explored in
statistical physics for well over a century. Indeed, the QRE is
essentially identical to the ``mean field approximation'' of
statistical physics. (See also~\cite{durl99}.)

In the context of game theory, the maxent Lagrangian was given an
{\it{ad hoc}} justification and investigated
in \cite{fukr93,fule93,shar04} and related work. The first attempt to
derive it in that context using first principles reasoning occurred
in \cite{megi76}. 


The use of the Boltzmann distribution mixed strategies also has a long
history in the Reinforcement Learning (RL) community, i.e., for the
design of computer algorithms for a player involved in an iterated
game with Nature~\cite{suba98,kali96}. Related work has considered
multiple computational players~\cite{crba96,huwe98a}. In particular,
some of that work has been done in the context of ``mechanism design''
of many computational players, i.e., in the context of designing the
utility functions of the players to induce them to maximize social
welfare~\cite{wotu99a,wotu01a,wotu02a,wolp03a}.  In all of this RL
work the Boltzmann distribution is usually motivated either as an
{\it{a priori}} reasonable way to trade off exploration and
exploitation, as part of Markov Chain Monte Carlo procedure, or by its
asymptotic convergence properties~\cite{wada92}.

The work in~\cite{drya00, aoki04, fash05} in particular, and
econophysics in general, also concerns the relation between
statistical physics and the social sciences. In particular, much of
that work considers the relation betweenn equilibrium distributions of
statistical physics and notions of equilibrium in social science
settings. None of it concerns game theory though. To relate that
domain to statistical physics one must drill deeper into statistical
physics, into its information-theoretic foundations as elaborated by
Jaynes.  The first relatively simple-minded work relating information
theory, statistical physics, and bounded rational game theory this way
was~\cite{wolp04aa}.

\section{Appendix 2 - Using the Entropic Prior to Derive the CE}
\label{sec:physics}
 
This appendix elaborates --- in a very detailed manner --- the
application of the entropic prior to statistical physics that results
in the {\bf{canonical ensemble}} (CE). The level of detail presented
borders on overkill.  However it turns out not to be trivial how to
set the analogous details arising in the application of the entropic
prior to game theory. In addition, the subtleties of how to use the
entropic pror to derive the CE are invariably slighted in the
literature.  Hence first working through the well-understood
statistical physics case can help hone intuition.
 
On the other hand though, it turns out that in the CE, the temperature
$T$ --- our prior knowledge concerning the system --- equals the
Lagrange parameter of a constrained optimization problem, rather than
the value of the constraint associated with that Lagrange
parameter. This is not the case with PGT, and introduces some
subtleties that are mostly absent from PGT. In addition, the PGT
analogue of what in the CE is the system's Hamiltonian function are
the players' utility functions. While there is a single Hamiltonian in
the CE, there are multiple utility functions in noncooperative
games. Associated with this, in PGT there are multiple analogues of
what in the CE is (global) temperature. All of this introduces
complications into PGT that are absent from the CE. Due to all this,
readers already comfortable with the entropic prior and how to apply
it in the CE may want to skip this appendix.

Write the precise microscopic state of a physical system under
consideration as $y$.  So for example, in classical (non-quantum)
statistical physics, $y$ is the set of positions and momenta of all
the particles in the system.  Arguments from physics are typically
invoked to justify a claim that the temperature $T$ of the system
``determines the expected energy of the system'' for the (known)
energy function of the system, $H(y)$.{\footnote{Properly speaking,
$H$ is the system's ``Hamiltonian''.}}

Note that for this conclusion of those arguments to be a falsifiable
statement, expectation values (``expected energy'') must be
meaningful.  So $T$ must be associated with a (falsifiable) physical
distribution over multiple $y$'s, $q(y)$, rather than with a single
$y$.  Typically this distribution arises by not fully specifying the
starting $y$ and by allowing unknown stochastic external influences to
perturb the system between when we acquire the value $T$ and any
subsequent observation of a property of the system. (It is implicitly
required that those external influences do not change the value that a
repeated temperature measurement would give.)  All that is fixed (in
addition to $T$) are some high-level aspects of how the system is set
up, and of how it is opened to the external world.  (For example, it
may be that the identity of the person performing the experiment, how
they physically hold the experimental instruments, etc., fixes those
physical details.) Accordingly, the state $y$ at that subsequent
observation can vary.

So physically, to falsify a prediction of what $q$ is associated with
a particular $T$, we can imagine repeatedly setting up our system in
the way specified and measuring the temperature, then opening it to
(unknown) external influences in the way specified, and after that
recording the resultant state $y$; the associated distribution across
$y$'s is the (falsifiable) $q$.  What we are interested in is the
relation between that $q$ and the measured $T$.
   
The aforementioned ``arguments from physics'' tell us that for any
specification of how the system is set up and then opened, there is
the same single-valued function of the measured $T$ to the expected
energy of the system under $q$. So via that single-valued function, a
particular value of $T$ picks out a unique set of $q$'s (namely those
with the associated expected energy).

Formally then, what we know is that the value of the temperature $T$
uniquely fixes the expected value under $q$ of a measurement of the
system energy $H(y)$, independent of the details of how the system is
set up and then opened to external influences, i.e., $T$ fixes $E_q(H)
\triangleq \int dy H(y) q(y)$. In general, for the same $T$, different choices of how
the system is set up and then opened to external influences will
result in a different one of the possible $q$ consistent with the
$T$-specified value of $E_q(H)$. However we don't know {\it{a priori}}
how the specification of the system's setup and opening chooses among
the set of all $q$ that are all consistent with a particular value of
$E_q(H)$.  So even if the precise value of $E_q(H)$ were given to us,
and even if how the system is set up and then opened were also
specified, {\it{for us}}, it is as though nothing is specified
concerning which of the $q$ consistent with $E_q(H)$ has been picked
out by what we know.  (It is in encapsulating this ignorance of the
distribution across $q$'s that the entropic prior will arise.)
 
Moreover, while for a particular choice of how the system is set up
and then opened up we can ascertain the expected energy of the system
by repeated experiments, often we cannot do this directly from a
$single$ one of those experiments. This means in particular that while
typically we can measure $T$ in such a single experiment, often we do
not know how that value $T$ fixes the expected energy under
$q$.{\footnote{Observationally, that expected energy is defined in
terms of a set of multiple experiments in addition to the current
one. Mathematically, even if we know the Hamiltonian function, often
we cannot evaluate its expected value for a particular $T$.}}  So
observing $T$ does not always allow $us$ to write down the expected
energy, only to know that it has been fixed. In such instances, having
observed $T$, we do not know what set of $q$'s that value of $T$ has
picked out, only that there is some such set.

The invariant ${\mathscr{I}}$ for this situation fixes $T$, and
therefore specifies that the distribution $q(y)$ must lie on a
hyperplane of the form $E_q(H) = h$.  But it does not specify the
value $h$. Nor does it specify anything concerning which $q$ goes with
any particular $h$, i.e., it tells us nothing concerning the
distribution of $q$'s across that hyperplane.  Our inference problem
is to circumvent this handicap: ${\mathscr{I}}$ is the value $T$,
together with the knowledge that it fixes $E_q(H)$, and we must use
this to say something concerning $q$, the quantity we wish to infer.
Formally, we wish to evaluate $P(q \mid {\mathscr{I}})$ for this
choice of ${\mathscr{I}}$.  (Note that this is a distribution across
distributions.)

Note that the distribution $q$ concerns the physical world. So in
particular, it is experimentally falsifiable.  In contrast, a
distribution $P(q \mid {\mathscr{I}})$ reflects $us$ (the researcher),
and our (in)ability to infer $q$ from $T$ and the specification of how
the system is set up and then opened. Although such a perspective is
not required, one can interpret $P(q \mid {\mathscr{I}})$ as a
subjective ``degree of belief'' in the objective (i.e., falsifiable)
distribution $q$. Alternatively, one can view $\mathscr{I}$ as picking
out a set of physical instances of our system that are consistent with
$\mathscr{I}$, and then interpret $P(q
\mid {\mathscr{I}})$ in terms of frequencies of those instances.

For the reasons elucidated above, it makes sense to use an entropic
prior over $q$'s for this ${\mathscr{I}}$.  With such a prior, the MAP
$q$ is the one that maximizes $S(q)$ subject to the constraint $E_q(H)
= h$. We just happen not to know $h$.

This is a constrained optimization problem with unknown constraint
value.  The associated Lagrangian is 
\begin{equation}
{\mathscr{L}}(\beta, q) \triangleq \beta [E_q(H) - h] - S(q) .
\label{eq:free_energy}
\end{equation}
Removing the additive constant $\beta h$ and dividing by the constant
$\beta$ gives $E_q(H) - \frac{S(q)}{\beta}$. This is known in
statistical physics as the {\bf{free energy}} of the system.  $\beta$
is the Lagrange parameter of our constraint. To solve our constrained
optimization problem, $q$ and $\beta$ are jointly set so that the
partial derivatives of ${\mathscr{L}}(\beta, q)$ are all
zero.{\footnote{Throughout this paper the terms in any Lagrangian that
restrict distributions to the unit simplices are implicit. The other
constraint needed for a Euclidean vector to be a valid probability
distribution is that none of its components are negative. This will
not need to be explicitly enforced in the Lagrangian here, since this
constraint is always obeyed for the $q$ optimizing
${\mathscr{L}}(\beta, q)$.}}  The minimizer of the free energy ---
the MAP $q$ --- is given by the Boltzmann distribution,
\begin{equation}
q(y) \propto \exp{(-\beta H(y))}.
\label{eq:statphysex}
\end{equation}
For macroscopically large systems, the posterior over $q$ is in
essence a delta function about the MAP solution, so the Bayes-optimal
solution for almost any loss function is given by
Eq.~ref{eq:statphysex}. 

$\beta$ turns out to be the (inverse of) the temperature of the
physical system (measured in units where Boltzmann's constant equals
1). In other words, the invariant of our problem is the value of the
Lagrange parameter, not of the associated constraint constant. (The
precise relationship between $\beta$ and $h$ depends on the function
$H$ in general.) 


This scenario and its solution $q$ is exactly the CE
discussed previously. It is the simplest of all scenarios considered
in statistical physics (hence its name).

$ $ 
 
{\bf{ACKNOWLEDGEMENTS:}} I would like to thank George Judge, Duncan
Luce, Herb Gintis, Larry Karp, Dan Friedman, Mike George, Wendy Cho,
Arnold Zellner, and Tom Vincent for helpful discussions.

\bibliographystyle{unsrt}


\begin{thebibliography}{10}

\bibitem{jabr03}
E.~T. Jaynes and G.~Larry Bretthorst.
\newblock {\em Probability Theory : The Logic of Science}.
\newblock Cambridge University Press, 2003.

\bibitem{gull88}
S.~F. Gull.
\newblock Bayesian inductive inference and maximum entropy.
\newblock In {\em Maximum Entropy and Bayesian Methods}, pages 53--74. Kluwer
  Academic Publishers, 1988.

\bibitem{lore90}
T.J. Loredo.
\newblock From laplace to sn 1987a: Bayesian inference in astrophysics.
\newblock In {\em Maximum Entropy and Bayesian Methods}, pages 81--142. Kluwer
  Academic Publishers, 1990.

\bibitem{zell04}
A.~Zellner.
\newblock Some aspects of the history of bayesian information processing.
\newblock {\em Journal of Econometrics}.
\newblock to appear.

\bibitem{wolp95c}
D.H. Wolpert.
\newblock The bootstrap is inconsistent with probability theory.
\newblock In {\em Maximum entropy and Bayesian Methods}. Kluwer publishers,
  1995.

\bibitem{brei96}
L.~Breiman.
\newblock Stacked regression.
\newblock {\em Machine Learning}, 24, 1996.

\bibitem{wolp96c}
D.~H. Wolpert.
\newblock Reconciling {B}ayesian and non-{B}ayesian analysis.
\newblock In {\em Maximum Entropy and {B}ayesian Methods}, pages 79--86. Kluwer
  Academic Publishers, 1996.

\bibitem{coth91}
T.~Cover and J.~Thomas.
\newblock {\em Elements of Information Theory}.
\newblock Wiley-Interscience, New York, 1991.

\bibitem{mack03}
D.~Mackay.
\newblock {\em Information theory, inference, and learning algorithms}.
\newblock Cambridge University Press, 2003.

\bibitem{stwo94}
C.E Strauss, D.H. D.H.~Wolpert, and D.R. Wolf.
\newblock Alpha, evidence, and the entropic prior.
\newblock In A.~Mohammed-Djafari, editor, {\em Maximum Entropy and Bayesian
  Methods 1992}. Kluwer, 1994.

\bibitem{wolp03b}
D.~H. Wolpert.
\newblock Factoring a canonical ensemble.
\newblock 2003.
\newblock preprint cond-mat/0307630.

\bibitem{wolp04b}
D.~H. Wolpert.
\newblock Generalizing mean field theory for distributed optimization and
  control.
\newblock 2004.
\newblock submitted.

\bibitem{mabi04}
W.~Macready, S.~Bieniawski, and D.H. Wolpert.
\newblock Adaptive multi-agent systems for constrained optimization.
\newblock Technical report IC-04-123, 2004.

\bibitem{lewo04b}
C.~Fan Lee and D.~H. Wolpert.
\newblock Product distribution theory for control of multi-agent systems.
\newblock In {\em Proceedings of AAMAS 04}, 2004.

\bibitem{wobi04a}
D.~H. Wolpert and S.~Bieniawski.
\newblock Distributed control by lagrangian steepest descent.
\newblock In {\em Proceedings of CDC 04}, 2004.

\bibitem{biwo04a}
S.~Bieniawski and D.~H. Wolpert.
\newblock Adaptive, distributed control of constrained multi-agent systems.
\newblock In {\em Proceedings of AAMAS 04}, 2004.

\bibitem{biwo04b}
S.~Bieniawski, D.~H. Wolpert, and I.~Kroo.
\newblock Discrete, continuous, and constrained optimization using collectives.
\newblock In {\em Proceedings of 10th AIAA/ISSMO Multidisciplinary Analysis and
  Optimization Conference, Albany, New York}, 2004.

\bibitem{anbi04}
N.~Antoine, S.~Bieniawski, I.~Kroo, and D.~H. Wolpert.
\newblock Fleet assignment using collective intelligence.
\newblock In {\em Proceedings of 42nd Aerospace Sciences Meeting}, 2004.
\newblock AIAA-2004-0622.

\bibitem{jayn57}
E.~T. Jaynes.
\newblock Information theory and statistical mechanics.
\newblock {\em Physical Review}, 106:620, 1957.

\bibitem{auma87}
R.~J. Aumann.
\newblock Correlated equilibrium as an expression of {B}ayesian rationality.
\newblock {\em Econometrica}, 55(1):1--18, 1987.

\bibitem{wolp04c}
D.~H. Wolpert.
\newblock What information theory says about best response, binding contracts,
  and collective intelligence.
\newblock In A.~Namatame et~al, editor, {\em Proceedings of WEHIA04}. Springer
  Verlag, 2004.

\bibitem{goho99}
T.~R.~Palfrey J.~K.~Goeree, C. A.~Holt.
\newblock Quantal response equilibrium and overbidding in private-value
  auctions.
\newblock 1999.

\bibitem{mcpa95}
R.~D. McKelvey and T.~R. Palfrey.
\newblock Quantal response equilibria for normal form games.
\newblock {\em Games and Economic Behavior}, 10:6--38, 1994.

\bibitem{chfr97}
H.~C. Chen and J.~W. Friedman.
\newblock {\em Games and Economic Behavior}, 18:32--54, 1997.

\bibitem{wolp04a}
D.~H. Wolpert.
\newblock Information theory - the bridge connecting bounded rational game
  theory and statistical physics.
\newblock In D.~Braha and Y.~Bar-Yam, editors, {\em Complex Engineering
  Systems}, 2004.

\bibitem{megi76}
J.~R. Meginniss.
\newblock A new class of symmetric utility rules for gambles, subjective
  marginal probability functions, and a generalized bayes rule.
\newblock {\em Proc. of the American Statisticical Association, Business and
  Economics Statistics Section}, pages 471 -- 476, 1976.

\bibitem{fukr93}
D.~Fudenberg and D.~Kreps.
\newblock Learning mixed equilibria.
\newblock {\em Game and Economic Behavior}, 5:320--367, 1993.

\bibitem{fule93}
D.~Fudenberg and D.~K. Levine.
\newblock Steady state learning and {N}ash equilibrium.
\newblock {\em Econometrica}, 61(3):547--573, 1993.

\bibitem{luce59}
D.~Luce.
\newblock {\em Individual Choice Behavior}.
\newblock Wesley, 1959.

\bibitem{wobi04b}
D.~H. Wolpert and S.~Bieniawski.
\newblock Adaptive distributed control: beyond single-instant categorical
  variables.
\newblock In A.~Skowron et~al, editor, {\em Proceedings of MSRAS04}. Springer
  Verlag, 2004.

\bibitem{mawo05}
William Macready and David~H. Wolpert.
\newblock Distributed constrained optimization with semicoordinate
  transformations.
\newblock submitted, 2005.

\bibitem{futi91}
D.~Fudenberg and J.~Tirole.
\newblock {\em Game Theory}.
\newblock MIT Press, Cambridge, MA, 1991.

\bibitem{auha92}
R.J. Aumann and S.~Hart.
\newblock {\em Handbook of Game Theory with Economic Applications}.
\newblock North-Holland Press, 1992.

\bibitem{baol99}
T.~Basar and G.J. Olsder.
\newblock {\em Dynamic Noncooperative Game Theory}.
\newblock Siam, Philadelphia, PA, 1999.
\newblock Second Edition.

\bibitem{binm92}
K.~Binmore.
\newblock {\em Fun and Games: A Text on Game Theory}.
\newblock D. C. Heath and Company, Lexington, MA, 1992.

\bibitem{lura85}
R.~D. Luce and H.~Raiffa.
\newblock {\em Games and Decisions}.
\newblock Dover Press, 1985.

\bibitem{grei99}
A.~Greif.
\newblock Economic history and game theory: A survey.
\newblock In R.~J. Aumann and S.~Hart, editors, {\em Handbook of Game Theory
  with Economic Applications}, volume~3. North Holland, Amsterdam, 1999.

\bibitem{besm00}
J.~Bernardo and A.~Smith.
\newblock {\em Bayesian Theory}.
\newblock Wiley and Sons, 2000.

\bibitem{berg85}
J.~M. Berger.
\newblock {\em Statistical Decision theory and Bayesian Analysis}.
\newblock Springer-Verlag, 1985.

\bibitem{wolp97}
D.~H. Wolpert.
\newblock On bias plus variance.
\newblock {\em Machine Learning}, 9:1211--1244, 1997.

\bibitem{pari94}
J.~B. Paris.
\newblock {\em The Uncertain Reasoner's Companion: A Mathematical Perspective}.
\newblock Cambridge University Press, 1994.

\bibitem{vanh03}
K.~S.~Van Horn.
\newblock Constructing a logic of plausible inference: a guide to cox's
  theorem.
\newblock {\em International Journal of Approximate Reasoning}, 34(1):3--24,
  2003.

\bibitem{came03}
C.F. Camerer.
\newblock {\em Behavioral Game theory: experiments in strategic interaction}.
\newblock Princeton University Press, 2003.

\bibitem{kahn03a}
D.~Kahneman.
\newblock A psychological perspective on economics.
\newblock {\em American Economic Review (Proceedings)}, 93:2:162--168, 2003.

\bibitem{kahn03b}
D.~Kahneman.
\newblock Maps of bounded rationality: Psychology of behavioral economics.
\newblock {\em American Economic Review}, 93:5:1449--1475, 2003.

\bibitem{wolp04aa}
D.~H. Wolpert.
\newblock Bounded rational games, information theory, and statistical physics.
\newblock In D.~Braha and Y.~Bar-Yam, editors, {\em Complex Engineering
  Systems}, 2004.

\bibitem{bova03}
S.~Boyd and L.~Vandenberghe.
\newblock {\em Convex Optimization}.
\newblock Cambridge University Press, 2003.

\bibitem{wotu02c}
D.~Wolpert and K.~Tumer.
\newblock Beyond mechanism design.
\newblock In H.~Gao et~al., editor, {\em International Congress of
  Mathematicians 2002 Proceedings}. Qingdao Publishing, 2002.

\bibitem{wotu01a}
D.~H. Wolpert and K.~Tumer.
\newblock Optimal payoff functions for members of collectives.
\newblock {\em Advances in Complex Systems}, 4(2/3):265--279, 2001.

\bibitem{tuwo03}
K.~Tumer and D.~H. Wolpert, editors.
\newblock {\em Collectives and the Design of Complex Systems}.
\newblock Springer Verlag, 2003.

\bibitem{tvka92}
A.~Tversky and D.~Kahneman.
\newblock Advances in prospect theory: Cumulative representation of
  uncertainty.
\newblock {\em Journal of Risk and Uncertainty}, 5:297--323, 1992.

\bibitem{wolp03a}
D.~H. Wolpert.
\newblock Theory of collective intelligence.
\newblock In K.~Tumer and D.~H. Wolpert, editors, {\em Collectives and the
  Design of Complex Systems}, New York, 2003. Springer.

\bibitem{wotu02a}
D.~H. Wolpert and K.~Tumer.
\newblock Collective intelligence, data routing and braess' paradox.
\newblock {\em Journal of Artificial Intelligence Research}, 2002.

\bibitem{aubr95}
R.~J. Aumann and A.~Brandenburger.
\newblock Epistemic conditions for nash equilibrium.
\newblock {\em Econometrica}, 63(5):1161--1180, 1995.

\bibitem{kala82}
B.~Kadane and P.~Larkey.
\newblock Subject probability and the theory of games.
\newblock {\em Management Science}, 28(2):113--125, 1982.
\newblock Same issue contains discussion of the article.

\bibitem{auma99}
R.~J. Aumann.
\newblock Interactive epistemology ii: Probability.
\newblock {\em Int. J. Game Theory}, 28:301--314, 1999.

\bibitem{hars73}
J.C. Harsanyi.
\newblock Games with randomly distributed payoffs: A new rationale for
  mixed-strategy equilibrium points.
\newblock {\em Int. J. Game Theory}, 2:1--23, 1973.

\bibitem{hase88}
J.C. Harsanyi and R.~Selten.
\newblock {\em A General Theory of Equilibriuum Selection in Games}.
\newblock MIT Press, 1988.

\bibitem{auma90}
R.~J. Aumann.
\newblock {\em Economic Decision Making: Games, Econometrics and Optimization}.
\newblock Elsevier, 1990.
\newblock The chapter ``Nash Equilibra are not Self-Enforcing''.

\bibitem{duha00}
R.~O. Duda, P.~E. Hart, and D.~G. Stork.
\newblock {\em Pattern Classification (2nd ed.)}.
\newblock Wiley and Sons, 2000.

\bibitem{fule98}
D.~Fudenberg and D.~K. Levine.
\newblock {\em The Theory of Learning in Games}.
\newblock MIT Press, Cambridge, MA, 1998.

\bibitem{durl99}
S.~Durlauf.
\newblock How can statistical mechanics contribute to social science?
\newblock {\em Proc. Natl. Acad. Sci. USA}, 96:10582--10584, 1999.

\bibitem{brdu01}
W.~A. Brock and S.~N. Durlauf.
\newblock Interaction-based models.
\newblock In {\em Handbook of Econometrics: Volume 5}, pages 3297--3380.
  Elsevier, 2001.
\newblock Chapter 54.

\bibitem{alla53}
M.~Allais.
\newblock {\em Econometrica}, 21:503--546, 1953.

\bibitem{liha05}
J.~A. List and M.~S. Haigh.
\newblock A simple test of expected utility theory using professional traders.
\newblock {\em Proceedings of the National Academy of Sciences}, 102:945--948,
  2005.

\bibitem{kuho05}
R.~Kurzban and D.~Houser.
\newblock Experiments investigating cooperative types in humans.
\newblock {\em Proceedings of the National Academy of Sciences},
  102(5):1803--1807, 2005.

\bibitem{luce03}
D.~Luce.
\newblock Whatever happened to information thoery in psychology?
\newblock {\em Review of General Psychology}, pages 183--188, 2003.

\bibitem{shar04}
J.S. Shamma and G.~Arslan.
\newblock Dynamic fictitious play, dynamic gradient play, and distributed
  convergence to nash equilibria.
\newblock submitted, 2004.

\bibitem{suba98}
R.~S. Sutton and A.~G. Barto.
\newblock {\em Reinforcement Learning: An Introduction}.
\newblock MIT Press, Cambridge, MA, 1998.

\bibitem{kali96}
L.~P. Kaelbing, M.~L. Littman, and A.~W. Moore.
\newblock Reinforcement learning: A survey.
\newblock {\em Journal of Artificial Intelligence Research}, 4:237--285, 1996.

\bibitem{crba96}
R.~H. Crites and A.~G. Barto.
\newblock Improving elevator performance using reinforcement learning.
\newblock In D.~S. Touretzky, M.~C. Mozer, and M.~E. Hasselmo, editors, {\em
  Advances in Neural Information Processing Systems - 8}, pages 1017--1023. MIT
  Press, 1996.

\bibitem{huwe98a}
J.~Hu and M.~P. Wellman.
\newblock Multiagent reinforcement learning: Theoretical framework and an
  algorithm.
\newblock In {\em Proceedings of the Fifteenth International Conference on
  Machine Learning}, pages 242--250, June 1998.

\bibitem{wotu99a}
D.~H. Wolpert, K.~Tumer, and J.~Frank.
\newblock Using collective intelligence to route internet traffic.
\newblock In {\em Advances in Neural Information Processing Systems - 11},
  pages 952--958. MIT Press, 1999.

\bibitem{wada92}
C.~Watkins and P.~Dayan.
\newblock Q-learning.
\newblock {\em Machine Learning}, 8(3/4):279--292, 1992.

\bibitem{drya00}
A.~Dragulescu and V.M. Yakovenko.
\newblock Statistical mechanics of money.
\newblock {\em Eur. Phys. J. B}, 17:723--729, 2000.

\bibitem{aoki04}
M.~Aoki.
\newblock {\em Modeling Aggregate Behavior and Fluctuations in Economics :
  Stochastic Views of Interacting Agents}.
\newblock Cambridge University Press, 2004.

\bibitem{fash05}
J.D. Farmer, M.~Shubik, and D.~E. Smith.
\newblock Economics: The next physical science?
\newblock SFI working paper 05-06-027.

\end{thebibliography}

\end{document}